\documentclass{sn-jnl}
\usepackage[utf8]{inputenc}
\usepackage{graphicx}
\usepackage{natbib}

\pdfoutput=1

\title{Ohm's Law, the Reconnection Rate, and Energy Conversion in Collisionless Magnetic Reconnection}
\author*[1]{\fnm{Yi-Hsin} \sur{Liu}}\email{Yi-Hsin.Liu@dartmouth.edu}
\author[2]{\fnm{Michael} \sur{Hesse}}
\author[3]{\fnm{Kevin} \sur{Genestreti}}
\author[4]{\fnm{Rumi} \sur{Nakamura}}
\author[5]{\fnm{Jim} \sur{Burch}}
\author[6]{\fnm{Paul A.} \sur{Cassak}}
\author[7,8]{\fnm{Naoki} \sur{Bessho}}
\author*[9]{\fnm{Jonathan P.} \sur{Eastwood}}\email{Jonathan.Eastwood@imperial.ac.uk}
\author[10]{\fnm{Tai} \sur{Phan}}
\author[11]{\fnm{Marc} \sur{Swisdak}}
\author[12]{\fnm{Sergio} \sur{Toledo-Redondo}}
\author[13]{\fnm{Masahiro} \sur{Hoshino}}
\author[14,15]{\fnm{Cecilia} \sur{Norgren}}
\author[16]{\fnm{Hantao} \sur{Ji}}
\author[4,17]{\fnm{Takuma K. M.} \sur{Nakamura}}

\affil[1]{Department of Physics and Astronomy, Dartmouth College, Hanover, New Hampshire 03750, USA}

\affil[2]{Armes Research Center, NASA, Moffett Field, California 94035, USA}

\affil[3]{Southwest Research Institute, Durham, New Hampshire 03824, USA}

\affil[4]{Space Research Institute, Austrian Academy of Sciences, Schmiedlstra{\ss}e 6, 8042 Graz, Austria}

\affil[5]{Southwest Research Institute, San Antonio, Texas 78238, USA}

\affil[6]{Department of Physics and Astronomy and Center for KINETIC Plasma Physics, West Virginia University, Morgantown, West Virginia 26506, USA}

\affil[7]{Goddard Space Flight Center, NASA, Greenbelt, Maryland 20771, USA}

\affil[8]{Department of Astronomy, University of Maryland, College Park, Maryland 20742, USA}

\affil[9]{Department of Physics, Imperial College, London, United Kingdom}

\affil[10]{Space Science Laboratory, UC Berkeley, Berkeley, California 94720, USA}

\affil[11]{IREAP, University of Maryland, College Park, Maryland 20742, USA}

\affil[12]{Department of Electromagnetism and Electronics, University of Murcia, Murcia, Spain}

\affil[13]{Department of Earth and Planetary Science, The University of
Tokyo, Tokyo, 113-0033, Japan}

\affil[14]{Swedish Institute of Space Physics, Uppsala, Sweden}

\affil[15]{Department of Physics and Technology, University of Bergen, Bergen, Norway}

\affil[16]{Department of Astrophysical Sciences, Princeton University, Princeton, New Jersey 08544, USA}

\affil[17]{Krimgen LLC, Hiroshima, 7320828, Japan}

\date{today}

\begin{document}

\abstract{Magnetic reconnection is a ubiquitous plasma process that transforms magnetic energy into particle energy during eruptive events throughout the universe. Reconnection not only converts energy during solar flares and geomagnetic substorms that drive space weather near Earth, but it may also play critical roles in the high energy emissions from the magnetospheres of neutron stars and black holes. In this review article, we focus on collisionless plasmas that are most relevant to reconnection in many space and astrophysical plasmas. Guided by first-principles kinetic simulations and spaceborne in-situ observations, we highlight the most recent progress in understanding this fundamental plasma process. 
We start by discussing the non-ideal electric field in the generalized Ohm's law that breaks the frozen-in flux condition in ideal magnetohydrodynamics and allows magnetic reconnection to occur. We point out that this same reconnection electric field also plays an important role in sustaining the current and pressure in the current sheet and then discuss the determination of its magnitude (i.e., the reconnection rate), based on force balance and energy conservation. This approach to determining the reconnection rate is applied to kinetic current sheets of a wide variety of magnetic geometries, parameters, and background conditions. We also briefly review the key diagnostics and modeling of energy conversion around the reconnection diffusion region, seeking insights from recently developed theories. Finally, future prospects and open questions are discussed.}

\maketitle

\tableofcontents

\section{Introduction}
\begin{figure}[ht]
\centering
\includegraphics[width=7cm]{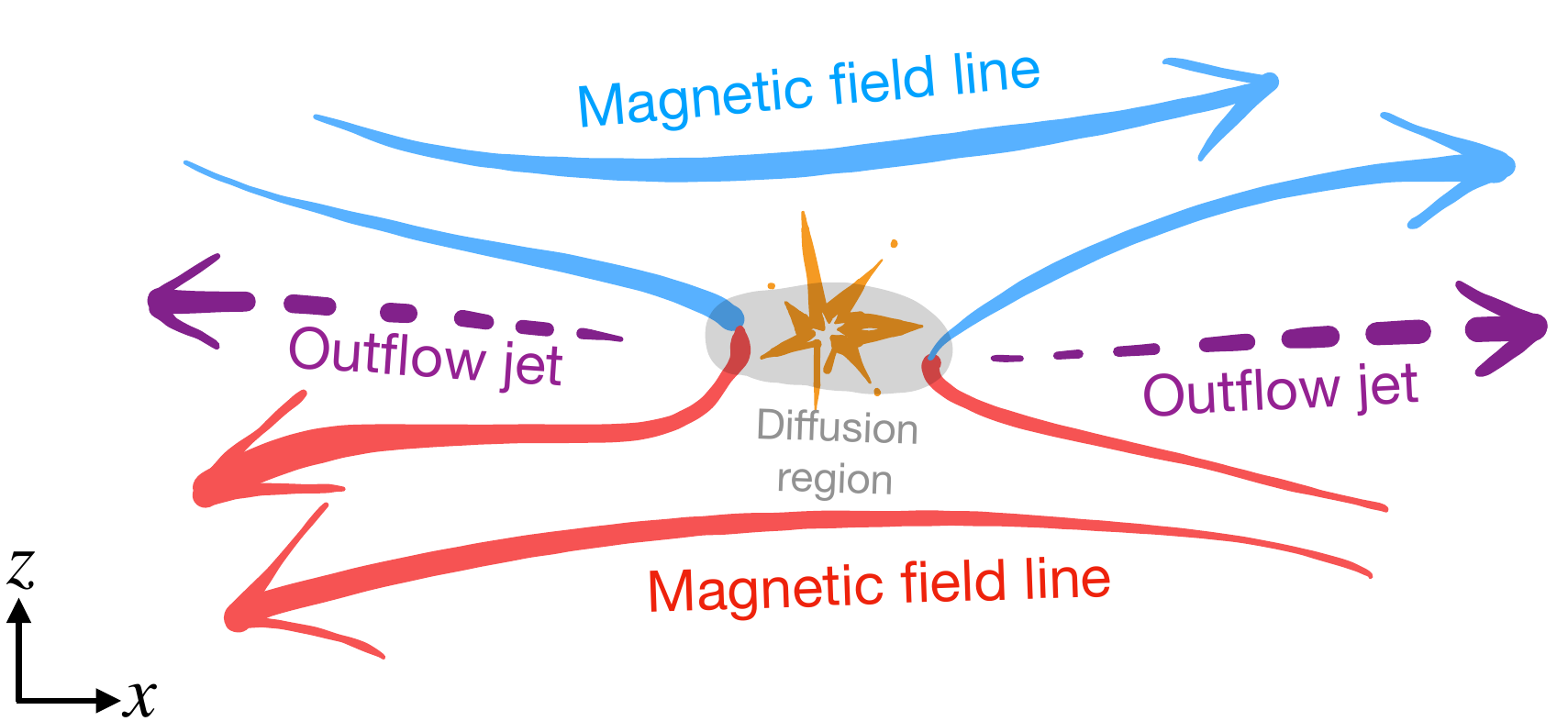} 
\caption {{\bf Artist's rendition of magnetic reconnection}. The breaking and rejoining of magnetic field lines (red and blue) in the diffusion region (gray) drive plasma outflow jets (purple arrows). 
} 
\label{reconnection}
\end{figure}

Magnetic reconnection is a ubiquitous process that converts magnetic energy into plasma thermal and kinetic energy in laboratories, space, and astrophysical plasmas \citep{zweibel09a,yamada2010magnetic}.
This efficient energy conversion process involves the effective ``breaking'' and ``rejoining'' of magnetic field lines (although note that reconnection does not violate Gauss' law, $\nabla\cdot {\bf B}=0$). 
%the terminology ``breaking'' refers to the unique properties of magnetic fields in ideal, i.e., perfectly conducting, plasmas.}. 
By altering their connectivity within the so-called ``diffusion region'' in the microscopic scale, the gray area in Fig.~\ref{reconnection}, this process imparts energy into outflow plasma jets, the purple arrows. While this picture captures the local process, the resulting change in the magnetic connectivity has far-reaching consequences as it can lead to energy release at large scales in the surrounding plasma systems, 
causing solar flares \citep{Carmichael64,Sturrock66,Hirayama74,kopp1976magnetic,Priest00}, planetary geomagnetic substorms \citep{dungey61a}, and superflares from other astrophysical objects, for example the Crab nebula \citep{Tavani2011,abdo11a,cerutti14a}.  

In a nutshell, magnetic reconnection is a nonlinear, dynamical process that involves electromagnetism, magnetic field geometry and topology, and complex charged particle motions in a multi-dimensional, multiscale system, where physics occurring at a singular point can lead to tremendous energy release at the macroscale. 
For these reasons, the study of magnetic reconnection has been a fascinating and challenging subject since it was first formulated in 1953 \citep{dungey53a}. Its study will continue to thrive with our increasing capability to observe electromagnetic phenomena in the universe [e.g., \cite{bale23a,burch16c,raouafi23b,muller20a}].
The development of reconnection theories is guided and constrained by a wealth of data from numerical simulations, in-situ and remote space observations, and laboratory experiments.
We do not intend to exhaustively include the many great efforts performed in 
%the laboratory plasma, space, planetary, and plasma astrophysics 
various communities over the past 70 years in this review paper. Here instead we focus on the progress in the past 20 years on collisionless reconnection, where our understanding has been accelerated by kinetic simulations and in-situ spacecraft observations of NASA's ongoing Magnetospheric Multiscale (MMS) mission \citep{burch16c}, THEMIS/ARTEMIS \citep{angelopoulos08b,sweetser11a}, and Cluster \citep{escoubet01a}. More exciting results are expected from the Parker Solar Probe \citep{raouafi23b} and Solar Orbiter \citep{muller20a} missions, but are not discussed here. It is worth noting that Earth's magnetosphere and the solar wind are the most ideal testing grounds for reconnection physics reachable by human probes with current space technology. Because the size of a single spacecraft is relatively small compared to the electron kinetic scale, and now the cadence of measurement well resolves the dynamic time scale of reconnection therein; see {\color{blue} Genestreti et al (2024)} (this collection) for the review on current sheets in geospace. A companion review of collisionless reconnection research in the laboratory over the past 20 years, in comparisons with kinetic simulations and space observations, is given by \cite{ji23a} (this collection). 

The fundamental questions of reconnection discussed in this review are: (1) what breaks the ideal-magnetohydrodynamic frozen-in flux condition, enabling reconnection to occur on a microscopic/kinetic scale? and what roles does the non-ideal electric field play (Sec.~\ref{Breaking the Flux Frozen-in Condition})? (2) what determines the rate at which reconnection processes the incoming magnetic flux (Sec.~\ref{reconnection_rate_models})? and (3) how plasmas are energized around the reconnection diffusion region (Sec.~\ref{energy_conversion})?
Each topic can be read independently, and we point out connections between different sections.
This article serves as a review but also, hopefully, a tutorial for graduate students and early career scientists.
%As you skim through this article, you will observe that the theoretical discussion of these topics basically follows the electron momentum equation (i.e., Ohm's law), the force-balance equation (i.e., the electron momentum equation $+$ the ion momentum equation), and the energy equation, toward a higher moment integral of the Vlasov equation. In particular, we show that the scaling of reconnection rates in these diverse regimes of interests can be coherently organized by the governing force-balance equation that determines the characteristic Alfv\'enic outflow speed, and also enables us to relate this to the inflow-outflow geometry. 

This review focuses on the fluid-type descriptions of reconnection physics within and around the diffusion region but is based on fully kinetic simulations and in-situ space measurements. %We also show local simulations and in-situ space observations that support some of the theoretical results.
It is not our intention to discuss all the details of each topic, but to integrate them into a bigger picture. Nevertheless, references that contain the full treatment are provided to interested readers.
The discussion of the rich kinetic features and particle distribution functions is beyond the scope of this paper but can be found in {\color{blue} Norgren et al. (2023, this issue)}. For discussions of a broader scope or emphasis on other areas of study, a variety of other papers complement this review \citep{vasyliunas75a,Priest00,birn07a,zweibel09a,mozer10d,yamada2010magnetic,gonzalez16a,burch16c,LCLee20a,ji22a,pontin22a,yamada22a}.

\section{Breaking the Frozen-in Flux Condition}
\label{Breaking the Flux Frozen-in Condition}

Alfv\'en's frozen-in flux theorem \citep{alfven42a} shows that perfectly conducting fluids, such as those in ideal magnetohydrodynamics (MHD), and embedded magnetic fields are constrained to move together. Mathematically, this occurs when ${\bf E} +{\bf  V} \times {\bf B}/c$ vanishes (e.g., \cite{stern66a}) \footnote{more precisely, any velocity $\bf V$ that satisfies $\nabla\times ({\bf E}+{\bf V}\times {\bf B}/c)=0$ can be regarded as the magnetic ``field line velocity'' \citep{vasyliunas72a}.}.  
In a hypothetical plasma for which the frozen-in flux theorem is satisfied, the total magnetic flux going through any close Amp\`erian loop in the plasma does not change in time. Note that the magnetic field self-consistently evolves with the moving plasma, which can generate currents that modify the magnetic fields. If the frozen-in condition works everywhere within the system of interest, the connectivity of magnetic field lines within this system cannot change because doing so would change the flux through a closed loop somewhere within the system.

The field line connectivity, nevertheless, can change when some dissipation breaks the frozen-in condition. For instance, the condition breaks down within the diffusion region (DR) in Fig.~\ref{reconnection} that is sandwiched by magnetic field lines that point in opposite directions. Within this diffusion region, the inflowing magnetic field lines are ``rewired'' to form highly curved (blue-red) field lines, which are again frozen to the plasma outside the diffusion region and act like a slingshot, shooting plasma out as outflow jets. Once the plasma is jetted out, the plasma pressure within the diffusion region drops, and plasma flows in from the top and bottom along with the magnetic field for further reconnection. It is thus a self-driven (i.e., spontaneous) non-linear process; once it starts, it does not want to stop as long as more magnetic field is available in the inflow region. In addition, because of Amp\`ere's law, the anti-parallel fields sandwich a current sheet where the DR resides. The singular point inside the DR where field lines reconnect is referred to as the ``X-point'' because the adjacent reconnected field lines form an X-shape. In 3D, the collection of these X-points extends in the out-of-plane direction to form an ``X-line''.

\subsection{The Generalized Ohm's Law}
\label{generalized_Ohms_law}
Magnetic reconnection is the process that changes the field line connectivity in plasmas, and it requires the existence of a reconnection electric field to break the frozen-in flux condition, either at a topological boundary \citep{vasyliunas75a}, or, more generally, in a localized region parallel to the magnetic field \citep{hesse88a,hesse05b}. This requirement is a simple consequence of Maxwell’s equations and the need to transport magnetic flux from the inflow to the outflow regions. While it has long been shown that reconnection cannot proceed without the presence of such a reconnection electric field, we only recently are understanding its full physical foundations.

\begin{figure}[ht]
\centering
\includegraphics[width=7cm]{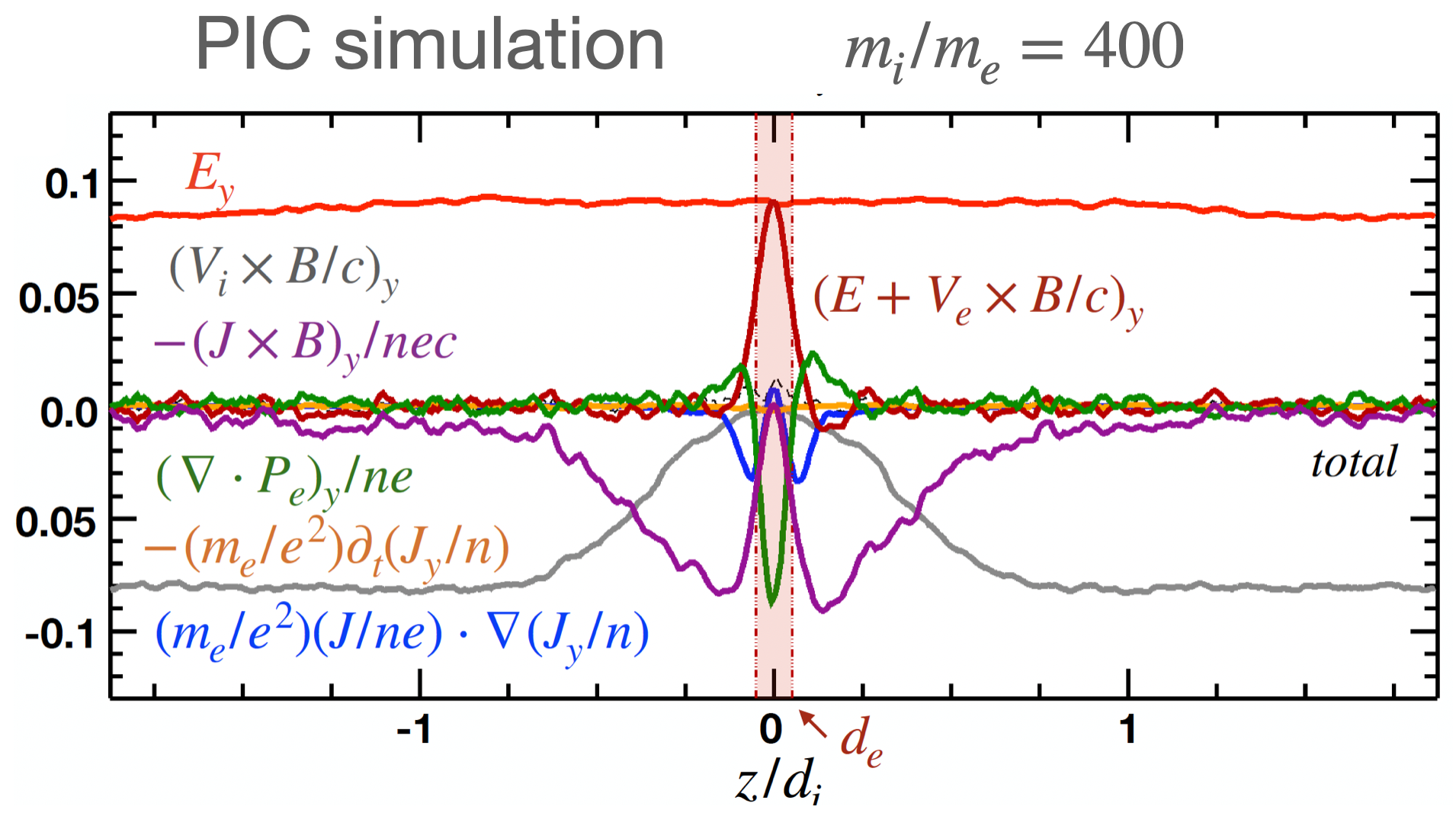} 
\caption {{\bf The generalized Ohm's law} in symmetric, antiparallel low-$\beta$ reconnection. The out-of-plane component of terms in the generalized Ohm's law (normalized by $B_{x0}V_{A0}/c$) across the x-line in the inflow ($z$) direction, based on a particle-in-cell simulation of reconnection. The vertical red transparent band marks the electron diffusion region (EDR), while the ion diffusion region (IDR) expands between $z\in [{-d_i,d_i}]$. Adapted from \cite{yhliu22a}, reproduced by permission of Springer Nature. 
} 
\label{Ohms}
\end{figure}

Beginning with \cite{vasyliunas75a}, it was recognized that the reconnection electric field has to be balanced by one or more terms in the generalized Ohm’s law \citep{vasyliunas75a,cai97a,hesse11a}. Writing the electron momentum equation in collisionless plasmas and solving for ${\bf E}$ gives, 
%The magnetic field line connectivity/topology can change only when some non-ideal effects break the frozen-in condition. This can be described by the generalized Ohm's law, which is basically the electron momentum equation 

\begin{equation}
{\bf E}+\frac{{\bf V}_{e} \times{\bf B}}{c}=-\frac{\nabla\cdot {\bf P}_e}{ne}-\frac{m_e}{e}({\bf V}_e\cdot\nabla){\bf V}_e-\frac{m_e}{e}\frac{\partial}{\partial t}{\bf V}_e, 
\label{e_momentum}
\end{equation}
where variables ${\bf E}$, ${\bf B}$, ${\bf V}_e$, ${\bf P}_e$, $n$, $e$, $m_e$ and $c$ are electric field, magnetic field, electron velocity, electron pressure tensor, density, proton charge, electron mass and the speed of light, respectively.

Since $m_e$ is small, the last two terms are only appreciable if the electron speed $V_e$ is much larger than the ion speed $V_i$, so in those terms we can replace ${\bf V}_e\simeq -{\bf J}/ne$, where ${\bf J}$ is the current density. Then, we obtain the generalized Ohm's Law close to that discussed in \cite{vasyliunas75a}, 
\begin{equation}
{\bf E}+\frac{{\bf V}_i \times{\bf B}}{c}=\frac{{\bf J}\times{\bf B}}{nec}-\frac{\nabla\cdot {\bf P}_e}{ne}-\frac{m_e}{e^2}\left(\frac{\bf J}{ne}\right)\cdot\nabla\left(\frac{{\bf J}}{n}\right)+\frac{m_e}{e^2}\frac{\partial}{\partial t}\left(\frac{{\bf J}}{n}\right), 
\label{Ohms_law}
\end{equation}
%where $d/dt\equiv \partial_t - ({\bf J}/ne)\cdot\nabla$.
The left-hand side (LHS) of Eq.(\ref{Ohms_law}) measures the ion frozen-in condition, which is violated when its value is non-zero. Terms on the right-hand side (RHS) contribute to this violation. In collisionless plasmas, it includes, from left to right, the Hall electric field [$({\bf J}\times {\bf B})/nec$], the divergence of electron pressure term, the spatial derivative of the electron inertia term, and the temporal derivative of the electron inertia term. With collisions, one also needs to include the resistive electric field $\eta {\bf J}$, but it is omitted from our treatment. 

Here we consider a symmetric, anti-parallel low-$\beta$ reconnection in a Particle-in-Cell (PIC) simulation. 
Figure~\ref{Ohms} shows the out-of-plane ($y$) component of the terms in the generalized Ohm's law (Eq.~(\ref{Ohms_law})) in a cut through the X-line in the inflow ($z$) direction.
Upstream of the ion diffusion region (IDR) at $\vert z \vert >d_i$ (the ion inertial scale $d_i\equiv c/\omega_{pi}$, where $\omega_{pi}=\sqrt{m_i/(4\pi n_i e^2)}$ is the ion plasma frequency), ion convection brings magnetic field in, inducing the motional electric field (in gray). 
%The ion inertial scale can also be viewed as the gyroradius of ions at the Alfv\'en speed.
The Hall electric field (in purple) becomes the dominant term supporting the reconnection electric field $E_y$ (in red) between the $d_i$ and the electron inertial scale ($d_e\equiv c/\omega_{pe}$, where $\omega_{pe}=\sqrt{m_e/(4\pi n_e e^2)}$ is the electron plasma frequency). The Hall term arises because of the decoupling of the relatively immobile ions from the motion of electrons that remain frozen to the magnetic field \citep{sonnerup79a}, which becomes significant beneath the ion inertial ($d_i$)-scale. %within which ions do not simply gyrate around one $\bf B$ field line.
The divergence of the electron pressure tensor is important within the electron gyro-scale because the off-diagonal component of a species' pressure tensor becomes pronounced only when the gradient scale of the magnetic field is small or comparable to particles' thermal gyro-radius ($\rho_e=m_e v_{the}/e B$) or bounce lengths \citep{hesse11a}.
The spatial derivative of electron inertia is important within the electron inertial scale. The maximum of the gyro-scale and $d_e$ determines the scale of the electron diffusion region (EDR). 
The time-derivative electron inertial term is negligible in the steady-state shown here, but it is significant in the initiation stage of reconnection, or in the presence of very fast fluctuations with time scales on the order of the electron plasma period \citep{vasyliunas75a}. 
%As we will see below, these are observed to be small or absent in the inner electron diffusion region.

%It is important to note that at the inflow-outflow stagnation point, which coincides with the x-line in the symmetric case, only the divergence of the pressure tensor can break the electron frozen-in condition.

By inspection of Eq.~(\ref{Ohms_law}), we see that the Hall term vanishes at the X-line, and so does the spatial-derivative inertia term in the symmetric case where the flow stagnation point (${\bf V}_{e,xz}=0$) coincides with the X-line. In addition, $\partial/\partial t=0$ in the steady state. These leave us with the divergence of the electron pressure tensor, $(\nabla\cdot {\bf P}_e)_y=\partial_x P_{exy}+\partial_z P_{ezy}$, which, at a quasi-2D reconnection X-line (i.e., $\partial/\partial y=0$), needs to have off-diagonal pressure components in order to balance the reconnection electric field. 
These off-diagonal terms around the X-line arise from the non-gyrotropic feature of the electron distributions.
Hence, it has been proposed that the electron pressure tensor term should be the main contributor to the reconnection electric field at the reconnection site, at least in 2D symmetric situations \citep{vasyliunas75a,Dungey88,lyons90a,cai97a,hesse99a}. The physical origin of the existence of a non-gyrotropic pressure tensor can be traced back to the free acceleration of electrons by the reconnection electric field but only within the unmagnetized EDR~\citep{kulsrud05,hesse11a}. 

In an asymmetric configuration (discussed further in Sec. \ref{asymmetic_MR}), the situation is slightly different in that the inertial term in Eq.~(\ref{Ohms_law}) does not necessarily vanish at the X point. Instead, it is possible that the inertial term contributes part of, or even the majority of the reconnection electric field at this location \citep{hesse14a}. However, we see from Eq.~(\ref{Ohms_law}) that non-gyrotropic pressure tensor effects still need to exist at the flow stagnation point \citep{hesse14a}, which is typically shifted toward the inflow region with a stronger magnetic field \citep{cassak07b,cassak08a}. A simple analysis shows that non-gyrotropic pressure effects are not only expected at the flow stagnation point, but are essential for consistent magnetic flux transport \citep{hesse14a}. Recent research has further indicated that the reconnection electric field is a consequence of the need to maintain the current density in the electron diffusion region, which would otherwise be reduced by non-gyrotropic electron pressure effects \citep{hesse18a}. These authors also showed that the thermal interaction of accelerated particles with the adjacent magnetic field, which gives rise to non-gyrotropic pressures and quasi-viscous current reductions, simultaneously leads to electron heating. This electron heating appears to be the key contributor to maintaining pressure balance in the electron diffusion region (see Sec.~\ref{nature_of_reconnection_electric_field} for more discussion).

\subsection{Observational Analysis of the Generalized Ohm's Law}
\label{Ohms_observation}
%discuss Egdal's MMS result \citep{egedal19a}; $\nabla \cdot {\bf P}_e$ term.

Determining which non-ideal terms are responsible for violating the frozen-in flux condition in EDRs was one of the major objectives of NASA's Magnetospheric Multiscale (MMS) mission \citep{Burch2016SSR}. Note that in the decades preceding MMS observations from many previous satellite missions had confirmed the predominance of the Hall term in the IDR  \citep{Nagai2001JGR,Oieroset2001Nat, Mozer2002PRL, Eastwood2010PRL}. The four identical MMS spacecraft are each capable of measuring the three-dimensional electromagnetic field vector \citep{Torbert2016SSR} and electron and ion velocity space distribution functions \citep{Pollock2016SSR} at very high time resolutions. The spacecraft orbits are typically maintained such that the fleet flies in a tightly-spaced tetrahedral formation with inter-spacecraft separations that can be on the order of the electron inertial length \citep{Fuselier2016SSR}. During crossings through an EDR, differences in the electron and ion fluid moments are obtained between spacecraft pairs, such that, for the first time, the gradient terms in Eq.~(\ref{Ohms_law}) can be approximated \citep{Chanteur1998ISSIR}. For more information on the methods, readers are directed to \cite{Hasegawa2024SSR} of this collection and \cite{paschmann98a}.

\begin{figure}[ht]
\centering
\includegraphics[width=17cm]{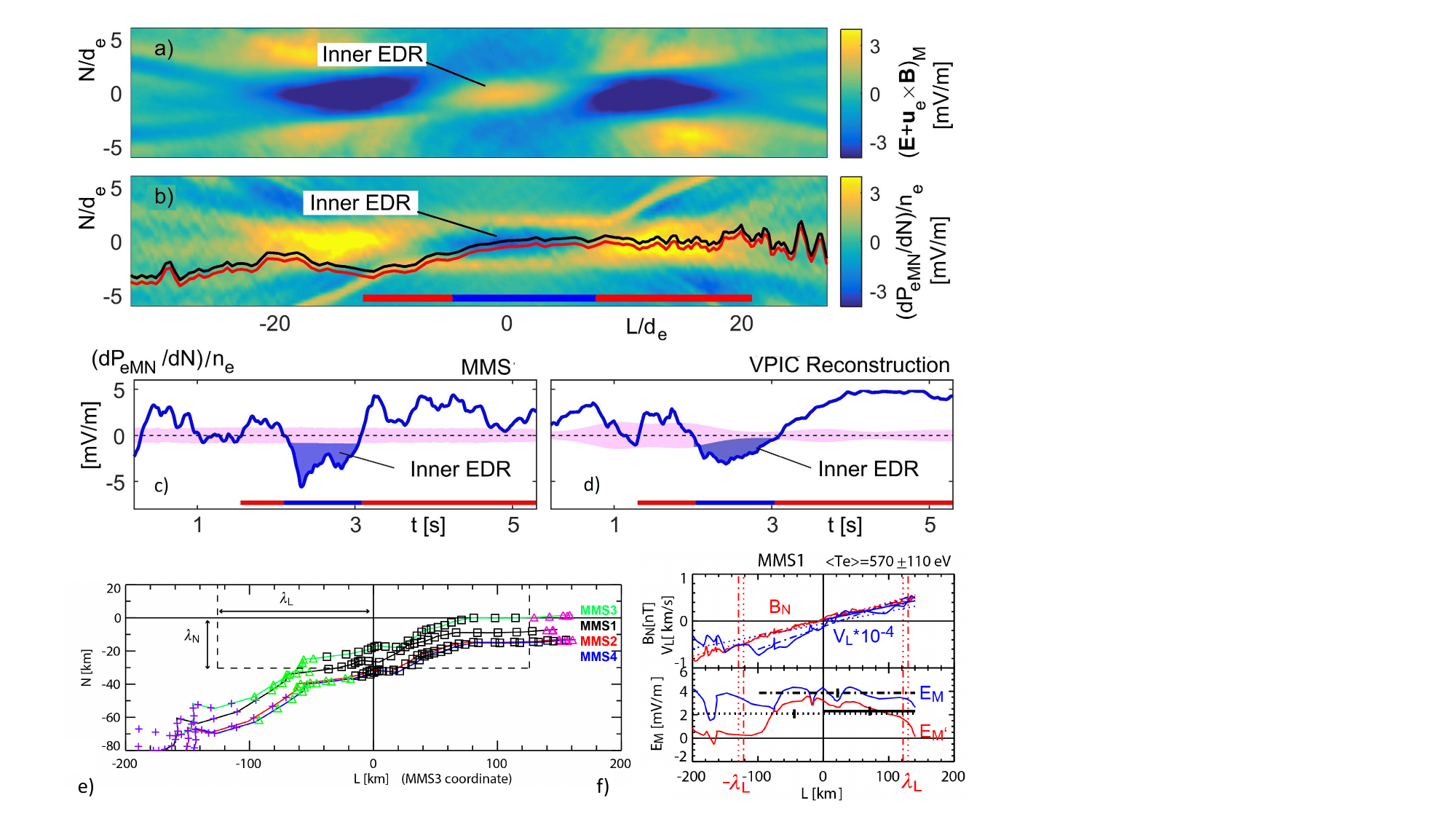} 
\caption {{\bf Non-gyrotropic pressure gradient in the inner EDR }
Simulated 2D profiles of (a) $({\bf E} + {\bf V}_e \times {\bf B}/c)_M$ and (b)  $\partial 
P_{eMN}/\partial N$. Based on the MMS data, the 1D profile of $\partial P_{eMN}/\partial N$ in 
(c) was obtained,  representing a close match to the simulated profile along the MMS trajectory 
in (d). Adapted from \cite{Egedal2019PRL}. (e) Four MMS spacecraft orbit relative to the inner diffusion region (bounded by dashed lines). (f) $B_N$ and $V_{eL}$ (top panel) and profile of $E_M$ and $E'_M$ around the X-line. The three horizontal lines (i.e., dash-dotted, solid, and dotted) in the bottom panel show the electric field value estimated from \cite{hesse99a} formula using $\partial_LV_{eL}$ obtained from data points between different intervals indicated by the length of these lines.
%22:34:01.7 UT and 22:34:03.1 UT (dotted horizontal line), near neutral sheet at $|N| < 50$ km (dash-dotted horizontal line), Earthward of X line near neutral sheet, i.e. $L>0$ and $|N| < 50$ (solid horizontal line). These data intervals are expressed as the length of the horizontal lines.
Adapted from \cite{NakamuraR2019JGR}. }
\label{Ohms_obs1}
\end{figure}

MMS has confirmed that the divergence of the electron pressure tensor dominates other non-ideal terms in EDRs near reconnection X-lines. This finding is consistent with the fact that most reconnection events observed by MMS have small or negligible electron flows in the reconnection plane at the X-line when measured in the co-moving frame of the X-line. \cite{Egedal2019PRL} analyzed a symmetric and nearly-anti-parallel EDR observed by MMS on 11 July 2017, evaluated the electron pressure gradient using MMS data, and compared it with a 2D PIC simulation that used initial conditions based on the observations. \citep{Egedal2019PRL} found that the non-gyrotropic pressure components $\partial_xP_{exy}+\partial_zP_{ezy}$ were predominantly responsible for balancing the reconnection electric field $E_y$, especially  the latter term.  Figure~\ref{Ohms_obs1}(a) and (b) show the numerical profile of $({\bf E} + {\bf V}_e \times {\bf B}/c)_M$ and the pressure gradient $\partial P_{eMN}/\partial N$ with the projected trajectory of the spacecraft, determined by matching the observed magnetic field data to the simulated profile of the current sheet \citep[see more detail in][]{Egedal2019PRL}. Note that the $LMN$ coordinate system is often used for reconnection observations once the quasi-2D reconnection plane is determined.  It corresponds to the $XYZ$ coordinate system shown in Fig.~\ref{reconnection}, used in most theoretical discussions of this review. The inner EDR is marked with a blue color in Fig.~\ref{Ohms_obs1}(c). The profile of $\partial P_{eMN}/\partial N$ is in excellent agreement with both theory and the simulation, being the main contribution to the non-ideal electric field within the inner electron diffusion region (EDR) where the electron frozen-in condition is broken.

An alternative method to calculate the non-gyrotropic pressure term at the inner EDR is to use the theory by \cite{hesse99a, hesse11a}, where the spatial scale of the electron diffusion is given by the electron orbit excursion in a field reversal, the so-called “bounce widths”, ${\lambda}_L$ and ${\lambda}_N$ in the $L$ and $N$ directions respectively, and expressed as ${\lambda}_{L,N} = [2m_eT_e/(e^{2}(\partial_{L,N}B_{N,L})^{2})]^{1/4}$. The electric field in the electron diffusion region, i.e., the non-gyrotropic pressure term, can then be expressed as $E_{\rm M,model}\simeq (1/e)(\partial_LV_{eL})(2m_eT_e)^{1/2}$. This result was confirmed by \cite{NakamuraR2019JGR} based on MMS analysis of the same event by determining the spacecraft orbit relative to an X-line as shown in Fig.~\ref{Ohms_obs1}(e). As expected for the inner EDR, the observed $E_{\rm M}$ and $E_{\rm M'}=({\bf E} + {\bf V}_e \times {\bf B}/c)_M$ in Fig.~\ref{Ohms_obs1}(f) coincide when the spacecraft was inside the inner diffusion region (bounded by the dashed lines in Fig.~\ref{Ohms_obs1}(e)). The dash-dotted horizontal line is the $E_{\rm M,model}$ calculated from the velocity gradients obtained from the upper panel of Fig.~\ref{Ohms_obs1}(d). The model shows a good agreement with the observed electric field in the inner EDR, indicating that the theoretical concept of the inner EDR for laminar reconnection presented by \cite{hesse99a, hesse11a} is well recovered for this event. This means that the divergence of the non-gyrotopic pressure term obtained with this model is also consistent with the reconnection electric field. The same scheme used by \cite{hesse99a, hesse11a} to determine the off-diagonal pressure gradient has been also applied to an EDR event during more turbulent magnetotail (symmetric) reconnection and also obtained good agreement with the observed electric field \citep{Ergun2022ApJ}, indicating that the Ohm's law for laminar reconnection can be maintained even in a turbulent environment.

%Nevertheless, some interesting differences between the force balance relationships are found for EDRs occurring in different background plasma conditions (see subsequent citations in this section).%; those differences are highlighted here, starting with EDRs at current sheets with (approximately) symmetric inflow conditions and anti-parallel inflow magnetic fields.

%During symmetric anti-parallel reconnection, the force exerted by the reconnection electric field is balanced predominately by gradients of the non-gyrotropic portion of the electron pressure tensor \cite{nakamura19,egedal19} within $\sim$1 thermal electron gyroradius around the X-line. In the same event studied by \citeA{nakamura19,egedal19}, a non-vanishing electron flow was observed at the X-line \cite{Hasegawa2018}. 

\begin{figure}[ht]
\centering
\includegraphics[width=12cm]{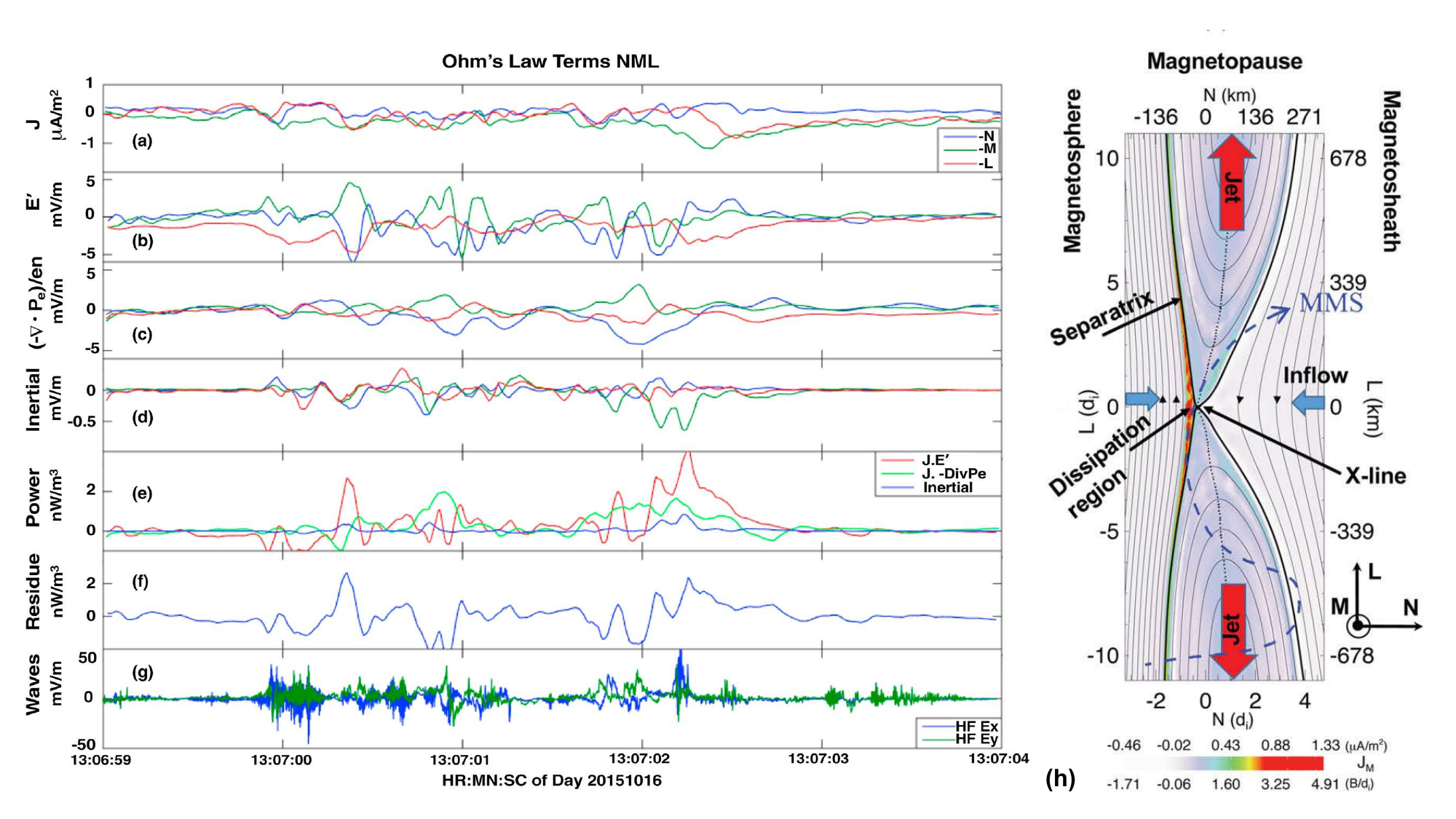} 
\caption {{\bf The generalized Ohm's law for an asymmetric reconnection event observed by MMS} (a) Three components of the magnetopause current density. (b–d) The comparison of terms in Ohm’s law for interval around 13:07:02 UT when the MMS fleet traversed an EDR (e) the total power dissipation from individual terms. (f) The residue, ${\bf J}\cdot\{{\bf E'}-(-\nabla\cdot{\bf P}_e/{en_e})-{m_e}\nabla\cdot{[n_e}({\bf V}_i{\bf V}_i-{\bf V}_e{\bf V}_e)]/e{n_e}\}$, (g) The full wave amplitude at frequencies up to 4 kHz, (h) The schematic of MMS path through EDR. Adapted from \cite{Torbert2016GRL} and \cite{burch16a}. 
}
\label{Ohms_obs2}
\end{figure}

During anti-parallel asymmetric reconnection, the non-ideal electric field is balanced by a combination of the electron inertial and pressure terms, as described at the end of Sec. \ref{generalized_Ohms_law} and confirmed by MMS observations, as shown in Fig.~\ref{Ohms_obs2}.
%adapted from \citep{Torbert2016JGR,burch16a}. 
MMS encountered the EDR at around 13:07:02 UT at a negative $J_M$ peak. It is seen that the contributions of the inertial term (Fig.~\ref{Ohms_obs2}d) are generally smaller than the pressure term (Fig.~\ref{Ohms_obs2}c), but at times can be comparable, in particular for the M (green) component only, which is primarily along the reconnection electric field. Overall both electron pressure gradients and electron inertial effects are important, with a ratio of about 4:1. Yet, there are residuals of a few mV/m (30-50\% of the $\bf E'$) during the encounters with the electron stagnation point (Fig.~\ref{Ohms_obs2}f) and it was also found that the error in the gradient approximation was considerable \citep{Torbert2016GRL}. \cite{Rager2018GRL} analyzed the same event with higher time resolution (7.5 ms) electron data and concluded that Ohm’s law could not be fully accurately resolved even with the 7.5 ms data due to time variability on the scale of the energy sweep of the particle instrument and smoothing of spatial structures by the four spacecraft gradient operator. One possibility of the violation of the Generalized Ohm’s law has been suggested to be evidence of anomalous resistivity \citep{Torbert2016GRL}. Yet, the results of the kinetic simulation performed for the event \citep{Torbert2016GRL} suggested that its effect is not significant. The small contribution from the anomalous resistivity to the Generalized Ohm's law is supported also by \cite{Graham2022Nat}  based on direct estimation of the anomalous resistivity, viscosity, and cross-field electron diffusion (see Eq.~(\ref{Ohms_averaged}) in Sec.~\ref{averaged_Ohms}) associated with lower hybrid waves during another asymmetric reconnection event measured by MMS. It was shown that
the anomalous resistivity is approximately balanced by anomalous viscosity. Hence, although waves do produce an anomalous electron drift and diffusion across the current layer associated with magnetic reconnection, their contribution to the reconnection electric field is considered to be negligible during this observation. More discussions of Ohm's law during the presence of 3D fluctuations will be deferred to Sec.~\ref{averaged_Ohms}.

The dominant role of the pressure divergence over the inertial term in Ohm's law has also been seen for cases of guide field reconnection in both a symmetric \citep{Wilder2017PRL} and asymmetric \citep{Genestreti2018JGR} current sheet. \cite{Genestreti2018JGR} also found that both out-of-the-reconnection-plane gradients $\partial_M$ and in-plane $\partial _{L,N}$ in the pressure tensor contribute to energy conversion near the X-point. A finite $\partial_M P_{eMM}\simeq \partial_\|P_{e,\|}$ near the X-line was also observed in 3D guide field reconnection simulations that have a significant 3D structure \citep{yhliu13a,stanier19a}.

\subsection{The Nature of the Reconnection Electric Field}
\label{nature_of_reconnection_electric_field}

%Reconnection electric field not only breaks the frozen-in condition, allowing the occurrence of reconnection but also plays a critical role in sustaining the current density and maintaining the thermal pressure at the x-line \citep{hesse11a,hesse20a}.

A question of more than just academic nature is why there is a reconnection electric field at all. This question transcends the simple conclusion from above that there has to be a reconnection electric field for flux to be transferred from inflow to outflow. This existence question was raised by \cite{hesse18a}, who investigated the current and energy balance in the electron diffusion region. %Specifically, \cite{hesse18a} asked the question, which forces act upon the electron current density, the main current density in this critical region? 

During the initial phase of an evolving symmetric reconnecting current sheet, the time-derivative of the electron inertia term, $(m_e/e)\partial {\bf V}_e/\partial t$ in Eq.~(\ref{e_momentum}), is the only non-ideal term available to break the frozen-in condition at the X-line. This dominance causes the continuous intensification of the current density (${\bf J}\simeq -en {\bf V}_e$) at the X-line, leading to a sharp current density peak around the electron gyro-scale $\rho_e$, generating a non-gyrotropic particle distribution \citep{hesse11a,aunai13d,zenitani16a} that eventually makes $\nabla\cdot {\bf P}_e$ the dominant non-ideal term in the quasi-steady ($\partial_t=0$) phase; note again that $\nabla\cdot {\bf P}_e$ is the only term available to support the reconnection electric field at the X-line in the steady state of a symmetric case. This transition of the dominant non-ideal term during this current density intensification is clearly demonstrated in PIC simulations \citep{yhliu14a}.

\begin{figure}[ht]
\centering
\includegraphics[width=11cm]{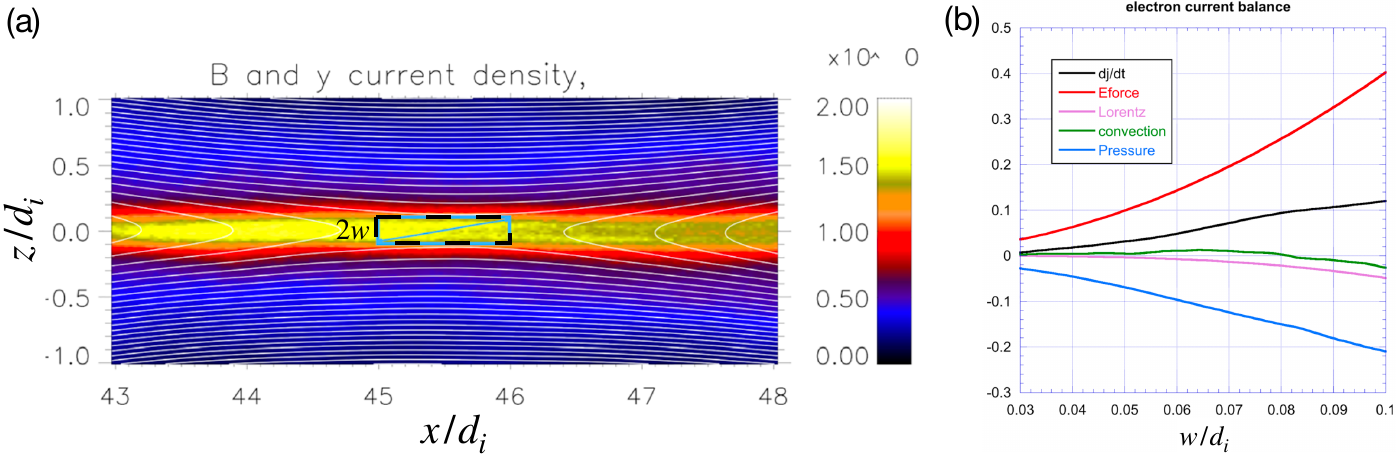} 
\caption {{\bf Terms that balance the current density in the quasi-steady state}. (a) The largest integration region centered on the X-point is shown as the blue-black rectangle with a thickness of 0.2 $d_i$ and a lateral extent of one $d_i$. The box size varies from an initially much smaller size while keeping the aspect ratio fixed. (b) Integration of the electron current balance equation as a function of the integration region half-thickness ``w''. The integrated time derivative of the out-of-place current density is indicated by the black curve, which is also equal to the sum of the four other curves in agreement with Eq.~(\ref{dJey_dt}). 
%It shows a small positive value for the current density derivative, indicative of a slow time dependence of the system. The convection term (green) is obtained by calculating the momentum flux across all four boundaries of the integration box. Like the Lorentz force term (purple), it is small. 
The major contributors are the current increase by the reconnection electric field (red), and the current reduction by pressure effects (blue). The dominance of these two terms holds over the entire range rather than at the X-point alone. Adapted from \cite{hesse18a}}.
\label{ER_current}
\end{figure}

Focusing on the quasi-steady state, \cite{hesse18a} investigated the electron momentum equation Eq.~(\ref{e_momentum}), rewritten in the form of a current evolution equation:
%\begin{equation}
%\frac{\partial J_{ey}}{\partial t} =\frac{e^2n_e}{m_e}E_y+\frac{e^2n_e}{m_e} (V_{ez}B_x-V_{ex}B_z)+\frac{e}{m_e}\left(\frac{\partial P_{eyz}}{\partial z}+\frac{\partial P_{exy}}{\partial x}\right)+e\nabla\cdot (n_e {\bf V}_e) V_{ey}.
%\label{dJey_dt}
%\end{equation}
\begin{equation}
\frac{\partial J_{ey}}{\partial t} =\frac{e^2n_e}{m_e}E_y+\frac{e^2n_e}{m_ec} ({\bf V}_e\times{\bf B})_y+\frac{e}{m_e}\left(\frac{\partial P_{eyz}}{\partial z}+\frac{\partial P_{exy}}{\partial x}\right)-\nabla\cdot ({\bf V}_e J_{ey}).
\label{dJey_dt}
\end{equation}
The terms on the right-hand side (RHS) of this equation describe, in order, the electric field force (due to the reconnection electric field $E_y$), conversion of in-plane to out-of-plane current by Lorentz forces, pressure gradient forces, and current convection into or out of the volume of interest.

To evaluate the balance of these terms over a larger domain that contains the singular X-line in the steady state, \cite{hesse18a} integrated the individual terms of this equation over rectangles of different sizes, centered about the X-line location in a PIC simulation of symmetric magnetic reconnection, as shown in Fig.~\ref{ER_current}(a). The results of this integration are displayed in Fig.~\ref{ER_current}(b). 
%which provides an alternative viewpoint: 
The only terms of importance are the non-gyrotropic pressure terms (i.e., $\partial P_{eyz}/\partial z+\partial P_{exy}/\partial x$), which act to reduce the current density, and the electric field term, which acts to increase it. These terms roughly balance each other, keeping the electron current density constant, or varying on very slow time scales to account for the overall system evolution.

\cite{hesse18a} further found that the source of the electron pressure, $P_e\equiv{\rm Tr}({\bf P}_e)/3$, within the EDR is essentially exclusively due to the non-gyrotropic effect, i.e., due to complex particle orbits rather than simple, gyrotropic behavior. 
%The study was extended to include an investigation of the thermal pressure at the X-line, which, as we know now \citep{yhliu22a}, is typically substantially reduced compared to what force balance in a one-dimensional current sheet requires. 
For this purpose, the equation
\begin{equation}
\frac{\partial P_e}{\partial t}=-\nabla\cdot({\bf V}_eP_e)-\frac{2}{3}\sum_l P_{e,ll}\frac{\partial V_{el}}{\partial x_l}-\frac{1}{3}\sum_{l,i}\frac{\partial Q_{e,lii}}{\partial x_i}-\frac{2}{3}\sum_{l,i\neq l}P_{e,li}\frac{\partial V_{el}}{\partial x_i}
\label{dp_dt}
\end{equation}
was integrated over the same, varying rectangle and analyzed in a way similar to Fig.~\ref{ER_current}(b). The result (not shown) was that among the terms on the RHS of Eq.~(\ref{dp_dt}), the last term provided a positive contribution, whereas negative contributions were provided by the first two terms. 
The heat flux term $Q_{e,lii}$ is negligible.
%the former term is referred to as ``quasi-viscous'' due to the involvement of cross-derivatives of the electron flow velocity, whereas the latter two terms are convection terms, generalized for anisotropic pressures. 
The dominant non-gyrotropic pressure $P_{e,li}$ contributions here appeared to be the same as the ones acting to reduce the current density in Eq.~(\ref{dJey_dt}), suggesting that the conversion of the current carrier motion to the plasma pressure plays an important role. This last term plus the second term on the RHS of Eq.~(\ref{dp_dt}) is basically the ``pressure-strain interaction'' \citep{Yang2017a,Yang2017b}, $-({\bf P}\cdot \nabla)\cdot {\bf V}$, that will be further discussed in Sec.~\ref{VGP}).

\begin{figure}[ht]
\centering
\includegraphics[width=7cm]{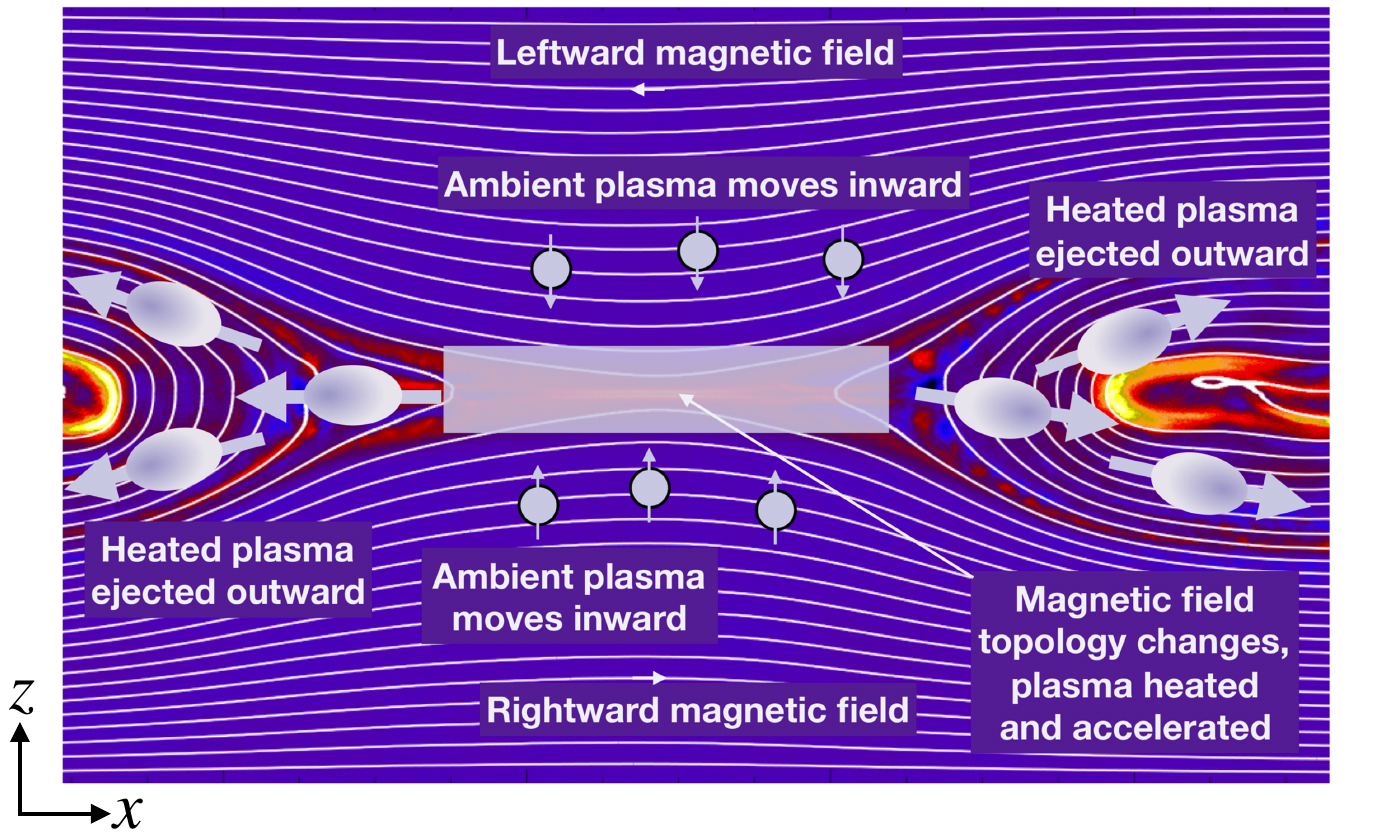} 
\caption {{\bf The nature of the reconnection electric field}. Shown are in-plane magnetic field lines (white), and the out-of-plane current density contour. In addition to breaking the frozen-in condition and transporting the flux into and out of the diffusion region, the reconnection electric field also sustains the electric current and increases the thermal pressure through the $-({\bf P}\cdot \nabla)\cdot {\bf V}$ term within the diffusion region. Reprinted from \cite{hesse20a}, with the permission of Wiley.}
\label{ER_nature}
\end{figure}

%In short, the reconnection electric field within the diffusion region converts incoming electromagnetic energy into the current carrier bulk kinetic energy through direct acceleration. The current at electron gyro-scale intensifies until the current sheet is thin enough to generate complex, non-gyrotropic particle distribution (i.e., non-zero $P_{eyz}$ and $P_{exy}$), which eventually ceases the thinning process and funnels the current carrier energy into the thermal energy to help maintain the current sheet pressure. This multifaceted nature of the reconnection electric field is highlighted in Fig.~\ref{ER_nature}.

%In short, the reconnection electric field within the diffusion region converts incoming electromagnetic energy into the current carrier bulk kinetic energy through direct acceleration. The current at electron gyro-scale is intensified until the current density gradient is strong enough to generate complex, non-gyrotropic particle distribution (i.e., non-zero $P_{eyz}$ and $P_{exy}$), making $\nabla\cdot {\bf P}$ force strong enough to counter-balance the reconnection electric field force (Eq.~(\ref{dJey_dt})). The same non-gyrotropic feature funnels the current carrier energy into the thermal energy through the $-({\bf P}\cdot \nabla)\cdot {\bf V}$ term (Eq.~(\ref{dp_dt})). This multifaceted nature of the reconnection electric field is highlighted in Fig.~\ref{ER_nature}. 

In short, the reconnection electric field within the diffusion region converts incoming electromagnetic energy into the current carrier bulk kinetic energy through direct acceleration. The current at the electron gyro-scale is intensified until the current density gradient is strong enough to generate complex, non-gyrotropic particle distribution (i.e., non-zero $P_{eyz}$ and $P_{exy}$), which funnels the current carrier kinetic energy into the thermal energy through the $-({\bf P}\cdot \nabla)\cdot {\bf V}$ term (Eq.~(\ref{dp_dt})). This makes the steady state possible, where $\partial_t J_{ey}$ vanishes and the $\nabla\cdot {\bf P}$ becomes strong enough to balance the reconnection electric field in the Ohm's law (Eq.~(\ref{dJey_dt}) or Eq.~(\ref{Ohms_law})). This multifaceted nature of the reconnection electric field is highlighted in Fig.~\ref{ER_nature}.

%making $\nabla\cdot {\bf P}$ force strong enough to counter-balance the reconnection electric field force (Eq.~(\ref{dJey_dt})). The same non-gyrotropic feature funnels the current carrier energy into the thermal energy through the $-({\bf P}\cdot \nabla)\cdot {\bf V}$ term (Eq.~(\ref{dp_dt})). This multifaceted nature of the reconnection electric field is highlighted in Fig.~\ref{ER_nature}. 

%which eventually ceases the thinning process and funnels the current carrier energy into the thermal energy to help maintain the current sheet pressure. This multifaceted nature of the reconnection electric field is highlighted in Fig.~\ref{ER_nature}.

%Returning to the question of why there is a reconnection electric field at all, 

An alternative viewpoint is also offered in \cite{hesse18a}, which argues that the electric field exists as a consequence of Maxwell’s equations, specifically, Amp\`ere’s law: 
\begin{equation}
\frac{\partial {\bf E}}{\partial t}=c\nabla \times {\bf B}-4\pi {\bf J}.
\label{ampere}
\end{equation}
%In a Gedanken experiment, 
Imagine that the current density is reduced by ``a mechanism'' below what is required to balance the $\nabla \times {\bf B}$ term. Then Amp\`ere’s law, Eq.~(\ref{ampere}), will immediately signal the need to increase the electric field, which accelerates the current carriers and re-establishes balance in the steady state.  

%such that the current density is increased by Eq.~(\ref{dJey_dt}) (i.e., the generalized Ohm's law) to re-establish balance. 

In other words, the steady-state reconnection electric field exists because there is a mechanism at work, which attempts to "dissipate" the current density. This conclusion holds irrespective of the dissipation mechanism – for example, classical collisions would have the same effect, as captured in Ohm's law ${\bf J}=\sigma {\bf E}$ where $\sigma$ is the conductivity determined by the collisions. In a collisionless plasma, however, the current dissipation is provided by non-gyrotropic pressure effects, which are a manifestation of complex particle orbits, which lead to the scattering of directed motion by the local magnetic geometry. \cite{hesse21a} extended this investigation to asymmetric systems (defined in Sec. \ref{asymmetic_MR}) and found that the overall conclusions also hold there, even though some of the current reduction was found to be due to convective effects, in addition to the above-discussed non-gyrotropic pressure effects.

%In principle, these additional roles played by the electric field within the diffusion region, in sustaining the electric current and in maintaining the pressure, may control the geometry of magnetic reconnection through force balance. A closure can determine the reconnection rate, and examples are shown in Sec.~\ref{localization} and electron-positron plasmas \citep{goodbred22a}. 
%also discussed in {\color{blue} F. Guo et al. (2023, this issue)}.  %As we will see later in Sec.~\ref{localization} and Paper 4.3, the self-consistent solution to reconnection rate in both space and astrophysical plasmas can be derived through these considerations \citep{yhliu22a,goodbred22a}. 

\section{Collisionless Magnetic Reconnection Rate}
\label{reconnection_rate_models}

In this section, we discuss how one can determine the magnitude of the reconnection electric field $E_R$ (i.e., the $E_y$ at the X-line), which is essentially the reconnection rate that measures how fast reconnection processes the incoming magnetic flux. We will organize the discussions of different regimes using the governing force-balance equation, as it determines the characteristic reconnection outflow speed, being critical to the rate. To avoid a common confusion in the normalization of reconnection rates, we normalize $E_R$ by the ``asymptotic'' value of reconnecting magnetic field component $B_R$ (or $B_{x0}$) and the associated proton Alfv\'en speed in the upstream region $V_A\equiv B_R/\sqrt{4\pi n m_i}$ to define the ``normalized reconnection rate'' $R\equiv cE_R/B_RV_A$. While we have aimed to unify the notation throughout this review, some differences in subsections are unavoidable in order to strike the balance between simplicity and consistency. New notations, if needed, are defined with respect to the coordinates shown in the relevant figures. 

\subsection{Standard Symmetric Anti-parallel Reconnection}\
\label{standard_symmetric_reconnection}
We begin with the simplest current sheet, one that has symmetric, antiparallel magnetic fields, as illustrated in Fig.~\ref{SweetParker}. Combining the electron and ion momentum equations, we can derive the MHD force balance equation. In the steady state, it reads
\begin{equation}
\frac{({\bf B}\cdot\nabla){\bf B}}{4\pi}\simeq \frac{\nabla B^2}{8\pi}+\nabla\cdot {\bf P}+nm_i({\bf V}\cdot \nabla){\bf V}.
\label{force_balance_eq}
\end{equation}
This force-balance equation works in most regions, including the ideal MHD region and the ion diffusion region, as long as the electron inertial term is negligible and the quasi-neutral condition holds; that is usually valid in the non-relativistic limit. As we will see, the scaling of reconnection rates in diverse regimes can be more or less captured by the force balance along the inflow and outflow symmetry lines.

\subsubsection{Sweet-Parker Scaling}
The first quantitative model of the magnetic reconnection rate was derived by Sweet and Parker \citep{parker57a,sweet58a}. From mass conservation $\nabla\cdot(n{\bf V})\simeq0$ in steady state and the incompressible assumption,
\begin{equation}
V_{\rm in}L \simeq V_{\rm out}\delta,
\label{continuity_SP}
\end{equation}
where $\delta$ and $L$ are the half-thickness and half-length of the diffusion region, respectively. From the momentum equation, balancing the magnetic tension and inertia force in Eq.~(\ref{force_balance_eq}), $({\bf B}\cdot \nabla) {\bf B}/4\pi\simeq n m_i ({\bf V}\cdot\nabla){\bf V}$, one gets
\begin{equation}
V_{\rm out}\simeq \frac{B_R}{\sqrt{4\pi n m_i}}=V_A.
\label{Vout_SP}
\end{equation}
Thus, the outflow speed is the characteristic Alfv\'en speed based on the upstream magnetic field $B_R$.
It then makes sense to define the {\it normalized reconnection rate} as 
\begin{equation}
R\equiv\frac{V_{\rm in}}{V_A},
\label{R_SP_scaling}
\end{equation}
which measures how fast plasma inflows bring in magnetic flux for processing. Combining Eqs.~(\ref{continuity_SP}) and (\ref{Vout_SP}), one realizes that the normalized reconnection rate is basically the aspect ratio of the diffusion region, i.e., $R\simeq \delta/L$. Note that $E_y$ is uniform in the 2D steady-state per Faraday's law and $E_y=V_{\rm in}B_R/c$ at the inflow boundary of the diffusion region, thus the definition of the reconnection rate can also be expressed as $R\equiv cE_R/(B_R V_A)$.

\begin{figure}[ht]
\centering
\includegraphics[width=12cm]{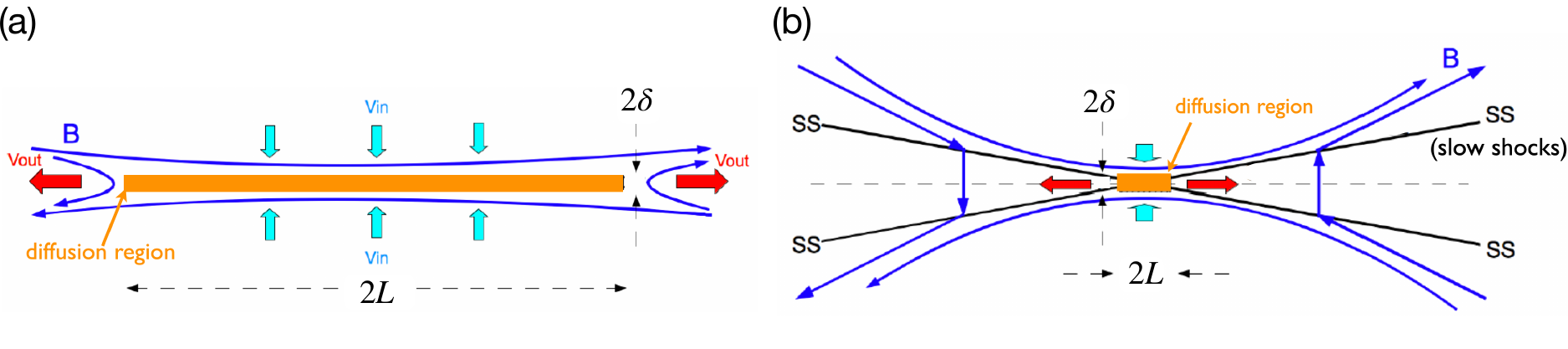} 
\caption {{\bf Classical reconnection models}. (a) Sweet-Parker solution. Adapted from \citep{sweet58a,parker57a}(b) Petschek solution. Adapted from \citep{petschek64a}
} 
\label{SweetParker}
\end{figure}

After coupling the inflow region to a diffusion region dominated by resistivity (which requires collisions), the full Sweet-Parker solution (omitted here) was derived in 1957. It has a system size long current sheet (Fig.~\ref{SweetParker}(a)), resulting in a low $\delta/L$ and thus a reconnection rate that is too low to explain the energy release during solar flares \citep{parker63a,parker73a}. \cite{petschek64a} proposed that standing slow shocks that bound the exhaust can resolve this challenge by opening out the outflow geometry, and localizing the diffusion region (Fig.~\ref{SweetParker}(b)). However, Petschek's solution is not self-consistent and tends to collapse into the long Sweet-Parker solution in (uniform) resistive-MHD simulations \citep{sato79a,biskamp86a}. Such an elongated reconnection layer can be unstable to the plasmoid instability (also called secondary tearing instability) if collisions are weak enough \citep{biskamp82a,Shibata01,Bhattacharjee09,loureiro07a,pucci14a,comisso16b}, but we will not discuss this resistive-MHD mode further.

\subsubsection{$R-S_{\rm lope}$ relation and the maximum plausible rate}
\label{maximum_rate}

For collisionless reconnection, the rate in Eq.~(\ref{R_SP_scaling}) is not bounded, since $\delta/L$ can, in principle, be any value.  To fix this problem, one needs to consider force balance along the inflow direction and recognize there is a scale separation between the regions ``immediately upstream'' and ``far (asymptotic) upstream'' of the ion diffusion region. %Note that, although Petschek also considered the force balance in the inflow direction, his treatment does not yield a bounded rate.
 
\begin{figure}[ht]
\centering
\includegraphics[width=11cm]{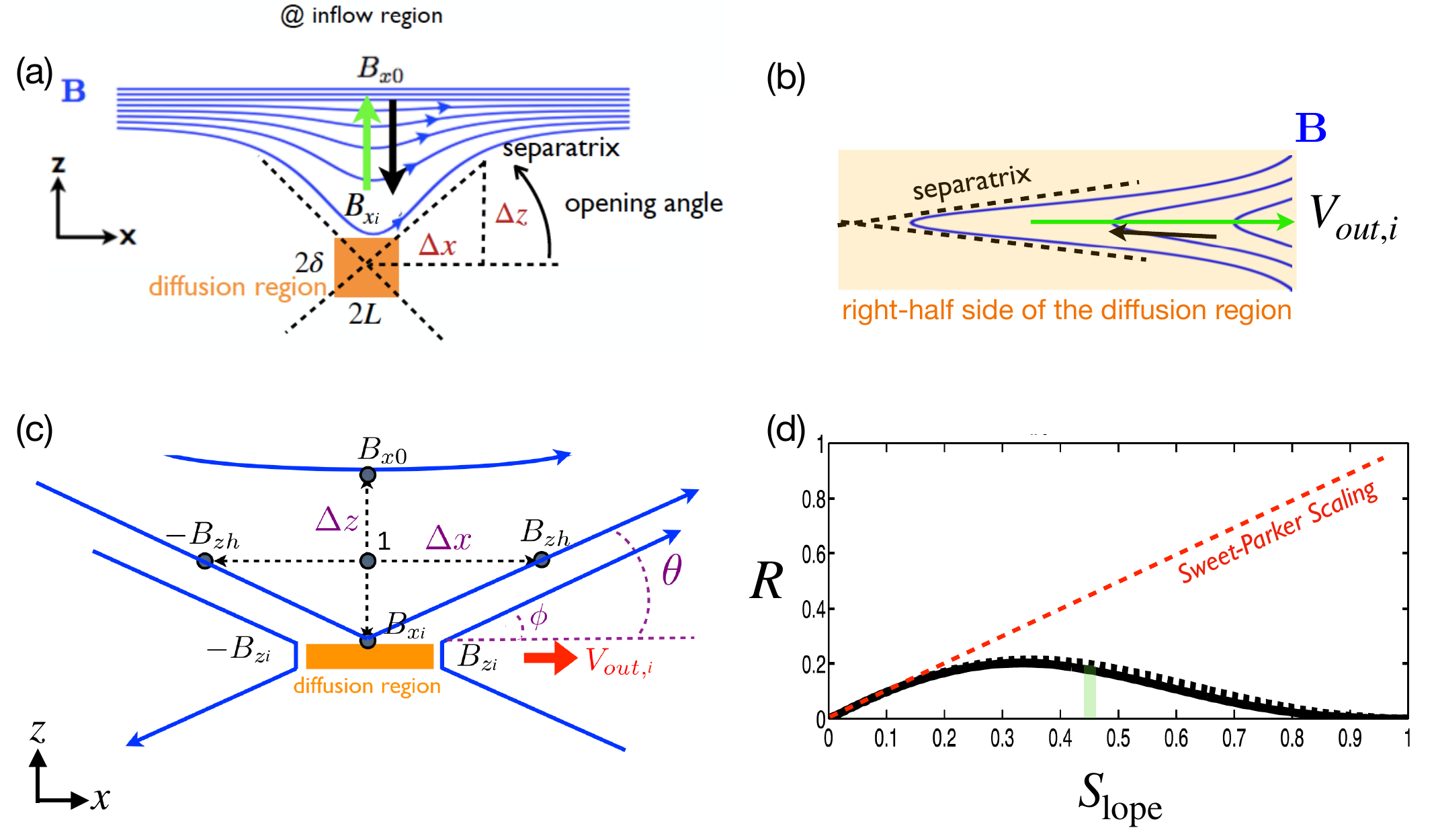} 
\caption {{\bf General constraints on the maximum plausible rate from mesoscale force-balance.} (a) Field line geometry and force balance upstream of the diffusion region. (b) Field line geometry and force balance along the outflow. (c) The scheme used to analyze the force balance. (d) The predicted reconnection rate $R$ as a function of the separatrix slope $S_{\rm lope}$. Adapted from \cite{yhliu17a}.} 
\label{mesoscale}
\end{figure}

Per geometry, this diffusion region aspect ratio $\delta/L$ is also the slope of the separatrix $S_{\rm lope}=\Delta z/\Delta x$, as shown in Fig.~\ref{mesoscale}(a). The $R\simeq \delta/L$ scaling only works when $S_{\rm lope}=\delta/L\ll 1$. In the $S_{\rm lope}\rightarrow 1$ limit (i.e., a localized diffusion region with an open outflow geometry as illustrated in Fig.~\ref{mesoscale}(a)), the upstream magnetic field is indented, which unavoidably induces a magnetic tension force $({\bf B}\cdot\nabla){\bf B}/4\pi$ pointing to the upstream region, as illustrated by the green arrow in Fig.~\ref{mesoscale}(a). In the low-$\beta$ limit, both the upstream $\nabla\cdot {\bf P}$ and $nm_i({\bf V}\cdot \nabla){\bf V}$ terms are negligible; thus, the only term that can counterbalance this tension force is the magnetic pressure gradient force ($-\nabla B^2/8\pi$, black arrow), which requires the reduction of the reconnecting field when it is convected into the diffusion region. Similarly, a finite magnetic pressure gradient force also arises in the outflow direction in the $S_{\rm lope}\rightarrow 1$ limit, as depicted by the black arrow in Fig.~\ref{mesoscale}(b), which slows down the outflow.

Quantitatively, through discretizing the inflow force-balance at point ${\bf 1}$ of Fig.~\ref{mesoscale}(c), one can relate the ratio of the magnetic field immediately upstream of the ion diffusion region $B_{xi}$ and the asymptotic value at far upstream $B_{x0}$ to the slope of the reconnection separatrix $S_{\rm lope}$ \citep{yhliu17a} as 
\begin{equation}
\frac{B_{xi}}{B_{x0}}\simeq \frac{1-S_{{\rm lope}}^2}{1+S_{{\rm lope}}^2}.
\label{Bxi_Bx0}
\end{equation}
In the large opening limit ($S_{\rm lope}\rightarrow 1$), the magnetic field $B_{xi}$ that actually reconnects is reduced, as is the rate $R$.

By analyzing the outflow force balance, including the $nm_i({\bf V}\cdot \nabla){\bf V}$ term at a point within the diffusion region (Fig.~\ref{mesoscale}(b)), one can derive the outflow speed at the outflow edge of the ion diffusion region, 
\begin{equation}
V_{{\rm out},i}\simeq V_{Ai}\sqrt{1-S_{{\rm lope}}^2},
\label{Vout}
\end{equation}
where $V_{Ai}\equiv B_{xi}/\sqrt{4\pi nm_i}$. Equation~(\ref{Vout}), coupled with Eq.~(\ref{Bxi_Bx0}), recovers the Alfv\'en speed $V_{A0}=B_{x0}/\sqrt{4\pi n m_i}$ in the small opening limit, as in the Sweet-Parker analysis. In the large opening ($S_{\rm lope}\rightarrow 1$) limit, the outflow speed is reduced, and so is the rate.

The reconnection rate is $R$$=$$cE_y/B_{x0}V_{A0}$$=$$(B_{zi}/B_{xi})(B_{xi}/B_{x0})(V_{{\rm out},i}/V_{A0})$. Using Eqs.(\ref{Bxi_Bx0}) and (\ref{Vout}), and noting  that $B_{zi}/B_{xi}\simeq S_{\rm lope}$, we then find the $R-S_{\rm lope}$ relation 
\begin{equation}
R=S_{{\rm lope}}\left(\frac{1-S_{{\rm lope}}^2}{1+S_{{\rm lope}}^2}\right)^2\sqrt{1-S_{{\rm lope}}^2}, 
\label{R}
\end{equation}
which is shown as the black solid curve in Fig.~\ref{mesoscale}(d).
Clearly, these two geometrical constraints along the inflow and outflow bring down the reconnection rate to zero in the $S_{\rm lope}\rightarrow 1$ limit, where the separatrix makes a right angle. The maximum plausible rate is around the value of $0.2$. For reference, the Sweet-Parker scaling is shown by the red dashed line, which is unbounded in the large $S_{\rm lope}$ limit.
Importantly, the profile of the predicted black curve is relatively flat for a wide range of $S_{\rm lope}$. Thus as long as there is some degree of localization, the predicted rate will be on the order of $\mathcal{O}(0.1)$. 
Note that this prediction does not depend on the dissipation physics or the thickness of the current sheet. Thus, this value likely also constrains the maximal plausible rate in theorized ``turbulent reconnection'' where the diffusion region is turbulent and thick, and has large-scale outflow exhausts \citep{Lazarian99}.
On the other hand, this maximum plausible reconnection rate of value $\simeq 0.2$ is clearly demonstrated using a large and spatially localized anomalous resistivity right at the X-line in MHD simulations \citep{SCLin21a, Jimenez22a}.

\subsubsection{Localization mechanism that leads to fast reconnection}
\label{localization}
\cite{petschek64a} provided the correct steady-state outflow solution of reconnection that predicts slow shocks and rotational discontinuities farther downstream, but the solution failed to capture the essential {\it localization mechanism}, that leads to the open geometry in the first place. While Eq.~(\ref{R}) provides the general $R-S_{\rm lope}$ relation, to determine the rate, we still need to identify the (primary) mechanism that localizes the diffusion region, determining the opening geometry captured by $S_{\rm lope}$. 

\begin{figure}[ht]
\centering
\includegraphics[width=12cm]{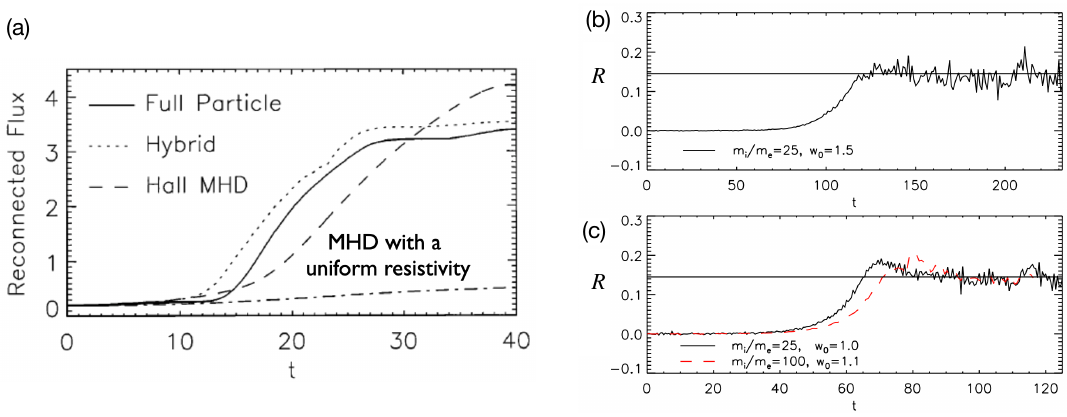} 
\caption {(a) The GEM reconnection challenge result that compares reconnection rates in four different numerical models. Reprinted from \cite{birn01a}, with the permission of Wiley. Panels (b) and (c) show the normalized reconnection rate in PIC simulations over a range of mass ratio $m_i/m_e$ and initial thickness $w_0$. Adapted from \cite{shay07a}. 
} 
\label{GEM_challenge}
\end{figure}

Kinetic simulations beyond the MHD model suggest that antiparallel reconnection with an open outflow geometry occurs when the current sheet thins down to the ion inertial scale \citep{bhattacharjee04a,cassak05a,daughton09a,jara_almonte21a}. When this occurs, the Hall term in the generalized Ohm's law \citep{vasyliunas75a,swisdak08a} dominates the electric field in the ion diffusion region (IDR), where the ions become demagnetized. The correlation between the Hall effect and fast reconnection was clearly demonstrated in the GEM reconnection challenge study \citep{birn01a}, as shown in Fig.~\ref{GEM_challenge}(a). This study showed that simulation models with the Hall term in the generalized Ohm's law (particle-in-cell (PIC), hybrid, and Hall-MHD) realize fast reconnection, while only the uniform resistive-MHD model, which lacks the Hall term, exhibits a slow rate \citep{parker57a,sweet58a}. The value of the fast rate in collisionless plasmas is on the order of $0.1$ over a range of electron-ion mass ratios and initial thicknesses, as shown in Fig.~\ref{GEM_challenge}(b) and (c). For decades, it had been unclear ``why'' and ``how'' the Hall term localizes the diffusion region, producing an open geometry. The dispersive property of waves arising from the Hall term was proposed as an explanation \citep{mandt94a,shay99a,rogers01a,drake08a}, but the role of dispersive waves derived from the linear analysis was called into question because reconnection can be fast even in systems that lack dispersive waves \citep{bessho05a,yhliu14a,tenbarge14a,stanier15a}.

\begin{figure}[ht]
\centering
\includegraphics[width=12cm]{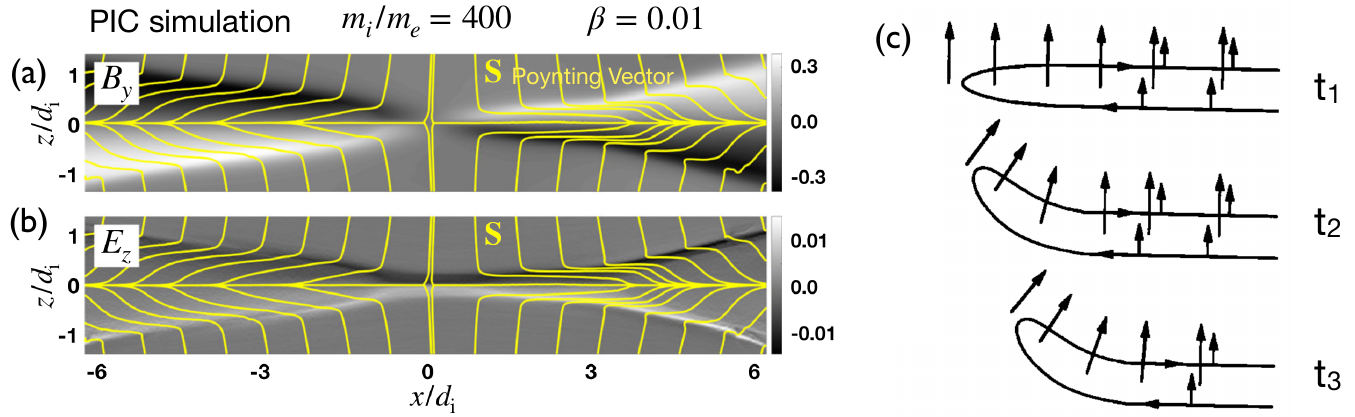} 
\caption {{\bf Hall electromagnetic fields} (a) The Hall magnetic field $B_y$ and (b) the Hall electric field $E_z$ (normalized by $B_{x0}$) overlaid with Poynting vector ${\bf S}$ streamlines (yellow). Adapted from \cite{yhliu22a}, reproduced by permission of Springer Nature. (c) Electrons drag the reconnected field out of the reconnection plane to form the Hall quadrupole magnetic field. Adapted from \citep{mandt94a}. 
} 
\label{Hall_fields}
\end{figure}

While the Hall electromagnetic fields were well-known for being the key feature of the ion diffusion region {\citep{sonnerup79a}}, their role in transporting the incoming magnetic energy was less recognized. Figure~\ref{Hall_fields}a shows the out-of-plane magnetic field $B_y$ in the non-linear stage. This out-of-plane quadrupolar Hall magnetic field arises because electrons, the primary current carrier within the IDR (i.e., ${\bf J} \simeq -e n {\bf V}_e$), drag both reconnected and not-yet reconnected magnetic field lines out of the reconnection plane \citep{mandt94a,ren05a,drake08a,burch16a}, as illustrated in Fig.~\ref{Hall_fields}(c); the laboratory evidence is reviewed in {\color{blue} Ji et al., 2023, this issue}.
Importantly, this Hall quadrupole magnetic field $B_y$ along with the inward-pointing Hall electric field $E_z\simeq V_{ey} B_x/c$, shown in Fig.~\ref{Hall_fields}b, constitute a Poynting vector $S_x=-cE_z B_y/4\pi$ in the $x$-direction. This component diverts the inflowing electromagnetic energy toward the outflow direction. This is shown by the streamlines of ${\bf S}=c{\bf E}\times{\bf B}/4\pi$ in yellow, which bend in the $x$ direction significantly before reaching the outflow symmetry line at $z=0$.

\begin{figure}[ht]
\centering
\includegraphics[width=12cm]{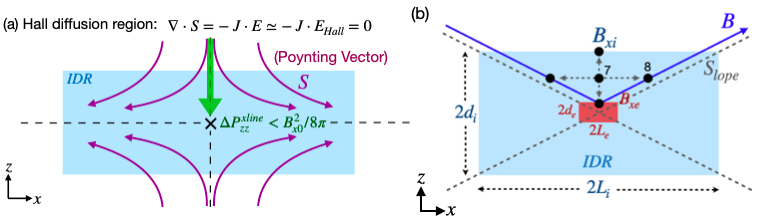} 
\caption {{\bf Transport patterns of electromagnetic energy in Hall reconnection and the diffusion region structure.} (a) The Hall effect results in this intrinsic, ``diverting'', Poynting vector ${\bf S}$ streamline pattern in purple, which limits the pressure increase of inflowing plasma (green arrow) that reaches the x-line. (b) The diagram used to derive the slope of the separatrix, $S_{{\rm lope}}$. The blue (red) box represents the IDR (EDR). The solid blue line depicts an upstream magnetic field $\bf B$ line adjacent to the separatrix shown by diagonal dashed lines. Adapted from \cite{yhliu22a}, reproduced by permission of Springer Nature.} 
\label{IDR_EDR}
\end{figure}

Since the Hall term dominates the electric field ${\bf E}\simeq {\bf E}_{{\rm Hall}}$ $=$ $ {\bf J}\times {\bf B}/nec$ inside the IDR, then $\nabla\cdot {\bf S}$ $=$ $-{\bf J}\cdot {\bf E}$ $ \simeq 0$ per Poynting's theorem in the steady state (n.b., further discussion of Poynting's theorem can be found in Sec.~\ref{Poynting}). 
When the divergence of a vector field vanishes, like that for magnetic fields (i.e., $\nabla\cdot{\bf B}=0$), Gauss' theorem indicates that the associated flux into a closed volume equals the flux out. The associated flux of ${\bf S}$ can be quantified by the number of streamlines equally spaced at the inflow boundary. These ${\bf S}$ streamlines (purple lines in Fig.~\ref{IDR_EDR}(a)) into the IDR (blue region) do not end within the region of $\nabla\cdot {\bf S}=0$. 
Approaching the X-line, the magnetic field strength decreases and eventually vanishes due to the symmetry of this system. The ${\bf S}$ streamlines thus need to get around this singular point and exit at the outflow direction. This results in an intrinsically ``diverting'' ${\bf S}$ streamline pattern around the X-line, consistent with the presence of $S_x$ $=$ $-cE_zB_y/4\pi$, as discussed in Fig.~\ref{Hall_fields}(a) and (b).
These streamlines play a role analogous to railroad tracks in guiding the transport of incoming (magnetic) energy through the IDR. From Fig.~\ref{IDR_EDR}(a), we realize that if ${\bf E}={\bf E}_{\rm Hall}$ none of the upstream magnetic energy can be transported to the X-line due to this diverting ${\bf S}$ streamline pattern. 
%Another way to understand this ``diverting pattern'' is realizing that the inflowing electrons will be deflected by the Hall quadrupole magnetic field (Fig.~\ref{Hall_fields}(a)) through the Lorentz force into the outflow direction, carrying the magnetic energy away from the x-line. i.e., since electrons remains {\it frozen-in} within the IDR outside the EDR. [Q: how about guide field case? inflow electrons all gyrate in the same way.]

As time proceeds, an energy void centered around the X-line develops. Without energy input, no pressure (either thermal or magnetic) can be built up at the X-line. The upstream magnetic pressure around the energy void will then locally pinch the upstream magnetic field lines. This is a localization mechanism needed for the open outflow geometry and fast reconnection \citep{yhliu22a}. Here we point out three more important observations. First, the diverted energy is deposited on the outflow symmetry line (i.e., $z=0$) downstream of the X-line, which helps establish the pressure balance across the exhaust (in the normal direction), keeping the exhaust open. This difference of energy content at the X-line and its downstream region itself also implies the localization of the diffusion region. Second, this diverting {\bf S} streamline pattern persists even in an (initially) elongated reconnection layer, but such a layer is not sustainable, as noted above. Third, in resistive-MHD, $\nabla\cdot {\bf S}$ $=$ $-{\bf J}\cdot{\bf E}$ $\simeq$ $-\eta J_y^2$ $<$ $0$ since the resistivity $\eta$ is always positive. Thus, diverting ${\bf S}$ streamlines are not required (i.e., $S_x\simeq 0$ is possible). The $\bf S$ streamlines can end and distribute energy uniformly on the outflow symmetry line, in favor of maintaining the pressure balance across the X-line. This is why the diffusion region in Sweet-Parker reconnection is not localized.

To quantify the degree of localization, we need to estimate the thermal pressure at the x-line. The key is that ${\bf J}\cdot{\bf E}\simeq 0$ inside the Hall-dominated IDR, which limits the energy conversion to particles and thus also limits the difference in the $zz$-component of the pressure tensor between the X-line and the far upstream asymptotic region 
$\Delta P_{zz}^{\rm xline} \equiv P_{zz}\vert_{\rm xline}- P_0$
(illustrated as the green arrow in Fig.~\ref{IDR_EDR}a). Given that magnetic pressure $B^2/8\pi=0$ at the antiparallel reconnection X-line, as long as $\Delta P_{zz}^{\rm xline} < B_{x0}^2/8\pi$, the inflowing reconnecting field bends toward the X-line to restore the force-balance condition $\nabla (P+B^2/8\pi)=({\bf B}\cdot \nabla){\bf B}/4\pi$ \citep{yhliu20a}. This bending makes the outflow exhausts open out. 

These observations can be used to determine the separatrix slope $S_{\rm lope}$, then using the $R-S_{\rm lope}$ relation discussed in the previous subsection (\ref{maximum_rate}); it will provide a first-principles prediction of reconnection rate in collisionless plasmas. Specifically, after recognizing that some limited energy still goes to the ballistically accelerated incoming ions \citep{wygant05a,aunai11a}, one can derive the pressure difference between the $d_e$- and $d_i$-scale \citep{yhliu22a},
\begin{equation}
P_{izz}\vert_{d_i}^{d_e} \simeq \frac{2}{3}\left(\frac{B_{xi}^2-B_{xe}^2}{8\pi}\right), 
\label{Pizz}
\end{equation}
where $B_{xe}$ is the reconnecting magnetic field at the inflow boundary of the EDR.
and the slope of the separatrix associated with the open outflow geometry can then be determined by analyzing the inflow force balance at point ${\bf 7}$ of Fig.~\ref{IDR_EDR}(b),
\begin{equation}
S_{{\rm lope}}\simeq \sqrt{\frac{1}{3}\left[\frac{1-(B_{xe}/B_{xi})}{1+(B_{xe}/B_{xi})}\right]},
\label{S}
\end{equation} 
where
\begin{equation}
\frac{B_{xe}}{B_{xi}} \simeq \left(\frac{m_e}{m_i}\right)^{1/4}
\label{Bxe_Bxi}
\end{equation}
is derived through coupling to the EDR. The cross-scale coupling from the mesoscale upstream region down to the IDR, and then the EDR is achieved by recognizing that the magnetic field line tends to straighten itself out (when it is possible). Thus, the separatrix slope $S_{\rm lope}$ is similar in these different regions.
For the real proton-to-electron mass ratio $m_i/m_e=1836$, the total pressure increase along the inflow symmetry line to the X-line was derived to be $\Delta P^{xline}_{zz}\simeq 0.25 B_{x0}^2/8\pi$, and the resulting reconnection rate $R\simeq 0.16$ from Eq.~(\ref{R}), consistent with numerical simulations in Fig.~\ref{GEM_challenge}, in-situ observations \citep{Genestreti2018JGR_tail,NakamuraT2018JGR,Torbert2018Sci,NakamuraR2019JGR} discussed in Sec.~\ref{subsubsec2}, and other examples discussed in a previous review \citep{cassak17a}. 

%\subsubsection{\color{red} Observational Evidence}\

\subsection{Asymmetric Reconnection}
While ``symmetric'' magnetic reconnection discussed in the previous subsection (Sec. \ref{standard_symmetric_reconnection}) is a reasonable approximation to the energy release process during geomagnetic substorms at Earth's magnetotail \citep[e.g.][]{angelopoulos08a,paschmann13a}, magnetic reconnection at Earth's magnetopause is ``asymmetric'', as it occurs at the boundary layer between the magnetosphere plasmas and magnetosheath plasmas \citep[e.g.][]{paschmann13a,paschmann05a,paschmann79a}, where the plasma and magnetic field conditions on two sides of the current sheet can be very different. In this subsection, we will generalize the theoretical modeling into this configuration.

\label{asymmetic_MR}
\subsubsection{Cassak-Shay Scaling}
To predict the rate of asymmetric reconnection in terms of upstream plasma parameters, we make the same simplifying assumptions typically made for such studies: two-dimensionality, steady state, upstream asymptotic magnetic fields are straight and anti-parallel, no bulk flow upstream except for the inflow, and the upstream plasmas are in local thermodynamic equilibrium.  The analysis is carried out in the reference frame in which the X-line is stationary.  We use subscripts ``1'' and ``2'' to denote the two upstream sides of the reconnection site, and the upstream reconnecting magnetic field strengths are $B$, number densities are $n$, and temperatures are $T$. For definiteness, if the magnetic field strength is stronger on one side than the other, we take the stronger magnetic field side to be ``2'', so that $B_2 \geq B_1$.  

First, we note that there must be a pressure balance in the MHD sense across the current sheet in the upstream asymptotic regions
\begin{equation}
    P_1 + \frac{B_1^2}{8 \pi} = P_2 + \frac{B_2^2}{8 \pi}. \label{eq:asympressbal}
\end{equation}
Here the total plasma pressure is $P= \sum_s^{i,e} n k_B T_s$. In writing Eq.~(\ref{eq:asympressbal}), we ignore the ram pressure due to the inflow speed $V_{\rm in}$. This is justifiable {\it a posteriori} because the resulting normalized reconnection rate $R$ is $\mathcal{O}(0.1)$ and the ratio of the inflow kinetic energy density $(1/2) nm_i V_{\rm in}^2$ to the upstream magnetic pressure $B^2/8 \pi$ scales like $R^2$, so the contribution of the ram pressure due to the inflow is at the $1\%$ level. Pressure balance follows from the momentum equation; if pressure balance were not satisfied, the current sheet would have a net force on it and would accelerate, which would violate the assumption that the system is in a steady state.

\begin{figure}[ht]
\centering
\includegraphics[width=3.2in]{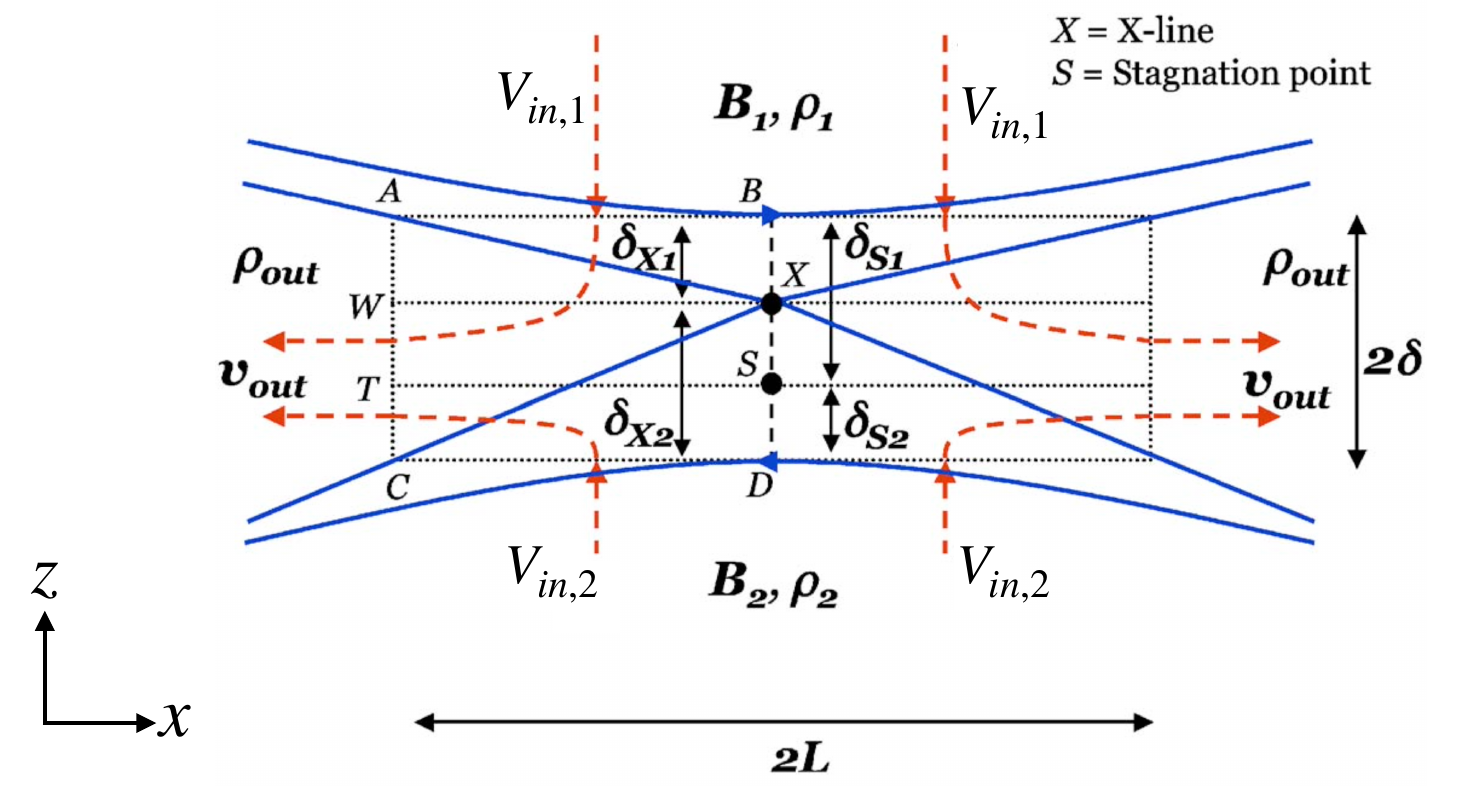} 
\caption { {\bf Sketch of asymmetric reconnection diffusion region.} Magnetic field lines (blue solid lines) and bulk flow streamlines (red dashed lines) in asymmetric reconnection.  The outer gray rectangle denotes the edge of the diffusion region.  $X$ denotes the location of the X-line, and $S$ denotes the location of the stagnation point, where the in-plane magnetic field and bulk flow go to zero, respectively.  
Reprinted from \cite{cassak07b}, with the permission of AIP Publishing.} 
%Reprinted from P.~A.~Cassak and M.~A.~Shay, ``Scaling of asymmetric magnetic reconnection:General theory and collisional simulations,'' Phys.~Plasmas, {\bf 14}, 102114 (2007) \citep{cassak07b}, with the permission of AIP Publishing.} 
\label{asymmetricDR}
\end{figure}

The most basic estimate of the asymmetric reconnection rate in terms of upstream parameters is obtained using a generalization of the classical Sweet-Parker analysis \citep{cassak07b}.  This approach simply relies on conservation laws, which impose that the flux of particles, energy, and magnetic flux coming in the upstream edge of the diffusion region must equal their fluxes leaving at the downstream edge in the steady state.  The diffusion region is assumed to be a rectangular box of half-thickness $\delta$ in the inflow direction and half-length $L$ in the outflow direction, as sketched in Fig.~\ref{asymmetricDR}.  We first treat the limit in which the process is incompressible \citep{cassak07b}.

In the steady-state in two dimensions, Faraday's law implies that the out-of-plane electric field $E_y$ must be uniform.  At the upstream edge of the (ion) diffusion region, the ideal-MHD Ohm's law is expected to be valid, so ${\bf E} + {\bf V} \times {\bf B} / c \simeq 0$.  An important result follows; defining the inflow speeds as $V_{{\rm in},1}$ and $V_{{\rm in},2}$, the constancy of $E_y$ implies 
\begin{equation}
    V_{{\rm in},1} B_{x1} \sim V_{{\rm in},2} B_{x2}. \label{eq:asymfluxbal}
\end{equation}
Since we assume $B_{x2} \geq B_{x1}$, this result implies the stronger magnetic field convects into the diffusion region more slowly.  By conservation of particles, the flux of particles entering the diffusion region must equal its flux as it leaves, which is quantified as 
\begin{equation}
    (n_1 V_{{\rm in},1} + n_2 V_{{\rm in},2}) L \sim 2 n_{{\rm out}} V_{{\rm out}} \delta, \label{eq:asymmassbal}
\end{equation}
where $n_{{\rm out}}$ and $V_{{\rm out}}$ are the number density and (outflow) bulk speed at the downstream edge of the diffusion region.  Similarly, the conservation of energy implies
\begin{equation}
\left(V_{{\rm in},1} \frac{B_{x1}^2}{8\pi} + V_{{\rm in},2} \frac{B_{x2}^2}{8\pi}\right) L \sim 2 \left(\frac{1}{2} n_{{\rm out}}m_i V_{{\rm out}}^2\right) V_{{\rm out}} \delta. \label{eq:asymenergybal}
\end{equation}
Finally, it was argued that the outflow number density scales as
\begin{equation}
n_{{\rm out}} \sim \frac{n_1 B_{x2} + n_2 B_{x1}}{B_{x1} + B_{x2}}, \label{eq:asymdensity}
\end{equation}
which follows from the plasmas mixing in proportion to the volume of the flux tubes on either upstream side, since the weaker magnetic field side reconnects more volume than the stronger magnetic field.  Putting the results of Eqs.~(\ref{eq:asymfluxbal}) through (\ref{eq:asymdensity}) together give predictions for the outflow speed $V_{\rm out}$ and the reconnection electric field $E_{R,asym}$:
\begin{eqnarray}
V_{{\rm out}} \simeq V_{A,{\rm asym}} & \sim & \sqrt{\frac{B_{x1} B_{x2}}{4 \pi n_{{\rm out}}m_i}},
%\left(\frac{B_{x1}+B_{x2}}{\rho_1 B_{x2} + \rho_2 B_{x1}}\right), 
\label{eq:asymoutflow} \\
E_{R,{\rm asym}} & \sim & 2\left(\frac{B_{x1} B_{x2}}{B_{x1} + B_{x2}}\right) \left(\frac{\delta}{L}\right) \frac{V_{{\rm out}}}{c}. \label{eq:asymreconrate}
\end{eqnarray}
This gives the desired asymmetric reconnection rate as a function of upstream parameters.  For each expression, the result reduces to the standard incompressible Sweet-Parker scaling $V_{{\rm out}} \sim V_A$ and $E_R \sim (\delta/L) V_A B_R / c$ in the symmetric limit.

The analysis described thus far did not take compressibility into account, which allows for the heating of the plasma as it passes through the diffusion region.  The analysis was extended \citep{Birn10} to include these effects.  The way to do so involves replacing magnetic energy $B^2/8\pi$ with magnetic enthalpy $B^2/4\pi$ and including the enthalpy flux $[\gamma/(\gamma-1)] P \equiv \kappa P$ (where $\gamma$ is the ratio of specific heats in the fluid description) in the energy flux balance in Eq.~(\ref{eq:asymenergybal}).  The predicted outflow speed ends up being unchanged from Eq.~(\ref{eq:asymoutflow}), but the predicted reconnection rate is multiplied by a factor of $r$ given by
\begin{equation}
r = \frac{\kappa (B_{x1} + B_{x2})}{\lambda_1 B_{x2} + \lambda_2 B_{x1}},
\end{equation}
where $\lambda_j = (1 + \kappa \beta_j)/(1 + \beta_j)$, and plasma $\beta_j = 8 \pi P_j / B_{xj}^2$ for $j = 1, 2$.  Taking the incompressible limit with either $\beta_j \rightarrow \infty$ or $\gamma \rightarrow \infty$ reproduces Eq.~(\ref{eq:asymreconrate}). A similar scaling analysis was extended to relativistic asymmetric magnetic reconnection \citep{mbarek22a}.

\subsubsection{$R-S_{\rm lope}$ relation and the maximum plausible rate}
Both results from the previous section predict the reconnection rate in terms of asymptotic upstream parameters but have a factor of $\delta / L$. It can be calculated for resistive reconnection analogously to the Sweet-Parker model \citep{cassak07b}, but for collisionless reconnection $\delta/L$ remained as a free parameter.  Empirically from two-fluid simulations, it was found that $\delta / L \sim 0.1$ \citep{cassak08a} for collisionless asymmetric reconnection, just like it does for symmetric reconnection \citep{shay99a}.  However, it is important to better understand why this is the case.  The analysis for symmetric reconnection used to show that 0.1 is approximately the maximum reconnection rate allowed \citep{yhliu17a} was extended to asymmetric reconnection \citep{Liu18}.

\begin{figure}[ht]
\centering
\includegraphics[width=10cm]{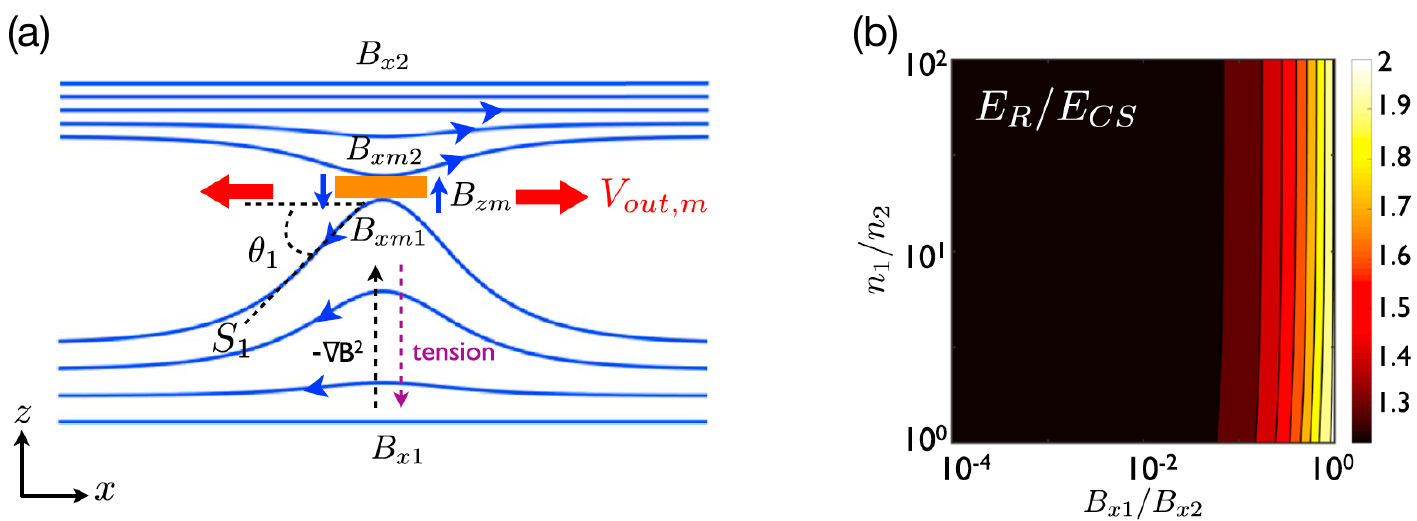} 
\caption { {\bf Estimation of the maximum plausible reconnection rate in asymmetric reconnection.} (a) The magnetic field geometry of asymmetric reconnection, where the field strength is different on two sides of the diffusion region (orange box). (b) The predicted ratio of the maximum reconnection electric field $E_R$ and $E_{R,asym}$ in Eq.~(\ref{eq:asymreconrate}) (with an effective $\delta/L=0.1$) over a wide range of magnetic field and density asymmetries is within a factor of 2. Reprinted from \cite{Liu18}, with the permission of Wiley.} 
\label{asymmetric}
\end{figure}

In this model, the reconnecting magnetic fields at the mesoscale bend in towards the reconnection site as they do in symmetric reconnection.  Force balance in the inflow direction is analogous to the symmetric reconnection case, where the magnetic curvature force opposes the magnetic pressure force between the X-line and the asymptotic region, reducing the magnetic field at the microscopic scale, as illustrated in Fig.~\ref{asymmetric}(a).  For asymmetric reconnection, however, the geometry on the two sides of the current sheet is different, specifically the slopes of the separatrix on the two sides of the current sheet.  Analogous to Eq.~(\ref{Bxi_Bx0}) for the magnetic field strength at the upstream edge of the diffusion region in symmetric reconnection, one gets
\begin{equation}
\frac{B_{xmj}}{B_{xj}}\simeq \frac{1-S_{{\rm lope},j}^2}{1+S_{{\rm lope},j}^2}, \label{eq:bmyihsinasym}
\end{equation}
for the two sides $j=1,2$ and the subscript ``$m$'' denotes the edges of the ``microscopic'' ion diffusion region. Here $S_{{\rm lope},j} = \delta_j / L$ is the slope made by the separatrix within the diffusion region, which can be different on each side and $\delta_1+\delta_2=\delta$. The reconnected magnetic field $B_{zm}$ at the downstream edge is assumed to be the same for each side.

A brief analysis predicts the outflow speed when taking into account the reduction of the upstream magnetic field and magnetic pressure, giving 
\begin{equation}
V_{{\rm out},m} \simeq \sqrt{\frac{B_{xm1}B_{xm2}}{4\pi n_{{\rm out},m}m_i}} \sqrt{1-4\frac{B_{xm1}B_{xm2}}{(B_{xm1}+B_{xm2})^2}\left(\frac{\delta}{L}\right)^2}, \label{eq:uoutyihsinasym}
\end{equation}
which generalizes Eq.~(\ref{eq:asymoutflow}) and captures that the outflow speed decreases when the exhausts open out. The subscript ``$m$'' again indicates quantities at the edges of the ion diffusion region. With Eqs.~(\ref{eq:bmyihsinasym}) and (\ref{eq:uoutyihsinasym}), an expression for the normalized reconnection rate $R$ can be obtained that is only a function of $S_{{\rm lope},1}$ (or $S_{{\rm lope},2}$) and the upstream plasma parameters, as was done for symmetric reconnection in Fig.~\ref{mesoscale}(d). The prediction of the maximum rate shows a similar scaling with magnetic field ratio and density ratio as Eq.~(\ref{eq:asymreconrate}), as shown in Fig.~\ref{asymmetric}(b).
Importantly, this analysis reveals that it is the reduction of the reconnecting magnetic field on the weak field side (i.e., $B_{x1}$ in Fig.~\ref{asymmetric})  that limits the reconnection rate. The slopes of separatrix on two sides are also predicted \citep{Liu18}.  
%Note that as long as there is some degree of diffusion region localization, the maximum reconnection rate is relatively insensitive to the opening angle.  
As of now, there has not been a first-principles calculation of the reconnection rate for collisionless asymmetric reconnection, generalizing the symmetric result in Sec.~\ref{localization}.

\subsubsection{Structure of the Diffusion Region During Asymmetric Reconnection}
\label{asym_structure}
In addition to the asymmetric conditions modifying the macroscale properties of the reconnection, such as the outflow speed and reconnection rate, they also impact the microscale physics within the diffusion region.  One key result is that the X-line (the location at which the magnetic topology changes) and the stagnation point (the location at which the in-plane bulk flow goes to zero) are not in the same location \citep{Hoshino83,Scholer89,LaBelleHamer95,nakamura00,Priest00b,Dorelli04,Mirnov06,cassak07b}. The reason follows from conservation laws \citep{cassak07b}.  From Eq.~(\ref{eq:asymfluxbal}), the inflow is slower on the high magnetic field side. Counter-intuitively, the rate at which the magnetic energy enters the diffusion region is higher on the high field side: $(V_{{\rm in},2} B_{x2}^2/8 \pi)/ (V_{{\rm in},1} B_{x1}^2/8 \pi) \sim B_{x2}/B_{x1}$.  Since no magnetic flux passes through the X-line, the X-line is displaced in the inflow direction toward the low magnetic field side so that the distance from the X-line to each side, $\delta_{X1}$ and $\delta_{X2}$ in Fig.~\ref{asymmetricDR}, has a ratio $\delta_{X1}/\delta_{X2}\simeq B_{x2}/B_{x1}$.

Similarly, the stagnation point has no particle flux across it.  The ratio of the incoming particle flux from the 2-side to the 1-side is $n_2 V_{{\rm in},2}/n_1 V_{{\rm in},1} \simeq n_2 B_{x1}/n_1 B_{x2}$ using Eq.~(\ref{eq:asymfluxbal}).  This implies that the stagnation point is offset from the center of the diffusion region toward whichever side has the smaller $n/B$ \citep{cassak07b} as is sketched in Fig.~\ref{asymmetricDR}.  It was further shown that this analysis implies the displacement of the X-line and stagnation point both in the ion diffusion region and the electron diffusion region during collisionless reconnection \citep{Cassak09c}.

The relative location of the X-line and stagnation point has important implications for the microphysics of reconnection, including the structure of the Hall fields in the ion diffusion region, transport of plasma through the diffusion region, and energizing the plasma.  The discussion above treated only asymmetries in the inflow direction.  It has been similarly shown that an asymmetry in the outflow direction leads to the X-line and stagnation point being displaced in the outflow direction \citep{Oka08,Murphy10}.

%\subsubsection{\color{red} Observational Evidence}\

\subsection{Guide field Reconnection}
Theories in the previous two subsections (the symmetric case in Sec.~\ref{standard_symmetric_reconnection} and the asymmetric case in Sec.~\ref{asymmetic_MR}) do not include the effect of an external guide field, $B_g$, that points out of the reconnection plane. However, in many situations, there is such a magnetic component during reconnection, and we call these cases ``guide-field reconnection''. For instance, solar wind magnetic fields (interplanetary magnetic field, IMF) in the magnetosheath plasma can touch Earth's magnetopause in all possible orientations, making a wide range of magnetic shear angles with respect to the magnetosphere magnetic fields. The strength of the guide field will depend on this magnetic shear angle and the X-line orientation, as illustrated in Fig.~\ref{angles}(b) in Sec.~\ref{diamagnetic_suppression}, and also discussed in \cite{gershman24a} (this issue).

\subsubsection{Theory and Simulations}
\label{guide_MR_thoery}

\begin{figure}[h]
\centering
\includegraphics[width=12cm]{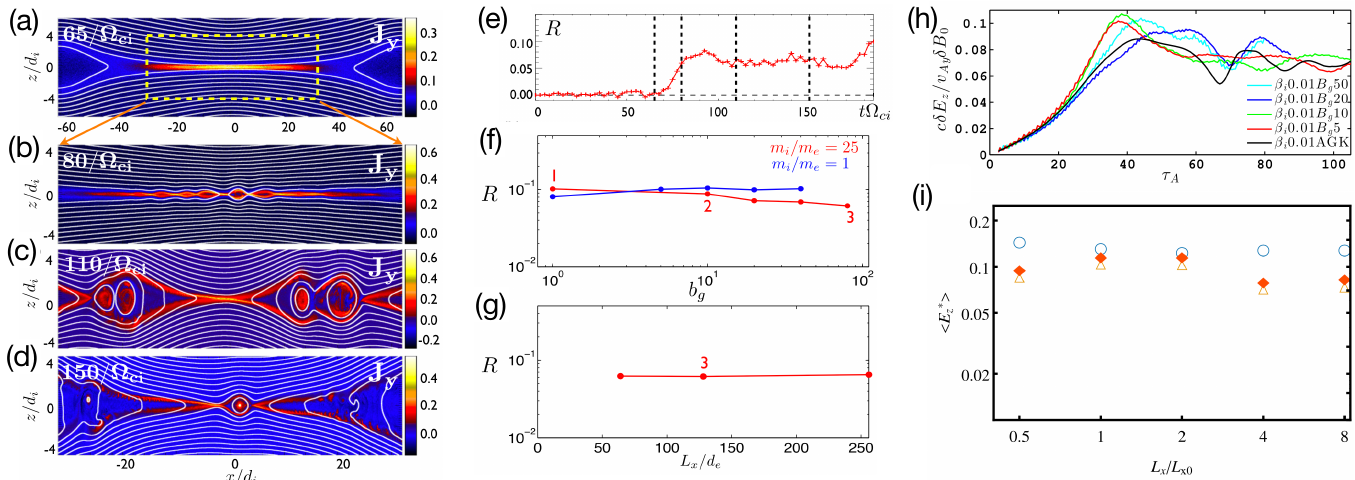} 
\caption {{\bf Guide field reconnection simulations}. (a)-(d) Evolution of an ion-scale current sheet with the guide field strength $b_g\equiv B_g/B_R=8$, white contours are flux surfaces. They correspond to the four different times indicated as vertical dashed lines in panel (e), where the time evolution of the normalized reconnection rate is shown. (a) shows the formation of the intensified out-of-plane current density arising from the initial perturbation before the onset of reconnection; (b) formation of secondary magnetic islands within this sheet; (c) coalescence of magnetic islands and their ejection from the X-line; (d) formation of another secondary island.
Panels (f) and (g) show the results of runs where the initial current sheets are on the electron scale. Panel (f) shows the rate $R$ as a function of guide field $b_g$ for a fixed system size $L_x/d_e=128$, while panel (g) shows the rate as a function of system size for fixed guide field $b_g=80$. Reprinted from \cite{yhliu14a}, with the permission of AIP Publishing.
Panel(h) shows the time-evolution of reconnection rate with different $b_g$ in gyrokinetic simulations. Reprinted from \cite{tenbarge14a}, with the permission of AIP Publishing.
Panel (i) shows the comparison of reconnection rates between PIC and the two-fluid model in different system sizes. Reprinted from \cite{stanier15b,stanier15a}, with the permission of AIP Publishing. They all show a reconnection rate $R$ on the order $\mathcal{O}(0.1)$.} 
\label{guide_Reconn}
\end{figure}

Early studies of guide field reconnection were actually motivated by the Sawtooth crashes in fusion devices \citep{von_goeler74a,kadomtsev75a,Aydemir91,aydemir92,Denton87,yamada94a,Biskamp94,beidler11a}. Reduced fluid simulations \citep{kleva95a} suggest the importance of the ion sound Larmor radius $\rho_s=\sqrt{k_BT_e/m_i}/\Omega_{ci}$, that is, the ion gyro-radius based on the electron temperature. (Some authors use the total temperature $T_e+T_i$ in the definition of the ion sound Larmor radius, e.g., \cite{rogers01a}). This kinetic spatial scale is a different ion length scale than what appears in anti-parallel reconnection, namely where  $\nabla_\| P_{e\|}$ contributes significantly to $E_\|$ in the generalized Ohm's law (Eq.~(\ref{Ohms_law})); more complete review on this scale and the relevant diffusion region signature can be found in \cite{ji23a}(this issue). Fluid simulations show that fast reconnection with an open geometry can be realized when $\rho_s$ is much larger than the resistive current sheet thickness of the Sweet-Parker solution and the electron inertial scale $d_e$ \citep{Aydemir91,Biskamp94,cassak07a}. 
If $\rho_s$ is smaller than these length scales, the current sheet tends to form an elongated Sweet-Parker-type layer but is often prone to secondary island generation, as seen in panels (a)-(d) in Fig.~\ref{guide_Reconn} \citep{Drake04,yhliu14a,stanier15b}. 
%It was later demonstrated that in a more complete fluid simulation, electron pressure anisotropy $P_{e\|}\neq P_{e\perp}$ in the generalized Ohm's law is required to realize fast reconnection \citep{cassak15a}. {\color{red}[ isotropic pressure is not enough.?]}

In the collisionless limit, there is no dispute that reconnection with a guide field of order $B_{x0}$ or smaller has a similar reconnection rate $R\sim \mathcal{O}(0.1)$ as antiparallel reconnection. However, no consensus has been reached on the large guide field limit.
It was suggested that the reconnection rate drops when the guide field weakens the dispersive property of the kinetic Alfv\'enic wave (KAW) \citep{rogers01a, tharp13a}, that drives the reconnection outflow. 
However, in several later simulations, including PIC \citep{yhliu14a}, gyrokinetic \citep{tenbarge14a}, and reduced two-fluid models \citep{huba05,stanier15a,stanier15b}, the reconnection rate $R\equiv c E_R/B_RV_A$, that is normalized to the reconnecting component, appears to be insensitive to a strong guide field that modifies the dispersive nature of the kinetic Alfv\'en wave (KAW). These results are highlighted in Fig.~\ref{guide_Reconn}. 
One should keep in mind that the relevant Alfv\'en speed $V_A\equiv B_R/\sqrt{4\pi n m_i}$ is based on the reconnecting magnetic field component only, independent of the guide field strength.
Even though empirically from simulations, the presence of a guide field may not affect the value of the collisionless reconnection rate significantly in both symmetric and asymmetric reconnection, a first-principles explanation of why it is this case remains missing. %In addition, the laboratory results suggested a lower reconnection rate with a stronger guide field \citep{tharp13a}. 

%\subsubsection{\color{red} Observational Evidence}\

\newpage

\subsection{Reconnection Rate Observation by the MMS mission}\
\label{Rate_Observation}

%NOTE: Yi-Hsin, would probably be best to have this section go after the reconnection rate models section.

The normalized reconnection rate has been determined from in-situ plasma particle and field measurements in a variety of ways \citep[e.g.][and references therein]{Hasegawa2024SSR}, e.g., the normalized reconnection electric field in the diffusion or inflow regions $E_y/B_{x0}V_{Ai0}$, the normal magnetic field component in the exhausts $B_z/B_{x0}$, the ion inflow speed in the asymptotic inflow region $V_{iz0}/V_{Ai0}$, the electron inflow speed at the inflow edge of the electron diffusion region (EDR) $V_{ez}/V_{Ae}$, the magnetic flux transport rate across the separatrices $\partial A_{y}/\partial t$, the opening angle of the separatrices (see Eq.~(\ref{R})), the aspect ratio of the EDR $(\partial B_z/\partial x)/(\partial B_x/\partial z)$, etc. Here, all quantities are evaluated in the co-moving frame of the X-line, and subscript ``0'' denotes quantities evaluated in the asymptotic inflow region. For most observations of reconnection, only a small number of these methods may be applicable, depending on where, relative to the X-line, the spacecraft collected measurements and the types of measurements that were made. 

Having a large number of rate measurements is a necessary foundation for determining how the background plasma conditions impact the rate. However, reconnection rate measurements typically are associated with large error bars, making comparative analysis difficult. Errors can arise from the determination of the appropriate coordinate system; for instance, $E_y$ in the EDR is significantly smaller than the normal electric field $E_z$, meaning that small errors in the coordinate axes can lead to large errors in the rate \citep[e.g.][]{Genestreti2018JGR_tail}. Additionally, errors may be introduced by the determination of the co-moving frame of the X-line; for instance, this frame velocity is often comparable to the upstream ion inflow speed at the magnetopause or magnetotail current sheets. Lastly, remote quantities that are determined by spacecraft far from the X-line (e.g., quantities determined in the inflow region that are used for normalization) are difficult to associate with measurements near the EDR when reconnection is time-varying and/or occurring in spatially inhomogeneous plasma conditions. 

Nevertheless, in many ways, MMS data are ideally suited for determining the reconnection rate. Unlike previous missions, MMS particle measurements are made at a rapid enough cadence to resolve the EDR, the spacecraft measures the full 3-D electric field vector, and the tightly-spaced tetrahedral formation of four spacecraft allows gradients of plasma quantities to be determined accurately. Below, we first review the reconnection rate observations derived by various methods listed above for symmetric antiparallel reconnection cases, followed by the reconnection rate observations for different background conditions, such as the asymmetry across the current sheet and the external guide field strength.  

\subsubsection{Rate observations for Symmetric Anti-parallel Reconnection}\label{subsubsec2}

The EDR crossing observation shown in Fig.~\ref{Ohms_obs1} of Sec.~\ref{Ohms_observation} took place when the average inter-probe separation was approximately 17 km. It is about half of the asymptotic electron inertial length, $d_e \simeq 30$ km. This close spacecraft distance enabled the application of multi-point analysis methods to determine the detailed characteristics of the current sheet for this event in a quantitative way, such as current sheet orientation, structure, and the spacecraft orbits within the EDR. \cite{Genestreti2018JGR_tail} estimated the reconnection rate $R = E_M/B_{L0}V_{Ai0}$ for this event by using several techniques to find the out-of-plane, $M$, direction along the reconnection electric field and estimated also the error bars using virtual data from a 2D PIC simulation \citep{NakamuraT2018JGR} performed using the initial conditions from the observation. 

\begin{figure}[ht]
\centering
\includegraphics[width=13cm]{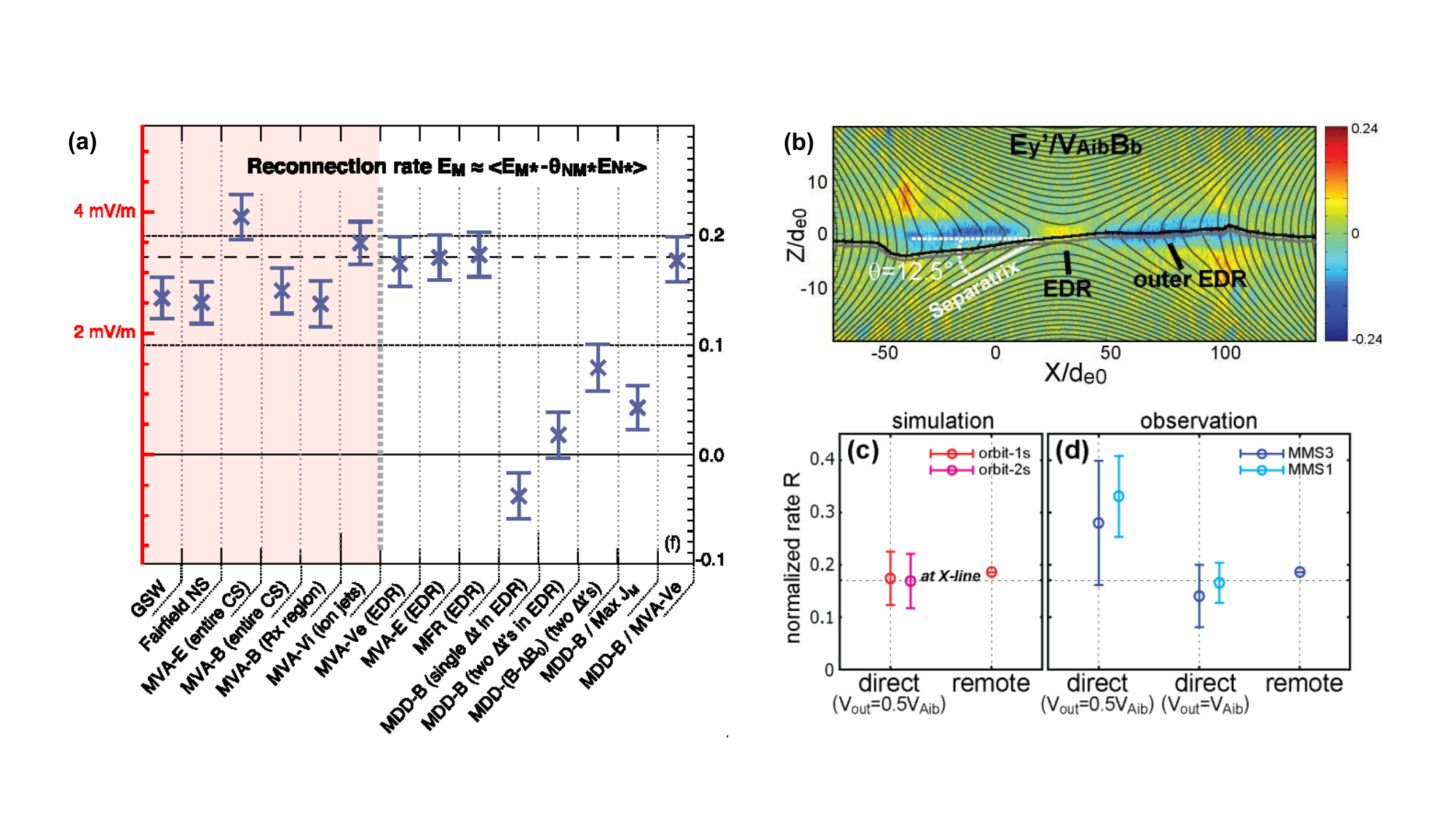} 
\caption {{\bf Reconnection rate estimation for symmetric anti-parallel reconnection event on 11 July 2017} (a) The reconnection rate in the X-line frame determined over the period 22:34:03–22:34:04 UT. The error bars mark the standard deviation of the reconnection rate over this period. The reconnection rate determined from near-EDR separatrix method   \citep{NakamuraT2018JGR} is marked by the long dashed horizontal line. The data in the red-shaded region are determined using coordinate systems that are not solely based on MMS data from within the EDR. GSW = solar-wind-aberrated geocentric solar magnetospheric; MVA = minimum variance analysis; MDD = maximum directional derivative; MFR = minimization of Faraday residue. Adapted from \cite{Genestreti2018JGR_tail}. (b) Simulated  $E_{y}'$(or $E_{R}$) with the in-plane field lines and the paths of two MMS virtual orbits. The opening angle $\theta$ of the separatrix from the simulation is estimated to be $12.5$ degree. The normalized reconnection rates $R$ directly obtained from the electric field near the EDR and remotely estimated at the separatrix  (c) for two virtual orbits (red and magenta) in the simulation shown in panels (b)  and  (d) from the MMS3 (blue) and MMS1 (cyan) observations. Adapted from 
\cite{NakamuraT2018JGR}. 
}
\label{Rate_obs1}
\end{figure}

Figure~\ref{Rate_obs1} shows the reconnection electric field $E_M$ (left axis) and normalized reconnection rate (right axis) estimated using different analysis methods to obtain the $LMN$ coordinate systems, as reviewed in \cite{Hasegawa2024SSR} (this collection). The average values for the upstream Alfv\'en speed and
lobe magnetic field, $B_{L0}$ and $V_{Ai0}$, were used for the normalization: $B_{L0}V_{Ai0}$ = 18.12 mV/m. $E_M$ is further corrected 
%by 
%{\color{red} angle theta ?} 
%by rotating the $MN$ coordinate by angle $\theta$, determinedd from the linear fit between the observed $E_M$ and $E_N$,   
to minimize the contamination from the large Hall field ($E_N$) in the estimation. A similar reconnection rate was obtained from different methods after reasonable adjustments were performed, and the normalized rate ranged between 0.14 and 0.22. The estimated reconnection rate is $E_M = 3.2$ mV/m $\pm$ 0.6 mV/m, which corresponds to a normalized rate of $R =$ 0.18 $\pm$ 0.035.  
This value well agrees with the normalized reconnection rate $R = E_M/B_{L0}V_{Ai0} = 0.18$ inside the simulated EDR by \cite{NakamuraT2018JGR} as shown in Fig.~\ref{Rate_obs1}c, which was obtained for the virtual MMS trajectory inside the simulation shown in Fig.~\ref{Rate_obs1}b. This value is also consistent with the reconnection rate, $R$ = 0.1-0.2, approximated using the aspect ratio, which is estimated from the scale-size of the current sheet from the spacecraft motion inside the EDR and the average current density \citep{Torbert2018Sci}.   

\cite{NakamuraT2018JGR} showed that the observed reconnection rate was consistent with that found in the simulation. Furthermore, the reconnection rate was estimated from the slope of the separatrix using a method introduced by \cite{yhliu17a} (see Sec.~\ref{maximum_rate}) giving R = 0.186 for the simulation and, from the MMS observations, R=0.17 as shown in (Fig.~\ref{Rate_obs1}c). \cite{Hasegawa2019JGR} obtained the opening angle of the separatrix field line from the 2D map of the magnetic field and electron streamlines by applying the electron-MHD (EMHD) reconstruction method (see details in \cite{Hasegawa2024SSR}) and obtained a similar reconnection rate, $R=$ 0.17.
The aspect ratio was also determined directly from the magnetic field gradients by \cite{Heur2022GRL} for three magnetotail symmetric anti-parallel reconnection events, including the 11 July 2017 event and a similar reconnection rate, $R=$ 0.1-0.2, was obtained.  

\begin{figure}[ht]
\centering
\includegraphics[width=12cm]{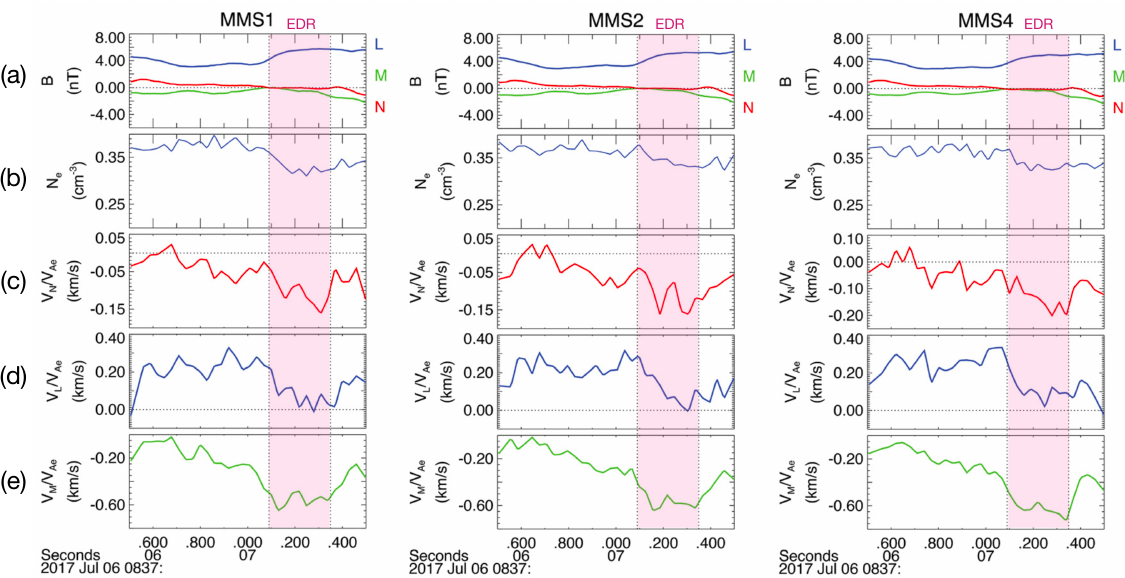} 
\caption {{\bf Plasma and field parameters near the EDR edge.}   %Vertical dotted lines marking the approximate edges of the EDR for MMS1, 2, and 4. 
The transparent red bands mark the EDR for MMS1, 2, and 4.  
(a) Magnetic field LMN components.  (b) Electron density. (c) $V_{eN}$/$V_{AeL}$. (d) $V_{eL}$/$V_{AeL}$. (e) $V_{eM}$/$V_{AeL}$, where $V_{AeL}$ is the mean electron Alfv\'en speed with $B=B_L$ for each spacecraft over the first half of each plot (08:37:06.5–08:37:07.0 UT).
%, which is an estimate of the inflow speed at the edge of the EDR. 
Adapted from \cite{Burch2022PhPl}. 
}
\label{Rate_obs2}
\end{figure}

\cite{Burch2022PhPl} determined the normalized reconnection rate from the inflow velocities normalized to the electron Alfvén speed ($V_{Ae}$) at the edge of the EDR, and compared with other methods during another magnetotail (symmetric) reconnection event on 6 July 2017.  Fig.~\ref{Rate_obs2} shows the plasma and field parameters near the EDR region. The vertical lines show the edge of the EDR (i.e., red transparent bands) for each spacecraft. For this event, the spacecraft was northward of the current sheet, and the converging inflow toward the current sheet center can be seen in the negative $V_N$. The normalized reconnection rates derived from the electron inflow velocity measurements, $V_N$/$V_{AeL}$, were 0.11–0.14 using average values of the inflow among 3 MMS spacecraft and 0.15-0.20 when maximum inflow velocity values were used. In comparison, $E_M$ normalized to the lobe inflow quantities $V_{iA}B_L$ indicates reconnection rates of 0.1-0.17. If $E_M$ is normalized to the EDR inflow quantities $V_{eA}B_L$, 
%{\color{red} [the same $B_L$ at IDR? or the $B_L$ at EDR?]} IT IS EDR inflow... in the inflow region, 
a lower reconnection rate, around 0.06, was found.
%suggesting different inflow regimes of the lobe and the EDR inflow region.  

\subsubsection{Current sheet structures and rate observations for Asymmetric and/or Guide field Reconnection}\label{rate_obs_asym}

\begin{figure}[ht]
\centering
\includegraphics[width=12.5cm]{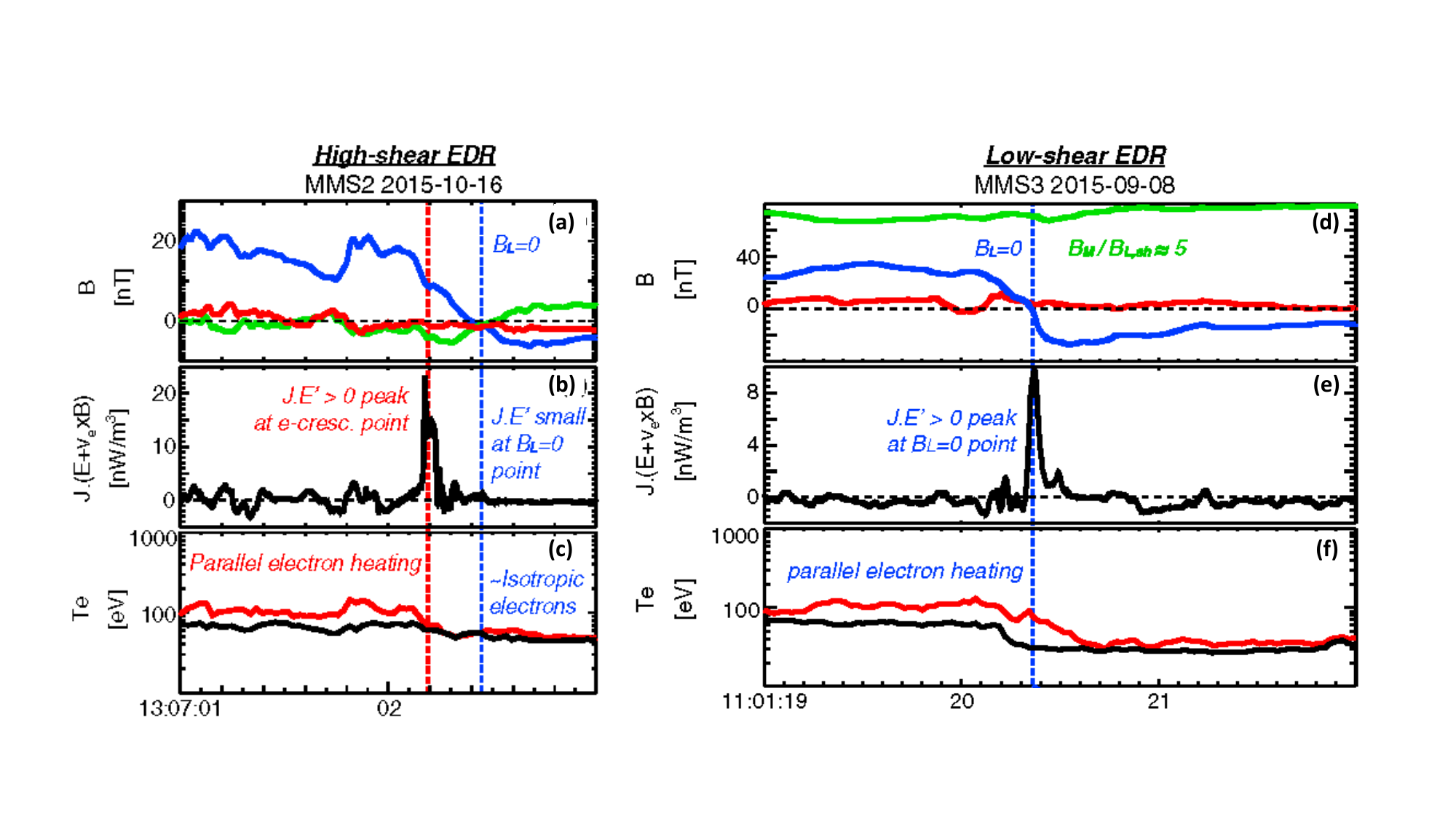} 
\caption {{\bf Observation of asymmetric reconnection events for antiparallel and guide field geometry.} (a,d) The  L(blue), M(green), N(red) components of magnetic fields, (b, e) the local energy conversion rate, ${\bf J}\cdot{\bf E'}$, (c, f) the parallel (red) and perpendicular (black) electron temperatures. The vertical dashed blue and red lines mark the $B_L = 0$ point and the electron ``crescent point''. 
Adapted from \cite{Genestreti2017JGR}. 
}
\label{Rate_obs3}
\end{figure}

As discussed in the two previous sections, the background asymmetry across the current sheet and/or the existence of the guide fields significantly modifies the structure of the reconnection current sheet, an effect that has been identified in observations. Fig.~\ref{Rate_obs3} shows two examples of MMS observations from magnetopause reconnection events. The left panels show an event on 16 October 2015 with anti-parallel field geometry, i.e., $B_M/B_L \simeq 0.1$, at the magnetosheath and a large asymmetry, i.e., the magnetosheath to magnetosphere density ratio $n_{sh}/n_{sp}= 16$. 
The right panels show an event on 8 September 2015 that has a strong guide field $B_M/B_L \simeq 5$ and a smaller density asymmetry, $n_{sh}/n_{sp}= 2.5$. 

The 16 October 2015 event (left panels) was first reported by \cite{burch16a}. The current sheet crossing took place from the magnetosphere (low density) to the magnetosheath (high density). The most pronounced feature of asymmetric reconnection is the deviation between the X-point where the $B_L=0$ (blue line) and the electron ``crescent point'' (red line), where the electron velocity distribution function (VDF) has a crescent shape indicating non-gyrotropic distribution in the flow stagnation point; this special point coincides with the peak ${\bf J}\cdot{\bf E'}= {\bf J}\cdot({\bf E} + {\bf V}_e \times {\bf B}/c)$, that was used to measure the dissipation \citep{Zenitani2011PhPl}.

The 8 September 2015 event (right panels of Fig.~\ref{Rate_obs3}), first reported by \cite{Eriksson2016PRL}, on the other hand, shows a clear peak in the energy conversion rate around the X-point, $B_L=0$ due to the dominant parallel components of the current and the electric field. Such features of the strong guide field event (i.e., $B_M/B_L>0.5$) were obtained also in a statistical study of the energy conversion rate in the magnetosheath reconnection by \cite{Wilder2018JGR}.  There was no significant non-gyrotropic electron distribution detected at the X-point during the 8 September 2015 event, indicating the effect of small gyroradius relative to the scale size of the current sheet. In modest guide field events, such as the one reported by \cite{Chen2017JGR}, energy conversion rate enhancement takes place both at the X-line and the flow stagnation point, and the parallel heating of electrons occurs at both locations. Overall, the separation of the flow stagnation point and the X point is found with density asymmetry and magnetic field asymmetry \citep{Genestreti2017JGR}, consistent with the prediction in Sec.~\ref{asym_structure}.

Although the asymmetry, as well as the guide field, significantly modifies the structure of the reconnection current sheet, the observed range of reconnection rates is similar to that in standard anti-parallel symmetric reconnection. \cite{Burch2020GRL} determined the normalized reconnection rate from the electron inflow velocities $V_{eN}$ for four MMS events, including three previously published crossings \citep{Chen2017JGR, phan18a, Pritchard2019GRL}, and obtained values between 0.05 and 0.25.
Among these four events, one event was an ``electron-only'' reconnection event in the magnetosheath \citep{phan18a} that will be discussed in the next section.

A survey of asymmetric reconnection rates has been performed by \cite{Pritchard2023JGR}, including seven magnetopause events that show values of $0.14\pm 0.09$ and seven magnetosheath events that show values of $0.16\pm 0.12$. There was no correlation between the normalized reconnection rate and guide field, as has been suggested by simulation (see Sec.~\ref{guide_MR_thoery}). A finite guide field has been also reported for reconnection events in the magnetotail in a current sheet with wave fluctuations \citep{Chen2019GRL} and varying guide field, 0.14–0.5; the normalized reconnection rate ranges between 0.05 and 0.3. 
A transient current sheet at the dipolarization front \citep{Hosner2024JGR} with a guide field 1.8 shows a normalized reconnection rate of 0.16-0.18, which is comparable to that observed during reconnection at the magnetopause and in the magnetosheath.

\subsection{Electron-only Reconnection}
\label{Electron_only_Reconnection}

%A new type of magnetic reconnection was recently identified in the magnetosheath downstream of Earth's bow shock \citep{phan19a}. These events were observed at the interface between turbulent eddies, within a small spatial domain (comparable to $d_i$ and $\rho_i$) and short time-scale (comparable to $\Omega_{ci}$ and $\omega_{pi}$).
%, resulting in ``electron-only'' reconnection.

\subsubsection{Observational Evidence}\

In turbulent plasma, reconnection has long been suggested to play a role in the dissipation of turbulent energy [e.g., \cite{Matthaeus86,servidio09a}]. The turbulent magnetosheath region downstream of Earth’s quasi-parallel bow shock often contains hundreds of small-scale current sheets in which reconnection could occur \citep{retino2007situ,sundkvist07a,yordanova16a,voros17a,Wilder2018JGR}. If standard reconnection were to operate in turbulent current sheets, the ion jets in the extended exhausts would be the easiest reconnection signature to detect. However, the ultra-high time resolution plasma and field measurements of MMS have revealed a lack of ion scale exhausts, although some electron jets were observed. It was suggested that this implies the existence of a new form of reconnection in which ions do not participate, but electrons do. This was dubbed ``electron-only'' reconnection \citep{phan18a,stawarz19a,stawarz22a}.

%Instead, the ultra-high time resolution plasma and field measurements of MMS have revealed that a novel form of reconnection in which ion jets do not form, known as ``electron-only'' reconnection, is the most common form of reconnection within magnetosheath turbulence \citep{phan18a,stawarz19a,stawarz22a}.

\begin{figure}[h]
\centering
\includegraphics[width=11cm]{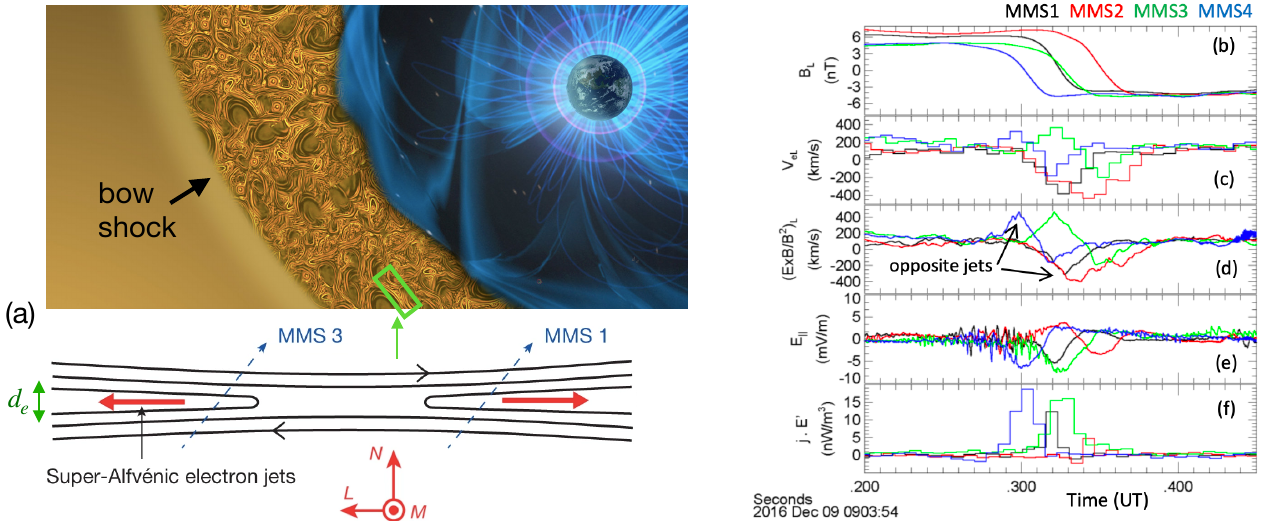} 
\caption {{\bf Observation of electron-only reconnection}. (a) Schematics showing electron-only reconnection in an electron-scale current sheet embedded in turbulent structures downstream of Earth’s quasi-parallel shock, (b) reconnecting magnetic field component (L), (c) electron outflow velocity, (d) $E\times B$ velocity in the outflow direction, (e) parallel electric field, and (f) non-ideal energy conversion ${\bf J}\cdot({\bf E}+{\bf V}_e\times {\bf B}/c)$. MMS 3 and 4 observed positive $V_{eL}$ outflow jets, while MMS 1 and 2 observed negative $V_{eL}$ jets inside the current sheet. Adapted from \cite{phan18a}, reproduced by permission of Springer Nature. The illustration in Panel(a) is credited to NASA's Goddard Space Flight Center.} 
\label{phan18}
\end{figure}

In this type of reconnection, the electron outflow jets from the reconnection X-line have speeds comparable to the electron Alfvén speed based on $B_L$ upstream of the electron diffusion region, and the current sheet width is substantially narrower than the ion Larmor radius or ion inertial scale. Importantly, in contrast to the electron diffusion region of standard reconnection, electron-only reconnecting current sheets are not embedded inside ion-scale current sheets \citep{phan18a}. 
Figure~\ref{phan18} shows a fortuitous event where pairs of MMS spacecraft simultaneously detected oppositely directed super-ion-Alfv\'enic electron outflow jets emanating from an X-line (Fig.~\ref{phan18}(c),(d)) in an electron-scale current sheet (Fig.~\ref{phan18}(b)) \citep{phan18a}. Strong parallel electric fields (Fig.~\ref{phan18}(e)) and enhanced energy conversion (Fig.~\ref{phan18}(f)) were present in the current sheet. This current sheet was one of hundreds of electron-scale current sheets in a 10-minute interval downstream of a quasi-parallel shock. Analysis of the statistical properties of this and other magnetosheath intervals measured by MMS reveals that the presence of electron-only reconnection is linked to the correlation length of the turbulence (i.e., the driving scale of the turbulence), with the correlation length of the electron-only events being several ion inertial lengths or less \citep{stawarz19a,stawarz22a}. These observations suggest that electron-only reconnection occurs in small-scale current sheets when there is insufficient space and/or time for the ions to couple to the reconnected magnetic field.

MMS has also detected sites of magnetic reconnection within the bow shock transition layer itself \citep{gingell19a,SWang19a}. The predominantly electron-only reconnection events in the shock disentangle the turbulent shock fields and may contribute to the overall shock heating. Along with complementary studies quantifying how current sheets and reconnection in the magnetosheath fit into the energy budget \citep{schwartz21a} and are influenced by the properties of the bow shock \citep{bessho19a,gingell20a,yordanova20a,bessho22a}, MMS has brought together three fundamental areas of plasma physics research – turbulence, shocks, and reconnection.
In addition to the bow shock and magnetosheath, MMS has measured electron reconnection in other contexts, most of which involve kinetic-scale turbulent structures. These included foreshock transients \citep{TZLiu20a}, electron-scale substructures inside macro-scale magnetic flux ropes \citep{man20a}, reconnection exhausts \citep{SYHuang21a,norgren18a}, magnetotail dipolarization fronts \citep{marshall20a}, and the early phase of magnetotail reconnection \citep{SLu20a}. See also the review by \cite{KHwang23a}(this issue) for further discussion of many of these phenomena. These findings suggest that electron-only reconnection is prevalent in kinetic-scale current sheets in space plasmas and could play an important role in the dissipation of turbulence energy. In the wake of this MMS discovery, electron-scale reconnection is now also studied in laboratory experiments \citep{PShi22a,greess22a,AChien23a,PShi23a}.

\subsubsection{Theory}
\begin{figure}[h]
\centering
\includegraphics[width=12cm]{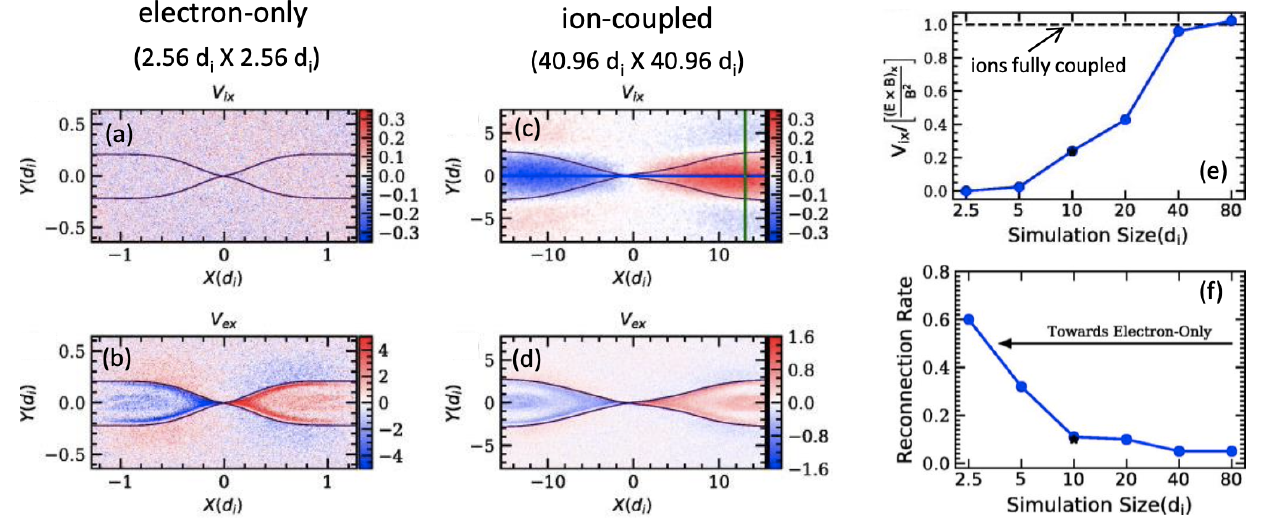} 
\caption {{\bf Transition from standard ion-coupled reconnection to electron-only reconnection}. (a,b) Ion and electron velocities in the exhaust (x or L) direction for a small 2D PIC simulation. Velocities are normalized to the asymptotic upstream proton Alfv\'en speed $V_{A}$, and $V_{Ae}=42.8V_{A}$ in this simulation. (c,d) ion and electron velocities for a larger simulation, (e) ion outflow velocity normalized to the $E \times B$ velocity as a function of simulation domain size, and (f) reconnection rate $R$ (normalized to $B_{x0}V_A$) versus simulation domain size. Adapted from \citep{pyakurel19a}.} 
\label{electron_only}
\end{figure}

A hybrid simulation study of resistive reconnection with no guide field by \cite{mandt94a} found that ions become decoupled from the magnetic field when the length of the current sheet (in the outflow direction) falls below $\sim 10 d_i$. Such reconnection without ion-coupling has also been modeled using electron magnetohydrodynamics (EMHD) simulations \citep{jain09a,jain15a}.  
Prompted by the MMS observation \citep{phan18a}, \cite{pyakurel19a} investigated the transition from ion-coupled to electron-only reconnection using particle-in-cell (PIC) simulations by varying the simulation domain size systematically. They found that the transition from fully ion-coupled to electron-only reconnection is gradual. This transition is characterized by a gradual increase in the degree to which the ions are frozen-in to the magnetic field as the simulation box size increases, with the ions being fully coupled (${\bf V}_{i\perp}\simeq {\bf E}\times {\bf B}/B^2$) when the box size reaches $40 d_i$ (Fig.~\ref{electron_only}(e)); note that the boundary conditions are periodic in the outflow direction. On the other hand, ion outflows are weakly coupled to the magnetic field when the domain size is below $20 d_i$ and clearly not coupled below $5 d_i$.
Another study suggested that it is the ion gyro-radius, instead of the inertial scale, that sets the transition to electron-only reconnection \citep{YGuan23a}.

Figures~\ref{electron_only}(a),(b) show an example of the ion and electron velocity along the outflow direction for the smallest simulated domain of $2.56 d_i \times 2.56 d_i$. The electron outflows are super-ion-Alfv\'enic outflows (Fig.~\ref{electron_only}(b)), while there are essentially no ion outflows (Fig.~\ref{electron_only}(a)). On the other hand, the simulation with a large domain size of $40.96 d_i \times 40.96 d_i$ exhibits standard ion-coupled reconnection with both ion and electron outflows (Fig.~\ref{electron_only}(c)-(d)).

Since ions are much more massive than electrons, they remain more or less immobile ($V_i\simeq 0$) compared to electrons within a small spatial and temporal scale, as shown in Fig.~\ref{electron_only}(a), and the rate of work done on ions ($en{\bf E}\cdot {\bf V}_i$) thus becomes negligible. The majority of magnetic energy is converted to electron energy. The dynamics are then described by the steady-state electron momentum equation,
\begin{equation}
\frac{({\bf B}\cdot\nabla){\bf B}}{4\pi} - e n{\bf E}\simeq nm_e({\bf V}_e\cdot \nabla){\bf V}_e,
\label{momentum_e_only}
\end{equation}
where the magnetic tension works to drive the electron outflow.
If we just consider tension and electron inertia, this equation results in electron Alfv\'enic jets with speed
\begin{equation}
V_{\rm out}\simeq V_{Ae}\simeq \frac{B_{x0}}{\sqrt{4\pi n m_e}}.
\label{Vout_e_only}
\end{equation}
The Hall electric field arising from the decoupling between the two species gives a positive $enE_x$ in Eq.~(\ref{momentum_e_only}), which slows down electrons while speeding up ions (i.e., tries to couple electrons and ions again). This partially explains why the peak electron outflow speed $V_{ex}$ is super-ion-Alfv\'enic but sub-electron-Alfv\'enic as shown in Fig.~\ref{electron_only}(b). The back-pressure (not included in Eq.~(\ref{Vout_e_only})) arising from the periodic boundaries in the outflow direction may also limit the outflow speed, especially within such a small system. 
For ions, \cite{pyakurel19a} show that the reduction of the ion outflow speeds as a function of the system's size compares well with the prediction from the ``standing wave'' approximation, where the Hall effect dominates \citep{mandt94a,rogers01a,drake08a}.%, as shown by the out-of-plane magnetic quadrupole field in Fig.~\ref{electron_only}(c).

%\begin{figure}[h]
%\centering
%\includegraphics[width=12cm]{electron_only} 
%\caption {{\bf Electron-only reconnection within a small system size} of $2.56 d_i \times 2.56 d_i$. (a) $V_{ix}$ shows now ion outflows. (b) $V_{ex}$ shows the electron outflow jets out of the EDR. Velocities are normalized to $V_{Ai}$, and $V_{Ae}=42.8V_{Ai}$ in this simulation. (c) The out-of-plane Hall quadrupole magnetic field. Panel (d) shows the scaling of the measured reconnection rate R with the system size, showing the transition from ion-coupled reconnection to ``electron-only'' reconnection. Adapted from \citep{pyakurel19a}.} 
%\label{electron_only}
%\end{figure}

Since magnetic flux remains frozen-in to electrons, the magnetic flux transport speed that determines the reconnection rate is now not limited by the ion Alfv\'en speed but by the faster electron Alfv\'en speed. 
The higher flux transport speed (Eq.~(\ref{Vout_e_only})) will make the normalized rate R (that is normalized to the proton Alfv\'en speed) higher than $\mathcal{O}(0.1)$, consistent with the simulated rate in the small system size limit, as shown in Fig.~\ref{electron_only}(f). 
If the EDR aspect ratio remains on the order of $\sim 0.1$, a rough estimate of the normalized rate is $R\le 0.1\sqrt{m_i/m_e}=\sqrt{1836}\times 0.1= 4.3$, which can only be regarded as the upper bound value because the simulated value appears to be smaller than unity (Fig.~\ref{electron_only}(f)). An analytical model better than this simple estimation needs to be derived.
%Such an ion-decoupled region could be ubiquitous in kinetic turbulence. In fact, if one zooms into the IDR of a standard ion-coupled reconnection, reconnection therein can also be viewed as electron-only reconnection. This is especially true during the onset phase of regular reconnection through electron-tearing instability before ions start to respond to the evolution of electromagnetic fields \citep{SLu20a}. 

%{\color{red} Phan:} 

%Reconnection rate. In terms of the reconnection rate, electron-only reconnection (at a small simulation domain size) exhibits much faster reconnection rates (normalized to the inflow ion Alfven speed) (Figure 2f), likely because the magnetic field motion is not limited by the ion Alfven speed [Shay and Drake, 1998; Pyakurel et al., 2019]. 

Given that electron-only reconnection tends to occur in plasma environments where the magnetic structure correlation length is small (several ion inertial lengths or less) \citep{stawarz22a}, such structures tend to be highly 3D in nature. \cite{pyakurel21a} found that the reconnection rate in 3D electron-only reconnection (with a finite X-line) is higher than in 2D. This is because, in addition to reconnection outflows in the standard exhaust direction, there is a differential mass flux out of the diffusion region along the X-line direction, enabling a faster inflow velocity and, thus, a larger reconnection rate. 
The theoretical findings of higher reconnection rates in 3D electron-only reconnection further suggest that it could play an important role in the dissipation of turbulence energy. 
%{\color{blue} The additional X-line outflow is due to the magnetic tension force, just as the exhaust tension force, allowing particles to leave the diffusion region efficiently along the X-line direction, unlike the 2D configuration.?? YL: not clear}

Observationally, \cite{Burch2022PhPl} reported a normalized reconnection rate for the \cite{phan18a} electron-only reconnection event using measurement of the inflow velocity and obtained a value of $0.25\pm 20\%$. 
\cite{Pritchard2023JGR} used measurements of the reconnection electric field for this event to determine a very similar normalized reconnection rate of $0.23\pm 43\%$. These values are at the high end of theoretical prediction, but more measurements are needed to determine whether the reconnection rates are significantly higher than for ion-coupled reconnection.

\subsection{Reconnection with Heavy and Cold Ions}

Space plasmas often have multiple ion populations, including, for instance, ions heavier than protons, or proton beams that are colder than the background protons. 
%When heavy ions are present, the overall plasma Alfv\'en velocity is reduced due to the increased mass density, resulting in a reduction of the effective reconnection electric field. Furthermore, heavy ions have larger characteristic spatial and time scales, resulting in an enlarged and multi-layered IDR. Such reconnection requires longer times to reach a steady state. On the other hand, cold ion populations have smaller gyroradii and also introduce an additional length scale to the IDR.  
In this subsection, we discuss the effects of multiple ion populations on the magnetic reconnection rate and the extension of the generalized Ohm's law to such plasmas.

\subsubsection{Theory}
It is of interest to understand how reconnection physics is affected when multiple ion species co-exist in a plasma. Similar to the treatment of a typical two-species (electron-proton) plasma, we can combine the momentum equations of multiple species into a single force balance equation,
\begin{equation}
\frac{({\bf B}\cdot\nabla){\bf B}}{4\pi}\simeq \sum_{s}^{1,2,...} n_{is} m_{is}({\bf V}_{is}\cdot\nabla){\bf V}_{is},
\end{equation}
where $m_{is}$ and $n_{is}$ represent the ion mass and density of species ``s'', respectively. The ion charge $q_{is}$ and density satisfy $\sum_s q_{is} n_{is}\simeq n_e\equiv n$ for quasi-neutrality.
Each ion species will have its own diffusion region \citep{shay2004three}.
Given a sufficiently large system, all ions will become frozen-in to the reconnection outflow outside the outermost diffusion region. With heavier ions, this will occur at larger spatial scales and longer timescales than that in the typical electron-proton plasma. The increased mass load at outflows can significantly limit the outflow speed; i.e., note that a proton is an ion of the lowest mass. The outflow speed scales as the Alfv\'en speed based on the effective mass density $\rho_{\rm heavy}=\sum_s n_{is}m_{is}$
\begin{equation}
V_{\rm out}\simeq V_{A,\rm heavy}= \frac{B_R}{\sqrt{4\pi \rho_{\rm heavy}}}.
\label{VA_heavy}
\end{equation}
Thus, based on this full mass load and the $\delta/L\sim \mathcal{O}(0.1)$ assumption, the reconnection rate normalized to $V_{A,heavy}$ is expected to be $\mathcal{O}(0.1)$. If one normalized $E_R$ to the proton Alfv\'en speed, a lower value is expected. 
However, if reconnection occurs within a small spatial domain and short time scale, the heavy ions can become decoupled, and the outflowing flux transport speed is not constrained by the Alfv\'en speed in Eq. (\ref{VA_heavy}). The reconnection electric field can thus be higher than the expected $0.1 B_R V_{A,\rm heavy}$.
This situation resembles ``electron-only" reconnection, as discussed in Sec.~\ref{Electron_only_Reconnection}, where protons are not fully coupled.

%For heavy ions to participate in reconnection, it requires a longer time and a larger system size to couple with ions, due to their larger kinetic time scale and spatial scale. If the steady state cannot be reached due to the rapid change of the system (or small system size), the mass-loading effect is not effective \citep{tenfjord2019impact}. A similar decoupling effect leads to a faster reconnection outflow speed (Eq.~(\ref{Vout_e_only})) during electron-only reconnection. On the other hand, if there is enough space to set the largest ion diffusion region in slowing varying background plasmas, the reconnection outflow speed will be the equivalent Alfv\'en speed, including all species.

\subsubsection{Results from Simulations and Observational Evidence}
%O+ multilayered

%\subsubsection{Heavy Ions}

\begin{figure}[h]
\centering
\includegraphics[width=12cm]{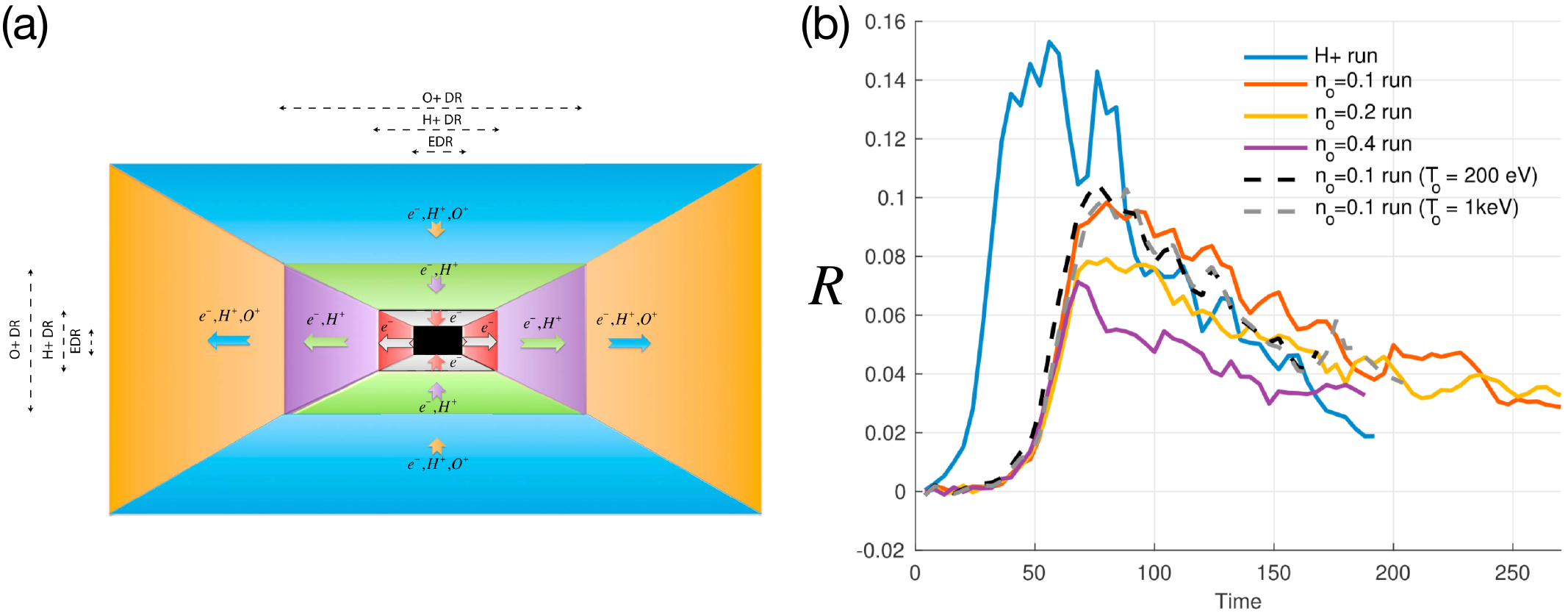} 
\caption {{\bf Reconnection rate for various amounts of oxygen (i.e., heavy ion) in PIC simulations}. (a) multi-layered diffusion regions. Adapted from \citep{liu2015heavy}. (b) Evolution of reconnection electric field in full PIC simulations of symmetric magnetic reconnection with different amounts of O$^+$ (i.e., $n_O$). Reprinted from \cite{tenfjord2019impact}, with the permission of Wiley.} 
\label{tenfjord2019}
\end{figure}

\paragraph{Heavy ions} Several PIC simulations that include three species (electrons, H$^+$, O$^+$) have shown the differential behavior of lighter and heavier ions near the x-line of magnetic reconnection, resulting in a multi-layered diffusion region with sizes related to the characteristic length-scales of each species, e.g., \citep{shay2004three, markidis2011kinetic, liu2015heavy}, as illustrated in Fig.~\ref{tenfjord2019}(a). Observational evidence of the multi-layered nature of the DR in the presence of oxygen has also been shown using Cluster \citep{escoubet01a} observations in Earth's magnetotail \citep{liu2015heavy}.

Full PIC simulations of magnetic reconnection in Earth's magnetotail that include O$^+$ have shown that the reduction in reconnection electric field does not really scale with the prediction based on the full mass-load \citep{shay2004three, markidis2011kinetic, tenfjord2019impact}.  
Figure~\ref{tenfjord2019}(b) shows the results of various full PIC simulations of magnetic reconnection with varying amounts of O$^+$, where the time evolution of the reconnection electric field is plotted. The maximum electric field drops with increasing amounts of oxygen. However, the reduction is consistent with
 \begin{equation}
%E_R \sim \frac{0.1}{1+\frac{n_O}{n_p}} B_{x0} V_{A,\rm proton},
%E_R \propto \frac{1}{1+\frac{n_{O^+}}{n_p}}, 
R \simeq \frac{0.1}{1+n_O/n_p},
 \end{equation}
rather than the reduction expected by the full mass load. i.e., $R\simeq 0.1/[1+m_O n_O/(m_pn_p)]^{1/2}$.  
This discrepancy can be explained by the fact that O$^+$ remains unmagnetized during the typical timescales of the simulations (which are related to the reconnection timescales in the Earth's magnetotail) \citep{tenfjord2019impact, kolsto2020collisionless}. Therefore, O$^+$ is ballistically accelerated by the non-ideal electric field within its diffusion region, and O$^+$ acts as an energy sink (reducing the reconnection rate), but the reduction is less severe compared with the prediction based on the full mass load. Another interesting conclusion that can be drawn from Figure~\ref{tenfjord2019}(b) is that changing the temperature of the O$^+$ population does not have an effect on the reconnection electric field. Finally, it is also noted that for larger domains and longer time scales, when all ions are magnetized outside of the diffusion region, the full mass-load scaling of the reconnection rate is expected.

\paragraph{Cold protons}
%\subsubsection{Cold Ions}

\begin{figure}[ht]
\centering
\includegraphics[width=11.5cm]{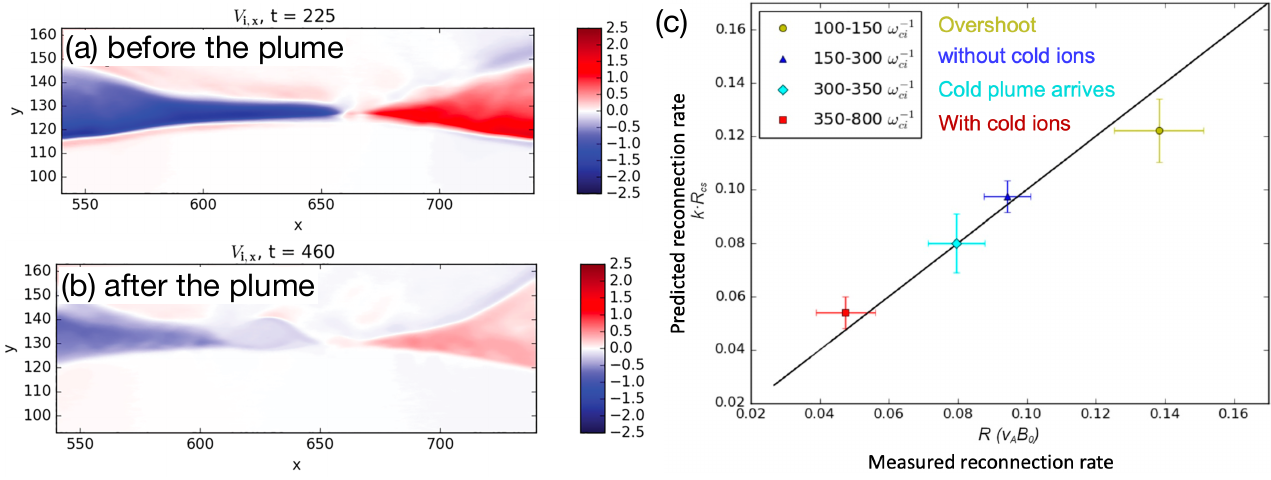} 
\caption {{\bf Reconnection electric field for various amounts of cold ions in a PIC simulation}. Panel (a) and (b) show the ion outflow speed $V_{ix}$ before and after the arrival of the cold plasma plume during asymmetric reconnection. (c) Predicted versus measured reconnection electric fields at various stages model the impact of a cold, dense plume on Earth's magnetopause reconnection. Except during the overshoot time, the scaling by \citet{cassak07b} explains the observed reductions of the reconnection electric field. Reprinted from \cite{dargent2020simulation}, with the permission of Wiley.} 
\label{dargent2020}
\end{figure}

Another common situation in magnetic reconnection in Earth's magnetosphere is having two distinct proton populations: a hot (keV-scale) component coming from the plasma sheet plus a cold (eV-scale) component arising from the Earth's ionosphere. In the following discussion, we refer to ``protons'' as simply ``ions''.

Ions demagnetize at length scales below the ion inertial length or the ion gyroradius. The ratio between the two is given by $\rho_i/d_i = \sqrt{\beta_i}$. For high-$\beta$ plasmas, the gyroradius is larger, while for low-$\beta$ plasmas, the inertial length is larger. 
At the Earth's magnetopause, the plasma beta is often of the order of 1, and therefore the two length scales are comparable. When hot and cold protons are present, cold protons have the ability to remain magnetized inside narrow structures such as the separatrix or the (hot) ion diffusion region \citep{toledo2015modification, toledo2018perpendicular, andre2010magnetic}. Under this situation, a multi-layered diffusion region is also generated due to the different gyroradius of the two proton populations. The multi-layered nature of the DR has been observed both using PIC simulations \citep{divin2016three, dargent2017kinetic, dargent2020simulation} and MMS observations \citep{toledo2016coldb}. 

%Simulations have also shown that changes in the ion temperature do not modify the reconnection rate \citep{tenfjord2019impact}. In the same way, 

\begin{figure}[ht]
\centering
\includegraphics[width=10cm]{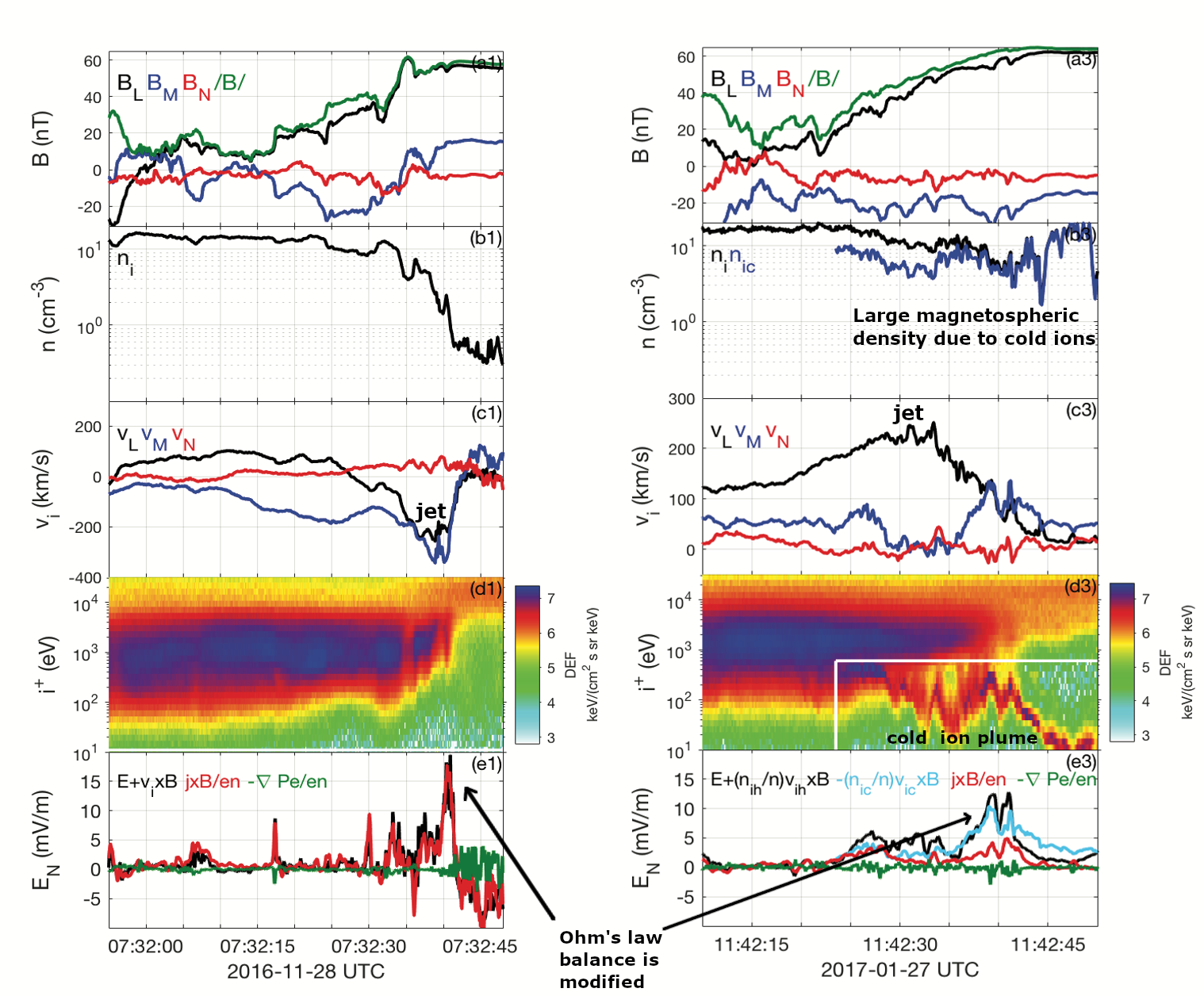} 
\caption {{\bf Generalized Ohm's law analysis including cold ions at the separatrix of dayside magnetic reconnection.} 
Comparison of two crossings of the magnetic reconnection separatrices. (a) Magnetic field in LMN coordinates. (b) total ion density and cold ion density. (c) Ion velocity in LMN coordinates. (d) Ion spectrogram. (e) Terms in the generalized Ohm's law (Eq.~(\ref{Ohms_multi})). Adapted from \cite{toledo2018perpendicular}, with the permission of Wiley.} 
\label{toledo2018}
\end{figure}

The inclusion of multiple proton populations leads to a mass-loading reduction of the reconnection electric field. However, the normalized reconnection rate ($E_R/V_{A,\rm proton}B_R$) remains unaffected \citep{divin2016three, dargent2017kinetic,dargent2020simulation,spinnangr2021micro}. Figure~\ref{dargent2020}(a)-(b) serves to illustrate the mass-loading effect by adding cold protons to magnetic reconnection at a later time. Figure~\ref{dargent2020}(c) compares the measured reconnection rate (horizontal axis) with the predicted reconnection rate using the full mass-load in the \citet{cassak07b} scaling, as discussed in Sec.~\ref{asymmetic_MR}. This setup of the PIC simulation mimics the impact of a cold, dense plume on the reconnecting Earth's magnetopause \citep{dargent2020simulation}. At $t < 150\:  \omega_{ci}^{-1}$, the maximum reconnection rate is reached in the simulation; this time is often referred to as the overshoot time. For $150 < t < 300 \:\omega_{ci}^{-1}$, reconnection proceeds, but the cold, dense plume has not yet reached the reconnection region (dark blue dot). At $t \simeq 300 \:\omega_{ci}^{-1}$, the cold, dense plume reaches the reconnection region, starts mass-loading the reconnecting flux tubes, and reduces the reconnection electric field (cyan dot). For $t > 350 \: \omega_{ci}^{-1}$, more cold ions have entrained reconnection, and the reconnection electric field is even smaller (red dot). Except during the overshoot time (yellow), all measured reconnection electric fields scale well with the predicted asymmetric reconnection electric field (Eq.~(\ref{eq:asymreconrate}) with the $\delta/L\sim 0.1$ assumption), indicating that the observed reduction corresponds to the effect of mass-load.

When both cold (eV) and hot (keV) proton populations are present in reconnection, Ohm's law can be expressed as \citep{toledo2015modification} 
\begin{equation}
{\bf E}+\frac{n_{ic}}{n}\frac{{\bf V}_{ic} \times{\bf B}}{c}+\frac{n_{ih}}{n}\frac{{\bf V}_{ih} \times{\bf B}}{c}=\frac{{\bf J}\times{\bf B}}{nec}-\frac{\nabla\cdot {\bf P}_e}{ne}+\frac{m_e}{e^2}\frac{d}{dt}\left(\frac{{\bf J}}{n}\right),
\label{Ohms_multi}
\end{equation}
where ``c'' and ``h'' indicate cold and hot populations, respectively. The electron density $n_e = n \simeq n_{ic} + n_{ih}$ , and ${\bf J} = en_{ic}{\bf V}_{ic} + en_{ih}{\bf V}_{ih} -en{\bf V}_e$.

Figure~\ref{toledo2018} shows two independent MMS crossings of the reconnecting dayside magnetopause. The magnetic field rotation can be observed in panels a1 and a2. The magnetic field amplitude at the two crossings, both at the magnetosheath and the magnetosphere, is of the same level. The magnetosheath densities are between 10 - 20 cm$^{-3}$ on the two crossings, but the density on the magnetosphere for crossing 1 ($\sim0.5$ cm$^{-3}$) is much smaller than for crossing 2 ($\sim11$ cm$^{-3}$), due to the presence of a cold ion plume in the latter (see panel d2). The ion velocities (panels c1 and c2) show ion jets consistent with the prediction in Eq.~(\ref{eq:asymoutflow}) \citep{cassak07b}. The normal component (N) of the Ohm's law terms is plotted in panels e1 and e2. For crossing 1, there is no cold ion population ($n_i=n_{ih}$, ${\bf V}_i={\bf V}_{ih}$) and the term ${\bf E}+{\bf V}_i \times {\bf B}/c$ is balanced by ${\bf J}\times {\bf B}/enc$, while for crossing 2 the term ${\bf E}+(n_{ih}/n_i){\bf V}_{ih}\times {\bf B}/c$ is mostly balanced by $-(n_{ic}/n_i){\bf V}_{ic}\times {\bf B}/c$, and ${\bf J} \times {\bf B}/enc$ contributes only a small fraction. The reason is that for crossing 2, the abundant cold ions remain magnetized inside the separatrix layer and reduce the perpendicular currents \citep{toledo2018perpendicular}. 

\subsection{High-$\beta$ Reconnection} 

While more magnetic energy is available in the low plasma $\beta\equiv P/(B^2/8\pi) \ll 1$ regime, reconnection also occurs in systems with high $\beta \gg 1$.
Such plasmas can be found in the outer heliosphere ($\beta$ up to 10) \citep{drake10a,schoeffler11a}, at the Galactic center ($\beta\sim 10^1-10^2$) \citep{marrone07a}, or in the hot intracluster medium (ICM) of galaxy clusters ($\beta\sim 10^2-10^4$) \citep{carilli02a,schekochihin06a}.

\begin{figure}[ht]
\centering
\includegraphics[width=12cm]{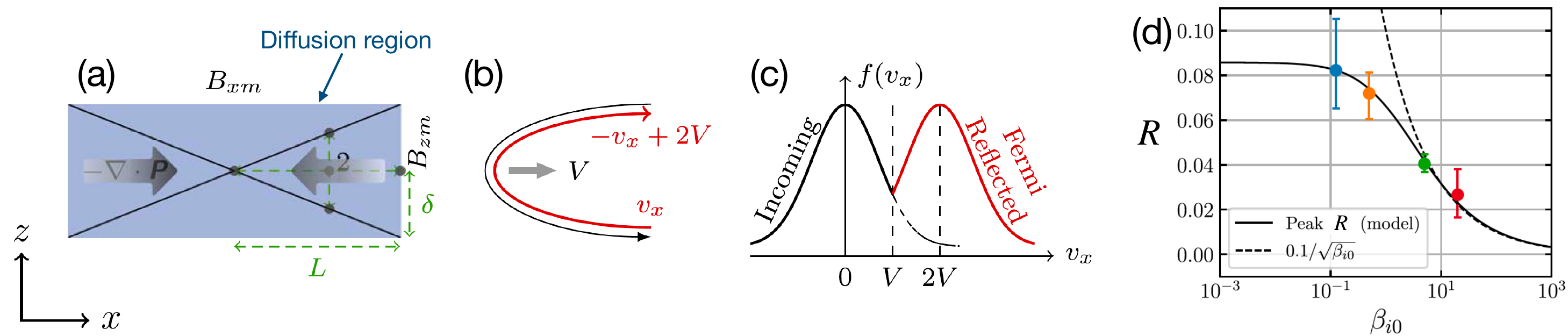} 
\caption {{\bf Including thermal pressure effect in the reconnection rate model.} (a) Illustrates the back-pressure (grey arrows) within the diffusion region, that opposes the outflow. (b) and (c) show how Fermi reflection is revealed in the particle distribution. (d) Shows that the predicted maximum plausible rate (solid curves) explains well the simulated rates (dots) in cases with different plasma-$\beta$. Adapted from \cite{XLi21a}, reproduced by permission of the AAS.} 
\label{high_beta}
\end{figure}

\subsubsection{Theory}
In this limit, self-generated pressure anisotropy and/or pressure gradients both upstream \citep{egedal13a} and downstream \citep{bessho10a,yhliu11b,yhliu12a,le14a,haggerty18a} of the diffusion region can affect the force balance that determines the maximum plausible reconnection rate. Thus, unlike in Sec.~\ref{maximum_rate}, we need to include plasma pressure effects in the mesoscale force balance,
\begin{equation}
\nabla\cdot \left(\varepsilon \frac{{\bf B}{\bf B}}{4\pi}\right)\simeq \frac{\nabla B^2}{8\pi}+\nabla P_\perp+nm_i({\bf V}\cdot \nabla){\bf V}
\end{equation}
where the pressure anisotropy (i.e., firehose) factor $\varepsilon=1-4\pi (P_\|-P_\perp)/B^2$ and $P_\|$ ($P_\perp$) refers to the pressure component parallel (perpendicular) to the local magnetic field. Using this force balance, one can follow the framework in Sec.~\ref{maximum_rate} to derive the general $R-S_{\rm lope}$ relation  \citep{XLi21a}. 

Specifically, the back-pressure $\nabla P_\perp$ can oppose the outflow, and a pressure anisotropy of $\varepsilon < 1$ can weaken the magnetic tension. It was shown that ion Fermi reflections of ions in the outflow region (illustrated in Fig.~\ref{high_beta}(b)) play the dominant role in increasing the back-pressure (illustrated in Fig.~\ref{high_beta}(c)) and reducing the reconnection outflow speed in the high-$\beta$ limit. 
The predicted maximum plausible reconnection rate scales as $R_{max}\simeq 0.1/\sqrt{\beta_i}$ in the high upstream ion beta ($\beta_i$) limit, comparing well with PIC simulations in Fig.~\ref{high_beta}(d).

\subsubsection{Observational Evidence}

\begin{figure}[ht]
\centering
\includegraphics[width=11.5cm]{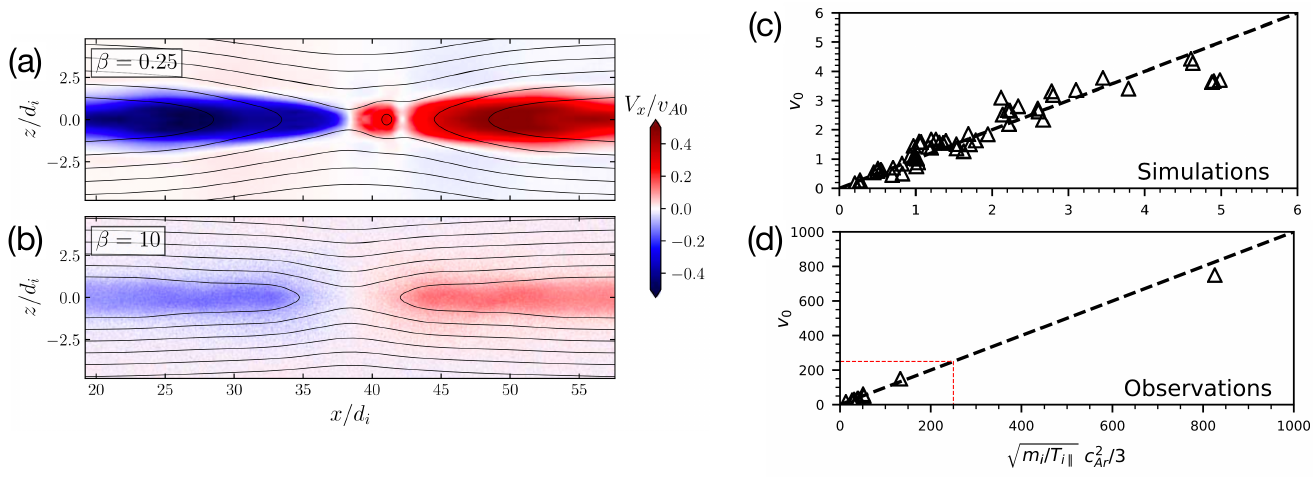} 
\caption {{\bf The scaling of outflow speed as a function of plasma $\beta$}. Ion outflow speeds in (a) low-$\beta$ and (b) high-$\beta$ plasmas Adapted from \cite{XLi21a}. In panel (c), the asymptotic $E\times B$ outflow velocity (presuming the ion outflow speed for magnetized ions) versus the outflow speed prediction for PIC simulations and for observations (d) in nearly anti-parallel events. Adapted from \cite{haggerty18a}} 
\label{high_beta_observation}
\end{figure}

The same theory \citep{XLi21a} also makes an important correction to the outflow speed that can be tested using observation. 
In the low-$\beta$ limit,
\begin{equation}
V_{\rm out}\simeq V_{{\rm out},m} \simeq 0.43 V_{A}.
\end{equation}
Here subscript ``$m$'' denotes the outflow edges of the ``microscopic'' ion diffusion region. The first equality holds in the small-aspect-ratio limit.
This prediction explains why the outflow speed is usually around half of the Alfv\'enic speed, as is often observed in space.
In the high-$\beta$ limit, the outflow speed is further reduced
\begin{equation}
V_{\rm out}\simeq V_{{\rm out},m}\simeq \frac{\sqrt{\pi}}{4}\frac{\varepsilon_mV_{A}}{\sqrt{\beta_{i}}},
\label{Vout_beta}
\end{equation}
where $\varepsilon_m$ is the anisotropic parameter at the inflow edge of the IDR. 
Equation (\ref{Vout_beta}) is almost identical to the expression obtained in \cite{haggerty18a}, $V_{\rm out}\simeq (1/3)\varepsilon_m V_{A}^2/(T_{i\|}/m_i)^{1/2}$, which adapted an empirical factor of $1/3$ in their model. Regardless of the difference in the approach, most importantly, both expressions agree well with 81 kinetic simulations and 14 in situ observations that span a wide range of parameter regimes, as shown in Fig.~\ref{high_beta_observation}(c)-(d).

\subsection{Reconnection Suppression by Diamagnetic Drifts and Sheared Flows}\

In the usual scenario, the outflow directions from a reconnection X-line ($\pm L$ in $LMN$ coordinates) are equivalent, and hence one expects the outflow jets to be symmetric.  However, in certain situations the X-line itself can move in one direction or the other.  Not only does this motion break the outflow symmetry, but if the motion is sufficiently fast, it can disrupt an outflow jet.  When this happens, reconnection itself can be suppressed.  This effect is particularly pronounced in two regimes: asymmetric reconnection in which a pressure gradient extends across the current sheet, such as planetary magnetopauses \citep[see e.g.][and references therein]{gershman24a,phan13a}, and reconnection in the presence of shear flows, such as events at the flanks of Earth's magnetopause \citep[see e.g.][and references therein]{KHwang23a}.

\subsubsection{Diamagnetic Suppression}
\label{diamagnetic_suppression}
In the presence of a magnetic field, any non-parallel pressure gradient produces a diamagnetic drift,
\begin{equation}\label{vstar}
{\bf V}^*_s = -c\frac{{\nabla} P_s {\times}{\bf B}}{q_snB^2},
\end{equation}
where $P_s=nk_BT_s$ is the thermal pressure and $q_s$ is the charge of species $s$. Somewhat famously, ${\bf V}^*_s$ is a fluid drift that does not correspond to actual particle motion; nevertheless, it can advect the magnetic field \citep{coppi65a,scott87a}. To see this, note that in a system with an invariant direction in $\hat{\bf y}$ (i.e., $\partial_y = 0$), one can write ${\bf B} = \hat{\bf y} \times \nabla \psi(x,z) + B_y(x,z)\hat{\bf y}$ where $\psi$ is the magnetic flux function. Taking the cross product of Faraday's law with $\hat{\bf y}$ yields $\partial_t \nabla \psi-c \nabla E_y=0$, or $E_y = \partial_t \psi/c$.

Next, dotting the electron momentum equation (Eq.~(\ref{e_momentum})) with $\hat{\bf y}$ gives
\begin{equation}
\label{ez}
%E_z = -\frac{1}{c} \hat{\bf z} \cdot ({\bf V}_e \times}{\bf B}) - \frac{m_e}{e}\frac{dV_{ey}}{dt}
E_y = -\frac{1}{c} \hat{\bf y} \cdot ({\bf V}_e \times {\bf B})+ E_{y, \rm non-ideal}.
\end{equation}
The last term represents the non-ideal electric field that breaks the electron frozen-in condition.  For simplicity, we ignore it and substitute for ${\bf B}$ to get $E_y = -{\bf V}_e \cdot \nabla \psi/c$ that leads to an advection equation for the flux \citep{coppi65a},
\begin{equation}\label{convection}
\partial_t \psi + {\bf V}_e \cdot \nabla \psi = 0. 
\end{equation}
Hence, the electron fluid velocity, which includes a diamagnetic component given by Eq.~(\ref{vstar}), advects magnetic structures.
(Note that if one retains the $E_{y, \rm non-ideal}$ term that can cause slippage between electrons and magnetic flux within the EDR, the result is the Magnetic Flux Transport (MFT) velocity $U_{\psi}$ \citep{yhliu16a,yhliu18c,TCLi21a}, that is basically the $E\times B$ drift speed based solely on the in-plane magnetic component; also discussed in {\color{blue} Hasagawa et al. (2024, this issue)}.)

\begin{figure}[h]
\centering
\includegraphics[width=11cm]{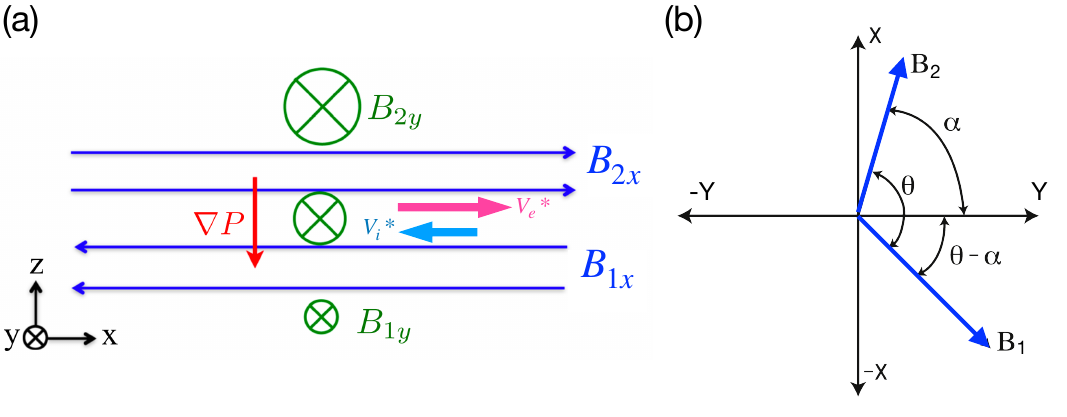} 
\caption{ \label{angles} {\bf Definition of the coordinate system}. (a) The guide field and pressure gradient give rise to diamagnetic drifts during reconnection on the $x$-$z$ plane. Adapted from \cite{yhliu16a}. (b) The view along the inflow ($z$) direction. Given the total magnetic shear angle $\theta$, the angle $\alpha$ that tells the orientation of the x-line is, in general, unknown.  Adapted from \cite{swisdak07a}, with the permission of Wiley.}
\end{figure}

Consider then, on a qualitative level, the effect of such a drift on a reconnecting X-line and specifically on the (ion) Alfv\'enic outflows (in the $\pm x$ direction in the current coordinate system), as shown in Fig.~\ref{angles}(a).  A diamagnetically drifting X-line will move in the same direction as one of the two outflows and, for certain parameters, the drift speed can exceed the outflow velocity.  This case is roughly analogous to shock propagation in that the X-line's motion is rapid enough that the X-line itself arrives downstream before any newly reconnected field lines.  As numerical simulations have shown \citep{swisdak03a}, the net effect is to choke off and suppress reconnection. The stability criterion is basically, 
\begin{equation}
\lvert V_e^*-V^*_i \rvert > V_A.
\end{equation}
where $V^*_e$ and $V^*_i$ are the electron and ion diamagnetic velocities, respectively.
A simple relationship quantifying when such suppression should occur-- particularly one that depends only on upstream parameters -- would be useful for spacecraft observations.  To derive such a condition, begin with a system characterized by magnetic field vectors ${\bf B}_1$ and ${\bf B}_2$, number densities $n_1$ and $n_2$ and pressures $P_1$ and $P_2$ on either side of a current layer.  The relation $\cos\theta = {\bf B}_1 \cdot {\bf B}_2/B_1B_2$ defines the angle $\theta$ between the asymptotic field, with $\theta = 180^{\circ}$ corresponding to anti-parallel reconnection.  The coordinate system, with the unknown angle $\alpha$, is shown in Fig. \ref{angles}(b): $B_{1x}$ and $B_{2x}$ are the reconnecting components of the field, and the respective guide fields are $B_{1y}$ and $B_{2y}$.  The X-line points parallel to $\hat{\bf y}$ while reconnection occurs in the $x-z$ plane.

The stability criterion is obtained in two steps.  The first requires determining the direction of the X-line (or, equivalently, the plane in which reconnection occurs) since this choice affects the field
components that enter the calculation of $V^*_e$.  Determining this orientation has been the subject of multiple papers \citep{swisdak07a,schreier10a,hesse13a,yhliu15b,yhliu18b} with no clear resolution, but while the exact results differ, there is general agreement that, to a reasonable approximation, the reconnection X-line bisects the angle $\theta$ between the two magnetic fields \citep{hesse13a,yhliu18b} (in other words, $\alpha=\theta/2$ in Fig. \ref{angles}(b)). The resulting outflow velocity from the X-line is given by the hybrid Alfv\'en speed,
\begin{equation}
V_{A, {\rm asym}}= \sqrt{\frac{B_1+B_2} {4\pi m_i\left(\frac{n_1}{B_1}+\frac{n_2}{B_2}\right)}} \sin{\frac{\theta}{2}}.
\label{ca}
\end{equation}
This equation agrees with Eq.~(\ref{eq:asymoutflow}).

Second, we calculate the component of the diamagnetic velocity along the outflow ($x$-) direction,
\begin{equation}
\left<V^*_x\right>=-\left<\frac{cB_y\partial_z P}{enB^2}\right>,
\end{equation}
where $\partial_z P$ is the derivative of the total pressure (electron plus ion) in the direction normal to the current layer. The angle bracket $\left< \right>$ indicates the spatial average across the current sheet of thickness $\delta$.

The stability condition (i.e., $\left<V^*_x\right> > V_A$) leads to
\begin{equation}\label{final}
\lvert\beta_1-\beta_2\rvert > \frac{2\delta}{d_i} \tan{\frac{\theta}{2}}
\end{equation}
which was first derived in 
\cite{swisdak10a}. Here $\beta_1$ and $\beta_2$ are the plasma betas on two sides.

Simulations suggest that the scale factor $\delta$ is $\approx d_i$ when the guide field is small, but $\approx \rho_s$, the sound Larmor radius, is in the opposite limit. However, the approximations made in deriving Eq.~(\ref{final}) mean that the pre-factor on the right-hand side is likely most accurately described as ``of order unity'' and so exactness is not expected.  Immediate consequences of Eq.~(\ref{final}) include: (1) Anti-parallel reconnection ($\theta = \pi$) is never subject to diamagnetic stabilization and (2) Stabilization is likely when the upstream fields nearly align (i.e., for $\theta$
small).

\begin{figure}
\begin{center}
\includegraphics[width=\textwidth]{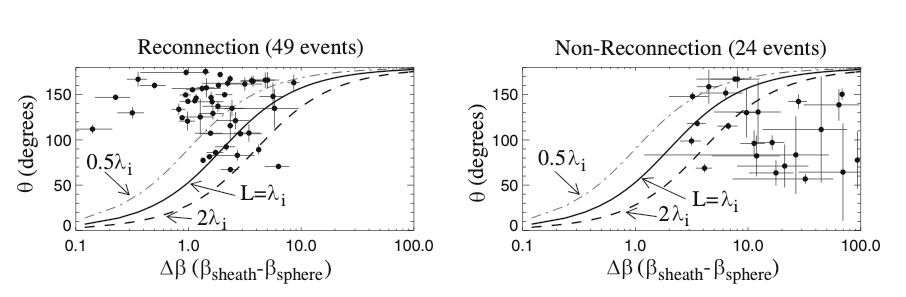}
\caption{\label{data} 
Results of statistical survey of reconnection (left) and non-reconnection (right) events. Scatter plot of magnetic shear versus $\Delta\beta$ at the magnetopause.  Reprinted from \cite{phan13a}, with the permission of Wiley.}
\end{center}
\end{figure}

\paragraph{Observational Evidence}
The condition expressed in Eq.~(\ref{final}) has been tested in a variety of locations, e.g., Earth's magnetopause \citep{phan13a}, the solar wind \citep{phan10a}, and the magnetospheres of Mercury \citep{dibraccio13a}, Jupiter \citep{desroche12a}, and Saturn \citep{masters12a}, with reasonable success.  Figure \ref{data} shows a representative example from \cite{phan13a} examining reconnection at the magnetopause.  Both panels show the $\Delta\beta-\theta$ plane.  The various curves divide the plane according to Eq.~(\ref{final}) with the upper/leftmost region where diamagnetic suppression should not occur and the lower/rightmost region where it should.  The left panel shows current sheet crossings for which reconnection was detected, while the right panel plots crossings for which no reconnection signals were observed.  To a large degree, the events are properly segregated by the diamagnetic suppression condition. However, it is important to recognize that non-reconnecting current sheets are, in general, observed even when diamagnetic suppression is not expected to operate, as other factors (e.g., current sheet thickness) may prevent reconnection onset.

\subsubsection{Sheared Flow Suppression}

\begin{figure}[h]
\centering
\includegraphics[width=10cm]{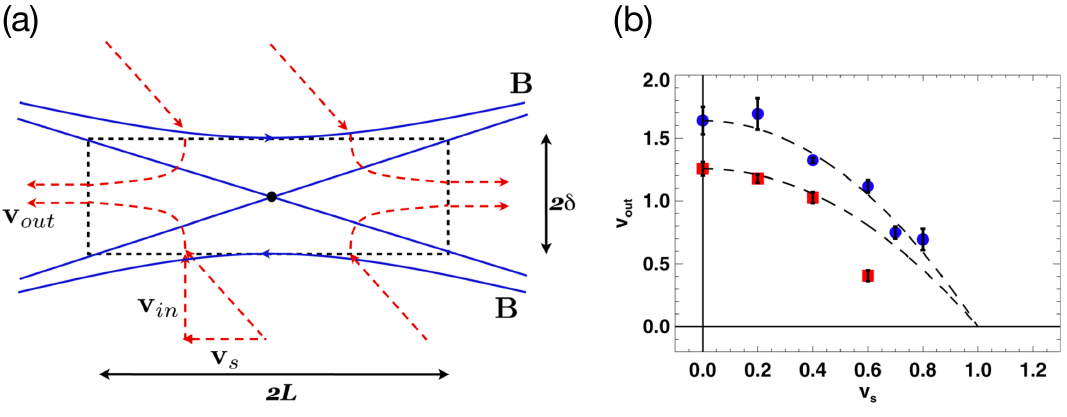} 
\caption{ (a) In-plane shear flow across the reconnection current sheet (Reprinted from \cite{cassak11a}, with the permission of AIP Publishing). (b) The outflow speed as a function of shear flow magnitude $V_{\rm shear}$ during symmetric reconnection. Reprinted from \cite{Cassak11b}, with the permission of AIP Publishing.}
%\caption{ (a) In-plane shear flow across the reconnection current sheet. Reprinted from P.~A.~Cassak and A.~Otto, ``Scaling of the magnetic reconnection rate with symmetric shear flow,'' Phys.~Plasmas, {\bf 18}, 074501 (2011) \citep{cassak11a}, with the permission of AIP Publishing. (b) The outflow speed as a function of shear flow magnitude $V_{\rm shear}$ during symmetric reconnection. Reprinted from P.~A.~Cassak, ``Theory and simulations of the scaling of magnetic reconnection with symmetric shear flow,'' Phys.~Plasmas, {\bf 18}, 072106 (2011) \citep{Cassak11b}, with the permission of AIP Publishing.}
\label{shear_suppression}
\end{figure}

Reconnection can also be suppressed by a background in-plane flow shear across the reconnection current sheet, a scenario in which there are different bulk flow speeds on either side of the reconnection site, as illustrated in Fig.~\ref{shear_suppression}(a)). The suppression criterion is qualitatively similar to that of the diamagnetic case \citep{Mitchell78,Chen90,LaBelleHamer95,cassak11a}. The presence of a flow shear $V_{{\rm shear}}$ opposes the development of reconnection outflows. If the shear flow velocity $V_{{\rm shear}} = (V_{x1} - V_{x2})/2$, where $V_{x1}$ and $V_{x2}$ are the bulk flow speed on either side of the reconnection site, is larger than the Alfv\'en speed, 
\begin{equation}
V_{\rm shear} > V_A,
\end{equation}
the reconnection outflow can not develop. Such outflow reduction was demonstrated using two-fluid simulations in Fig.~\ref{shear_suppression}(b).
A similar conclusion is extended to relativistic magnetic reconnection \citep{peery24a}, but the critical velocity is set by the relativistic Alfv\'en speed, like that discussed in Sec.~\ref{rel_reconn}.

Since magnetopause reconnection, where the upstream plasma can have a bulk flow because the magnetosheath plasma moves due to the solar wind, is asymmetric, we will discuss how flow shear impacts asymmetric reconnection. If asymmetric reconnection occurs in a region where there is a bulk flow $V_x$ in the $x$-direction (along or against the reconnecting magnetic field) that is different on either side of the diffusion region, reconnection can slow down. If the flow shear $V_{{\rm shear}} = (V_{x1} - V_{x2})/2$ is large enough, it can fully suppress reconnection \citep{doss2015asymmetric}, just as in the symmetric reconnection case.  
%{\color{red} This is because the action of the flow shear is to promote the Kelvin-Helmholtz instability.? From energetic consideration...} 
The reconnection electric field $E_{R,{\rm asym, shear}}$ was shown to scale as
\begin{equation}
E_{R,{\rm asym, shear}}\sim E_{R,{\rm asym}} \left[1-\frac{V_{{\rm shear}}^2}{V_{A,{\rm asym}}^2} \frac{4 n_1 B_{x2} n_2 B_{x1}} {(n_1 B_{x2} + n_2 B_{x1})^2} \right],
\label{eqn:asymshearrecon}
\end{equation}
where $V_{A,{\rm asym}}$ and $E_{R,{\rm asym}}$ are defined in Eq.~(\ref{eq:asymoutflow})-(\ref{eq:asymreconrate}). The critical flow shear at which reconnection is suppressed, when $E_{R,{\rm sym, shear}}$ goes to zero, is
\begin{equation}
  V_{{\rm shear,crit}} \sim V_{A, {\rm asym}} \frac{n_1 B_{x2} + n_2 B_{x1}} {2 \sqrt{n_1 B_{x2} n_2 B_{x1}}}.
\label{eqn:vcrit}
\end{equation}
In the symmetric limit, the critical flow shear is simply $V_A$, as has been well known \citep{Mitchell78,Chen90,LaBelleHamer95,cassak11a}. Interestingly, for asymmetric reconnection, the critical flow shear is faster. This implies that it is more difficult to suppress asymmetric reconnection by flow shear than symmetric reconnection \citep{doss2015asymmetric}.  It follows that magnetopause reconnection is not expected to be suppressed by flow shear \citep{doss2015asymmetric}.

\paragraph{Observational Evidence} During radial IMF conditions, the magnetosheath flow and the direction of reconnection jets become roughly aligned at the magnetopause flanks.  \citet{toledo2021solar} took advantage of MMS-Cluster conjunction to investigate the mesoscale of magnetic reconnection along the magnetopause. The MMS fleet was located near the subsolar region, while the Cluster fleet was located in the dusk flank (X$_{GSE}\sim 0$). Based on seven simultaneous crossings (magnetopause current sheet observation within 5 minutes at the two locations), the expected reconnection electric field was $E_{R,asym,shear}/E_{R,asym}=$ 0.71 - 0.98 in the flank,  based on Eq. (\ref{eqn:asymshearrecon}), and thus the effect was negligible for the seven crossings near the subsolar region. While these observations indicate that the shear flow suppression mechanism may have some impact in regulating the global coupling of the magnetosphere to the solar wind during radial IMF conditions, more observations are needed to quantify this impact.

\subsection{Reconnection Driven by Converging Flows}
While flows that oppose the development of outflows can suppress reconnection, external flows that converge into the current sheet can, in contrast, drive the reconnection process.

\begin{figure}[h]
\centering
\includegraphics[width=11.5cm]{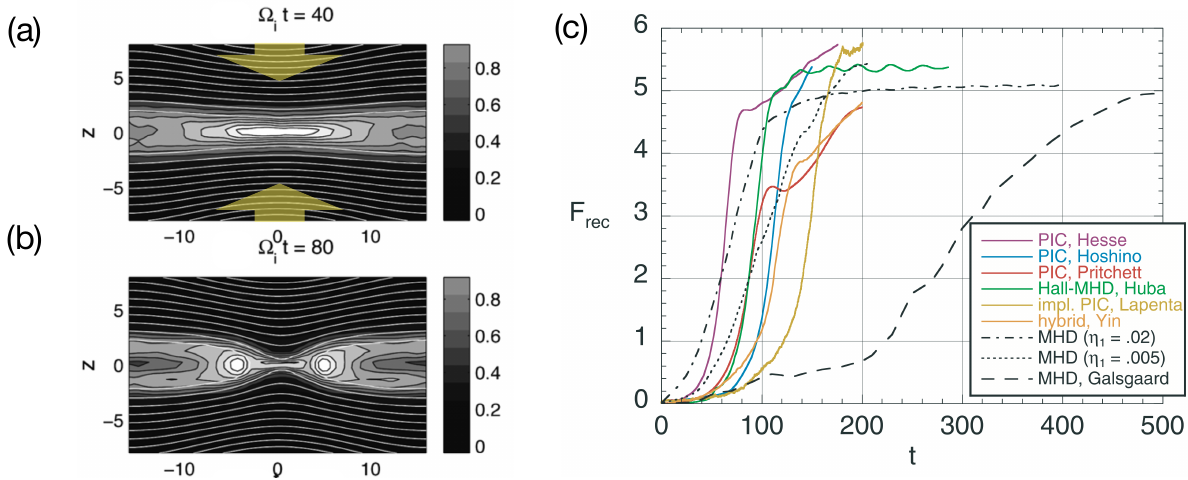} 
\caption {{\bf Newton Challenge}. (a)-(b) Reconnection driven from the top and bottom inflow boundaries within a PIC simulation. Adapted from \cite{pritchett05a}. (c) The evolution of the reconnected magnetic flux. PIC, Hall, and hybrid simulations show similar reconnection rates (given by the slope of each curve), while resistive MHD simulations show lower reconnection rates. Reprinted from \cite{birn05a}, with the permission of Wiley.} 
\label{Newton_Challenge}
\end{figure}

To compare the results of driven reconnection in various different types of simulations, a series of studies called the Newton Challenge [\cite{birn05a,pritchett05a,huba06a}; see Fig.~\ref{Newton_Challenge}] was conducted. Similar to the GEM Reconnection Challenge \citep{birn01a}, these authors used two-dimensional resistive MHD (with a localized resistivity), Hall MHD, full PIC, and hybrid simulations. Boundary inflows were imposed on both the top and the bottom boundaries with the functional form $V_{\rm in}=2a\omega \mbox{tanh}(\omega t)/\mbox{cosh}^2(\omega t)$ with $a=2d_i$ and $\omega=0.05\Omega_i$, giving the maximum inflow speed $0.08V_A$, where $V_A=B_{x0}/\sqrt{4\pi n_0 m_i}$ with $B_{x0}$ being the asymptotic reconnecting field and $n_0$ being the initial peak density in the current sheet. These inflows drive the boundary electric field $E_y=B_xV_{\rm in}/c$ out of the reconnection plane, where the maximum electric field reaches $0.1B_{x0}V_A/c$. Note that the reconnecting component $B_x$ at the boundary increases due to $V_{\rm in}$, with the maximum value around 1.1 to 1.2 times larger than the initial asymptotic value $B_{x0}$. Such a ``pileup'' of the upstream magnetic flux may compensate for any local reduction in the reconnection rate due to a weak current sheet dissipation \citep{dorelli19a}, resolving the debate of whether magnetopause reconnection rates are controlled by the solar wind driving or local reconnection physics \citep{borovsky08a,lopez16a}. 

The PIC simulation results of the Newton Challenge by \cite{pritchett05a} show that the reconnection electric field $E_R$ at the X-line increases with time, reaching a maximum of $\sim 0.12 B_{x0}V_A/c$, slightly larger than the maximum of the boundary $E_y$, after which $E_R$ decreases to values less than $0.05 B_{x0}V_A/c$. In the later stage, reconnection reaches a quasi-steady state, during which the magnetic field geometry shows an almost equilibrium state with two large magnetic islands. In all of the physical models, the final state reaches a similar equilibrium, even though the reconnection rate (the reconnection electric field) in the resistive MHD simulations in the earlier fast stage is slightly smaller than that in simulations with Hall physics. 

Although the final magnetic configurations are the same in all the simulation methods, the ``two-stage'' evolution of reconnection (the fast phase followed by the slow phase) is observed only in PIC and hybrid simulations, where particle kinetics is included. \cite{pritchett05a} explains that the following slow phase in the Newton Challenge appears after the outflows reach the periodic boundaries, and at that time, the system does not reach equilibrium with a large magnetic island. In contrast with kinetic simulations, Hall MHD \citep{huba06a} and resistive MHD \citep{birn05a} simulations do not show the two-step evolution, and the fast reconnection phase continues until the system reaches the equilibrium with a large magnetic island. These results suggest that the kinetics of ions and electrons play a role in the late-stage evolution of driven reconnection. 

While the Newton Challenge was primarily a study of 2D-driven reconnection, 3D-driven reconnection was also considered. \cite{sullivan08a} used Hall MHD simulations with external inflows similar to those in the Newton Challenge to compare the results in 2D and 3D simulations. The external inflows in 3D cases are not uniform in the $y$ (electric current) direction but are localized around $y=0$ in the simulation box. In 2D simulations, they observed the reconnection electric field scaled as predicted, $E_R\sim (\delta/L) V_{Am} B_m/c$, where $V_{Am}$ and $B_m$ are the Alfv\'en speed and the magnetic field at the edge of the diffusion region, and the aspect ratio of the diffusion region $\delta/L$ is 0.1-0.2. In contrast, in 3D runs, the reconnection rate is a factor of 2 larger than the prediction of $(\delta/L) V_{Am} B_m/c$, which is attributed to the fact that the diffusion region is localized in the $y$ direction and not uniformly distributed as in \cite{pritchett05a}. 
%{\color{red} It remains interesting to see how the reconnection change under a very strong driving.?}

\begin{figure}[ht]
\centering
\includegraphics[width=12cm]{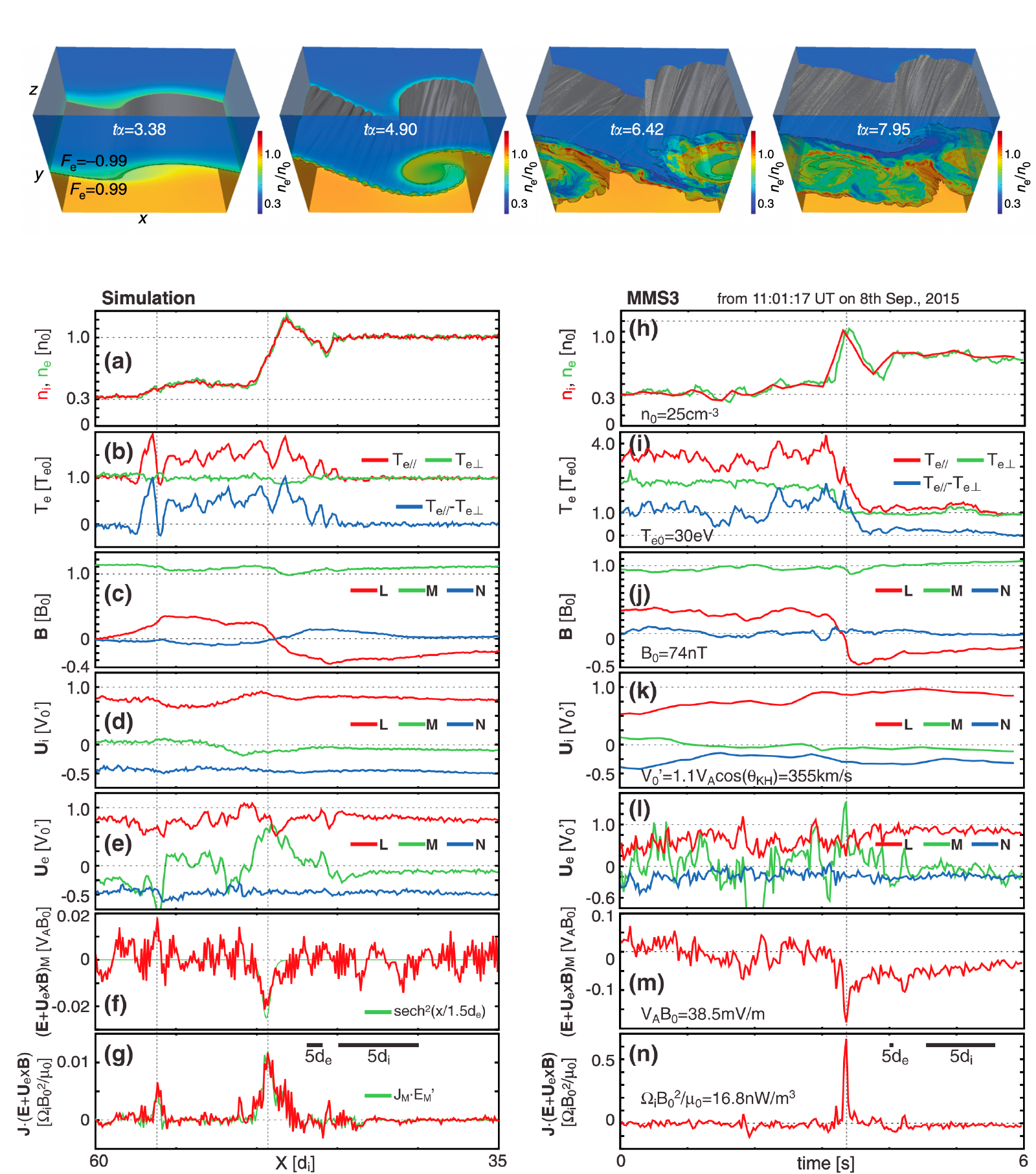} 
\caption {Top panels show the time evolution of Kelvin-Helmholtz vortex-induced reconnection (VIR) in a 3D PIC simulation \citep{TKMNakamura17b}. The rest of the panels show its comparison with MM3 data obtained on September 8, 2015. Especially, panels (f) and (m) compare the non-ideal electric field in the M direction, which shows $0.5V_{AL}B_L$ and $1V_{AL}B_L$, respectively, due to reconnection driven by the vortex flow. Reprinted from \cite{TKMNakamura17a}, with the permission of Wiley.} 
\label{KH_driven_reconnection}
\end{figure}

\subsubsection{Reconnection within Vortices, Turbulence and Shocks}
MMS has observed evidence of driven reconnection in the flanks of the Earth's magnetopause. During northward IMF reconnection occurs in the high-latitude regions, which transfer the accumulated magnetic flux to the low-latitude magnetopause, forming the so-called ``low-latitude boundary layer'' (LLBL), in which most of the plasma originates from the magnetosheath. 
In the flank side of the LLBL, strong velocity shear is unstable to the Kelvin-Helmholtz (KH) instability. The super-Alfv\'enic shear flows associated with this instability can drive reconnection within these vortices \citep[see e.g.][and references therein]{KHwang23a}. MMS detected such KH-driven reconnection \citep{eriksson16a,WLi16a,TKMNakamura17a,NakamuraT2018JGR,JHwang20a,kieokaew20a,JHwang21a}. 

\cite{TKMNakamura17a} presented 3D PIC simulations of a KH-driven reconnection event observed by MMS on 8 September 2015, as illustrated in Fig.~\ref{KH_driven_reconnection}. 
The right panels indicate that MMS3 crossed a current sheet in the reconnection region from the magnetospheric side to the magnetosheath side, during which it detected a large non-ideal electric field $E'_M\equiv({\bf E}+{\bf V}_e\times {\bf B}/c)_M=$-7 mV/m; although this value corresponds to $\sim 0.2 V_{A}B_{x0}$, it is $\sim 1V_{AL}B_L$ if one normalizes it to the $L$ component of the local magnetic field \footnote{Note that this location, where $B_L=-35nT\sim 0.5 B_{0}$, is considered to be close to the edge of the diffusion region} $B_L\sim 35$ nT $\times$ the Alfv\'en speed $V_{AL}$ based on $B_L$ and the local density 17 $\mbox{cm}^{-3}$. This $\vert E'_M \vert \sim V_{AL}B_L$ is 10 times larger than a typical standard laminar reconnection value of $0.1V_{AL}B_L$, perhaps due to the fact that reconnection is driven by the strong flows generated by the KH instability. The observational data are consistent with the 3D PIC simulation in the left panels, where the peak of $\vert E'_M \vert$ (panel (f)) is $0.5 V_{AL}B_L$, which is also 5 times larger than the standard reconnection rate.

Driven reconnection can also occur in turbulent environments. The strong flows therein can force reconnection to occur at a variety of rates.
\cite{haggerty17a} performed 2D PIC simulations of turbulent reconnection and observed normalized reconnection rates distributed between 0 to 0.5, suggesting that reconnection rates are not limited to the order of 0.1. \cite{bessho20a,bessho22a} demonstrated using 2D PIC simulations that reconnection in turbulence associated with Earth's bow shock is strongly driven by super Alfv\'enic flows, and normalized reconnection rates in both electron-only reconnection and ion-coupled reconnection can be of the order of unity.

\subsection{Turbulent 3D Reconnection}
While kinetic-scale reconnection (either ``electron-only'' or ion-coupled) within turbulent plasmas can be affected by turbulence driving, it has been proposed that the dissipation mechanism of reconnection itself may be affected by turbulence or instabilities within the diffusion region. 
Turbulent reconnection operates in large, 3D systems, where the additional degree of freedom introduces various types of instabilities \citep[e.g.][]{daughton11a}, leading to turbulence. Unlike laminar reconnection, it was proposed that turbulence may produce ``turbulence-diffusion'' [e.g., \cite{higashimori13a}] or ``anomalous dissipation'' [e.g., \cite{che11a,price16a}], modifying the diffusion region physics that breaks the MHD frozen-in condition. Other competing ideas also exist, including coupling to the Goldreich-Sridhar-like turbulence spectrum \citep{Lazarian99}, field-line super-diffusion \citep{eyink11a}, and fast field-line separation \citep{boozer12a}.

\begin{figure}[h]
\centering
\includegraphics[width=12cm]{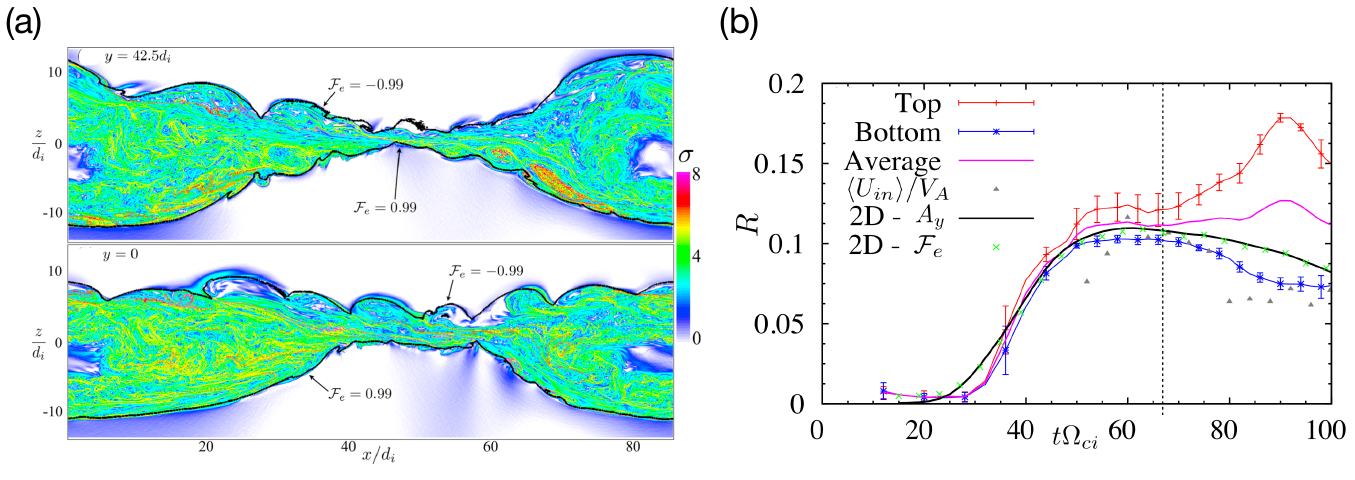} 
\caption {Panel (a) shows the field line exponentiation factor $\sigma$ at two $y$-locations within a 3D PIC simulation. The solid black lines mark the boundaries of the electron mixing fraction $\vert \mathcal{F}_e\vert=0.99$. Panel (b) shows the reconnection rate computed based on the top (red), bottom (blue), and average (purple) magnetic fluxes and using $\vert \mathcal{F}_e\vert=0.99$. Grey triangles are the inflow rates applied to the bottom region. The black curve is the corresponding 2D reconnection rate measured from the flux function $A_y$, while the green crosses are the 2D rate obtained from the same electron mixing approach. Reprinted from \cite{daughton14a}, with the permission of AIP Publishing.} 
\label{3D_Reconn}
\end{figure}

Figure~\ref{3D_Reconn} shows an example of 3D magnetic reconnection in a large PIC simulation. The entire reconnection layer becomes turbulent because of self-driven instabilities within the current sheet. Unlike in 2D models, an in-plane flux function does not exist that allows a straightforward calculation of the reconnection rate. Instead, \cite{daughton14a} devised an approach based on the electron mixing across the separatrix in full 3D systems. 
The measured reconnection rate is shown in Fig.~\ref{3D_Reconn}(b), and interestingly, the 3D reconnection rate appears to be similar to its 2D laminar counterpart [also in \cite{le18a}].
This 3D rate may still be bounded by the same geometrical constraints discussed in Sec.~\ref{maximum_rate} if the force balance is taken to work in an average sense. In addition, a broad turbulent reconnection layer is often dominated by a few active diffusion regions at the kinetic scale where the dissipation mechanism may be similar to that in Sec.~\ref{Ohms}. A thorough investigation is required to validate these assertions. 
To date, it remains challenging to model turbulent reconnection rate from first principles; more discussion on this topic can be found in {\color{blue}  Stawarz et al. (2024, this issue)}, {\color{blue} Graham et al. (2024, this issue)} and {\color{blue} Guo et al. (2024, this issue)}.

%MMS observations point out that electron-only reconnection also exists in turbulence, for example, the turbulent magnetosheath downstream of Earth's bow shock \citep{phan??}. 

%How turbulence affects the dissipation and reconnection rate during turbulent reconnection remains an interesting question. More discussion on this topic can be found in paper ??.
%The relevant works include \cite{price16a,che??}....... previous studies ..... Caution needs to be taken because turbulence inside the simulation domain can be amplified by the periodic boundary \citep{yhliu18b}.

%{\color{blue} A pretty much open question. We could just spend a few paragraphs discussing the role of unsteady physics, secondary tearing, 3D effects......etc.}\\

\subsubsection{Averaged 3D Ohm's Law}
\label{averaged_Ohms}
Other than those kinetic terms discussed in Sec. \ref{generalized_Ohms_law}, an alternative view holds that fluctuating electric fields, generated by kinetic instabilities such as Buneman modes or lower-hybrid drift effects, can effectively scatter electrons in the electron diffusion region and consequently lead to effective resistance to the reconnection electric field. The effects of such fluctuations are captured by time- or spatial averaging of the microscopic electron momentum equation (Eq.~(\ref{e_momentum})):

%\begin{equation}
%\begin{split}
%e\left<n_e\right>\left<{\bf E}\right>=-e\left<n_e{\bf V}_e\right>\times \left<{\bf B}\right>-\left(\frac{\partial\left<P_{xye}\right>}{\partial x}+\frac{\partial\left<P_{xze}\right>}{\partial z}\right)-m_e\frac{\left<n_e V_{ey}\right>}{\partial t} \\
%+m_e\nabla\cdot(\left<n_e {\bf V}_e{\bf V}_e\right>)-e\left<\delta n_e \delta {\bf E}\right>-e\left<\delta (n_e {\bf V}_e)\times \delta {\bf B}\right>+m_e \nabla\cdot \left<\delta (n_e{\bf V}_e)\detla {\bf V}_e\right>
%\end{split}
%\end{equation}

\begin{equation}
\begin{split}
e\left<n_e\right>\left<{\bf E}\right>
=-\frac{e}{c}\left<n_e\right>\left<{\bf V}_e\right>\times \left<{\bf B}\right>
-\nabla\cdot \left<{\bf P}_e\right>
-m_e\nabla\cdot(\left<n_e {\bf V}_e\right>\left<{\bf V}_e\right>)
-m_e\frac{\partial \left<n_e {\bf V}_e\right>}{\partial t}\\
-e\left<\delta n_e \delta {\bf E}\right>
-\frac{e}{c}\left<\delta (n_e {\bf V}_e)\times \delta {\bf B}\right>
+m_e \nabla\cdot \left<\delta (n_e{\bf V}_e)\delta {\bf V}_e\right>
\end{split}
\label{Ohms_averaged}
\end{equation}
Here, $\left< \right>$ denotes spatial averages \citep{le18a} (a similar equation can be obtained for time averaging, with a different form of the time derivative of the electron momentum density). Terms involving the delta symbol are fluctuating quantities, with vanishing spatial averages. The last three terms on the RHS of Eq.~(\ref{Ohms_averaged}) are often referred to as anomalous drag, anomalous momentum transport, and anomalous viscosity, respectively [e.g., \cite{buchner98a,che11a,che11c,price16a,le18a}]. It should be noted that Eq.~(\ref{Ohms_averaged}) does not contain any new or additional information: all information is already included in the microscopic description (Eq.~(\ref{e_momentum})), while Equation~(\ref{Ohms_averaged}) is obtained by spatial averaging of this microscopic equation and hence contains less information.

Translationally invariant models demonstrate, without exception, that non-gyrotropic pressure tensor effects dominate at the X-line for symmetric configurations (or, more generally, at the flow stagnation point) [e.g., \cite{pritchett01b, schmitz06a}]. Three-dimensional models of collisionless reconnection can show, however, when averaged, significant contributions of the anomalous terms and the presence of substantial fluctuations at the X point [e.g., \cite{buchner98a,che11a,che11c,KFujimoto12a,price16a,munoz16a}]. However, some local analyses continued to show the dominance of non-gyrotropic pressure terms \citep{hesse05b,YNLiu24a}, and the magnitude of these anomalous terms are sensitive to how one averages the Ohms' law \citep{le18a}. A recent, very large simulation demonstrated the near-absence of significant fluctuations at the X-line if effects of periodic boundaries can be excluded \citep{yhliu18b}. 

Prior to the Magnetospheric Multiscale mission, these two theories (i.e., anomalous dissipation versus non-gyrotropic electron pressure) were competing, and MMS had, as a key goal, to determine which of these theories was matched by reality. Beginning with the first key observation of an electron diffusion region at the magnetopause \citep{burch16a}, observations have shown remarkably quiescent electron diffusion regions, whether they are asymmetric with \citep{burch16b} or without a guide field \citep{burch16a}, or whether they are in the tail’s plasma sheet \citep{Torbert2018Sci}. While it has been difficult to measure electron pressure tensor effects directly, there has been some indication that these are indeed important \citep{Genestreti2018JGR}, and a recent observation even shows that the analytic prediction of \citep{hesse99a,hesse11a} provides a reasonable match to the observed reconnection electric field \citep{NakamuraR2019JGR}. Furthermore, a tailored, translationally invariant, numerical simulation \citep{NakamuraT2018JGR} provides an exceptionally good match between observations and model results. While observations around the outflow region show significant fluctuations and turbulent effects \citep{ergun16a,ergun18a,burch18a}, there is rapidly increasing evidence that the central electron diffusion region is indeed relatively quiescent and properly described by the quasi-viscous, electron nongyrotropy-based model \citep{hesse99a}. Therefore, it appears that MMS has accomplished its primary objective: to determine the physics behind the electron diffusion region \citep{Torbert2018Sci}.

%\subsubsection{{\color{red} Observation.?}}

\subsection{Relativistic Reconnection}
\label{rel_reconn}
In plasmas near compact astrophysical objects, such as neutron stars and black holes, the magnetic field strength is extremely strong [e.g., \cite{uzdensky11b,ripperda20a} and references therein], and the plasma flow speed can become relativistic. Under this condition, assuming an anti-parallel magnetic geometry, the relevant force balance equation becomes
\begin{equation}
\frac{({\bf B}\cdot\nabla){\bf B}}{4\pi}\simeq n'm_i({\bf U}\cdot\nabla){\bf U},
\label{force_balance_rel}
\end{equation}
where ${\bf U}=\Gamma {\bf V}$, $\Gamma \equiv [1-(V/c)^2]^{-1/2}$ is the Lorentz factor, and $n'$ is the plasma proper density. 
The resulting outflow speed (in the $x$-direction) is the relativistic Alfv\'en speed \citep{yhliu17a}, 
\begin{equation}
V_{\rm out}\simeq V_{Ax}=c\sqrt{\frac{\sigma_R}{1+\sigma_R}},
\label{rel_VA}
\end{equation}
which can approach the speed of light $c$ when the magnetization parameter $\sigma_R= B_R^2/8\pi n mc^2 \gg 1$.
With an external guide field $B_g$, the Alfv\'enic outflow speed becomes
\begin{equation}
V_{\rm out}\simeq V_{Ax}=c\sqrt{\frac{\sigma_R}{1+\sigma_R+\sigma_g}},
\label{rel_VA_guide}
\end{equation}
where $\sigma_g= B_g^2/8\pi n mc^2$. This expression can be formally derived after considering the additional momentum carried by the outflowing Poynting vector $S_x=-E_zB_y/4\pi$ (that is not included in Eq.~(\ref{force_balance_rel}), but considered in \cite{peery24a}), where the motional electric field $E_z=-V_{\rm out} B_g/c$ is associated with the convection of the guide field. 
It is interesting to note that the guide field can significantly slow down the outflow speed, unlike in the non-relativistic case. 
To comprehend this fact in another way, we see that with a guide field the total Alfv\'en speed (i.e., Eq.~(\ref{rel_VA}) with $\sigma_R$ replaced by $\sigma_R+\sigma_g$) is still limited by the speed of light $c$ due to the special relativity, and Eq.~(\ref{rel_VA_guide}) is the projection of this total Alfv\'en velocity along the magnetic field to the outflow direction \citep{melzani14a,yhliu15a}, thus its magnitude is expected to be lower than $c$. 
%Note that this discussion applies to the simple anti-parallel reconnection. A more general case with a guide field can be found in Ref. \ref{perry23a}.

Relativistic magnetic reconnection has been proposed to explain the superflares observed in the Crab Nebula and argued to cause fast radio bursts (FRBs) from neutron stars and magnetars \citep{philippov19a,mahlmann22a}. Interested readers are referred to the discussion in {\color{blue} Guo et al. (2024, this issue)}.

\section{Energy Conversion within the Diffusion Region}
\label{energy_conversion}

Aside from changing the large-scale magnetic connectivity/topology, perhaps the most important consequence of magnetic reconnection is converting magnetic energy into plasma kinetic energy and thermal energy. In this section, we collect approaches being used to quantify the energetics and energy conversion processes around the diffusion region, with a particular focus on progress enabled by MMS observations as well as recent advances in simulation capabilities. For the discussion of non-thermal particle accelerations during reconnections, a complimentary review can be found in \cite{oka23a} (this collection).

\subsection{Energy Conservation and Energy Fluxes} 
\label{Energy_Conservation_and_Energy_Fluxes}
The second moment of the Vlasov equation gives the energy equation, which in the conservative form \citep{birn10b} reads
\begin{equation}
\frac{\partial u_{\rm total}}{\partial t}+\nabla\cdot \left({\bf S}+{\bf H}+{\bf K}+{\bf q}\right)=0.
\label{energy}
\end{equation}
Here, $u_{\rm total}\equiv  \sum_s^{i,e} \left(\mbox{Tr}({\bf P}_s)/2+nm_s V_s^2/2\right)+(B^2+E^2)/8\pi $ is the total energy density with $\mbox{Tr}({\bf P}_s)\equiv \sum_j^{x,y,z}{\bf P}_{s,jj}$ being the trace of the pressure tensor, ${\bf S}\equiv c{\bf E}\times {\bf B}/4\pi$ is the Poynting vector, ${\bf H}\equiv\sum_s^{i,e} \left[(1/2)\mbox{Tr}({\bf P}_s) {\bf V}_s+{\bf P}_s\cdot {\bf V}_s \right]$ is the enthalpy flux, and ${\bf K}\equiv\sum_s^{i,e}(1/2)nm_s V_s^2 {\bf V}_s$ is the bulk-flow kinetic energy flux. ${\bf q}\equiv \sum_s^{i,e} (m_s/2)\int{\vert{\bf v}_s-{\bf V}_s\vert^2({\bf v}_s-{\bf V}_s)f_s d^3v_s}$ is the heat flux, where ${\bf v}_s$ is the particle velocity, and $f_s$ is the particle distribution function of species ``s''.

\begin{figure}[h]
\centering
\includegraphics[width=8cm]{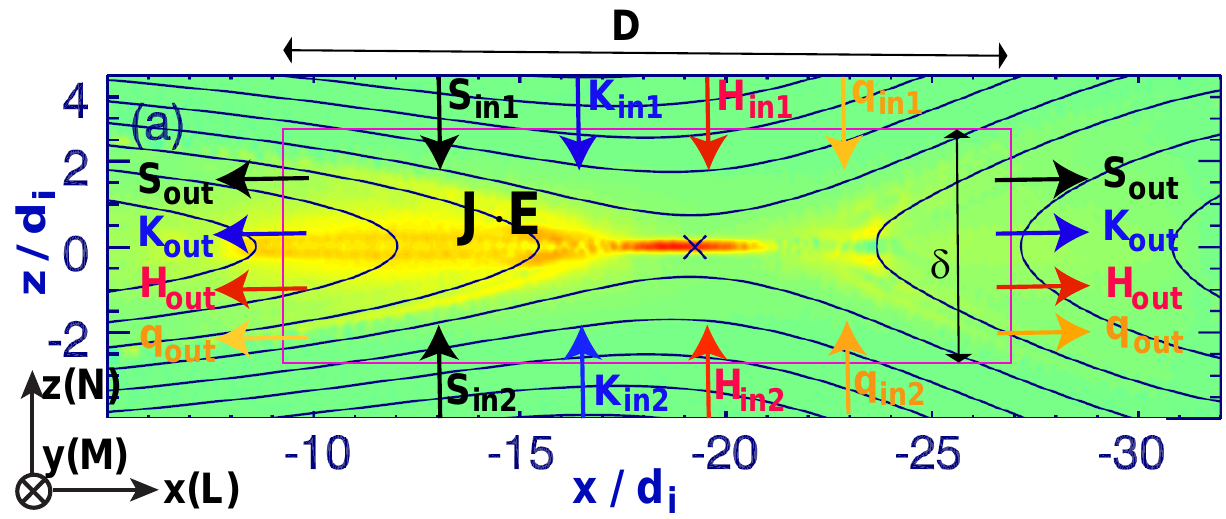} 
\caption{ {\bf Schematic of the energy conversion in the reconnection region}, modified from \citep{eastwood13a}. The color represents ${\bf J}\cdot {\bf E}$ from a PIC simulation. Reprinted from \cite{SLu18b}, with the permission of AIP Publishing.} 
\label{1illustration}
\end{figure}

Figure~\ref{1illustration} shows schematically the energy fluxes into and out of a reconnection site that is treated as invariant in the y-direction (out-of-plane). Here, for simplicity, the reconnection process is implicitly taken as being in a steady state ($\partial/\partial t\simeq 0$). We now discuss the nature of the energy fluxes in reconnection before turning to look more closely at the problem of energy conversion.

Based on magnetotail observations in the IDR by the Cluster spacecraft, in anti-parallel, symmetric reconnection the outflowing energy flux is dominated by $H_{ix}$, taking up $\sim 50 \%$ of the total, followed by $H_{ex}$ ($\sim 20 \%$) and $K_{ix}$ ($\sim 10\%$); the outflowing $S_x$ ($\sim 10\%-20\%$) is comparable to $H_{ex}$ and $K_{ix}$, and even dominates in certain regions in the IDR, where the Hall term dominates \citep{eastwood13a}. The relative rankings between different forms of energy flux are qualitatively consistent with other subsequent Cluster observations (even when considering the energetics of $O^+$ \citep{tyler16a}), kinetic simulations (e.g., \cite{birn10b,lapenta20a}) and laboratory experiments (e.g., \cite{yamada16a}; see also \cite{yamada18a} and Table II in \cite{ji23a}, this issue for a brief summary of progress in this area). 

\begin{figure}[h]
\centering
\includegraphics[width=8.7cm]{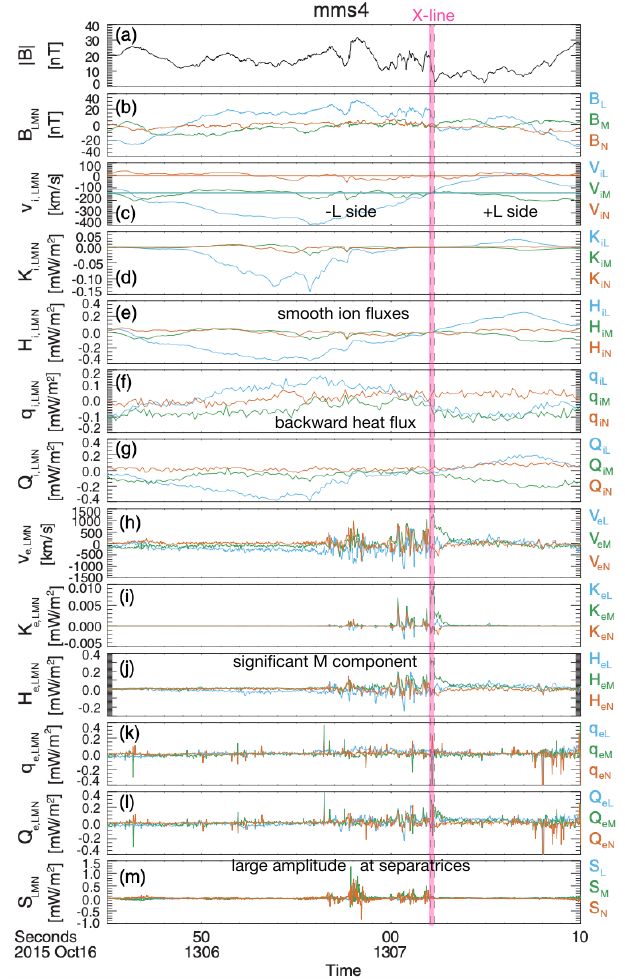} 
\caption {{\bf MMS observations of energy fluxes around an EDR of dayside magnetopause reconnection}, first reported by \cite{burch16a}. Panels (d,e,f,g) show the kinetic, enthalpy, heat, and total ion flux. Panels (i,j,k,l) show equivalent electron fluxes. MMS reveals that the ion fluxes are smoothly varying, whereas the electron fluxes are structured and variable. The MMS data shows the existence of a significant out-of-plane electron energy flux at the X-line (marked by vertical pink lines), discussed in more detail in the text. Modified from \cite{eastwood20a}.} 
\label{4Flux_Eastwood2020}
\end{figure}

Prior to MMS, it was not possible to access the dynamics of the EDR with sufficient resolution to determine the detailed properties of the energy fluxes. However, recent efforts have now enabled such analysis around the EDR of asymmetric reconnection at the dayside magnetopause \citep{eastwood20a}, as shown in Fig.~\ref{4Flux_Eastwood2020}. The study also confirms previous ion-scale observations, such as smoothly varying ion energy fluxes dominated by the ion enthalpy flux in the exhausts, and demonstrates the influence of the large-scale asymmetries introduced by the magnetopause, finding, for example, the peak of the total ion energy flux to be displaced towards the magnetospheric side.

In the case of the ions, the heat flux was observed to be directed back towards the X-line, a feature also seen in symmetric reconnection simulations \citep{SLu18b} and can be explained with non-Maxwellian distributions \citep{hesse18a}. It should, therefore, be emphasized that the ``standard'' decomposition of energy flux (which is the relevant parameter for energy transport considerations) in the presence of non-Maxwellian distributions or specifically collections of beams/multiple populations should be interpreted with care (see. e.g. \citep{goldman20a}).

In the case of the electrons, the results from MMS are more surprising. At the EDR, it would be expected to observe an enhanced out-of-plane kinetic energy flux because of the enhanced current density at the X-line. However, the small mass of the electrons renders $K_e$ negligible. MMS showed that the combination of electron heating at the EDR together with fast electron motion leads to an out-of-plane electron enthalpy flux density, which is comparable to the ion flux densities in the exhaust \citep{eastwood20a}. This may have an important impact on the plasma dynamics, particularly in driving electron-scale instabilities out of the plane. The MMS observations raise further questions about the ultimate source and sink of this out-of-plane energy flux at the EDR, and how it varies along the X-line across the magnetopause or in the magnetotail. Answering this requires a more detailed experimental study of both the energy equation (Eq.~(\ref{energy})) as well as the transfer of energy from fields to particles, the latter being controlled by ${\bf J}\cdot {\bf E}$ as we now discuss.

\newpage

\subsection{Poynting's Theorem and ${\bf J}\cdot {\bf E}$} 
\label{Poynting}
To understand the transfer of energy between electromagnetic fields and particles during magnetic reconnection, we can use Poynting's theorem, 
\begin{equation}
\frac{\partial u_{EM}}{\partial t}+\nabla\cdot{\bf S}=-{\bf J}\cdot{\bf E},
\end{equation}
where $u_{EM}=(B^2+E^2)/8\pi$ is the energy density of electromagnetic fields and ${\bf S} = c({\bf E}\times{\bf B})/4\pi$ is the Poynting vector.
Since the left-hand side of this equation describes the continuity of the electromagnetic energy, the source term on the right-hand side, ${\bf J}\cdot {\bf E}$,  will measure the energy conversion from electromagnetic energy to plasma energy. A similar equation can be written for the particles, and the sum of these two equations reduces to Eq.~(\ref{energy}), i.e., conservation of total energy. The signature of ${\bf J}\cdot {\bf E}$ in anti-parallel, symmetric reconnection is shown in Fig.~\ref{1illustration} based on PIC simulation results \citep{SLu18b}. It is most enhanced within the $d_e$-scale EDR, but positive values (i.e., energy transfers to the plasma) extend further downstream within the outflow exhaust. ${\bf J}\cdot {\bf E}$ can also be decomposed according to the electric field components to assist in understanding the energization mechanisms. The reconnection electric field ($E_y$) is along the reconnection X-line. The electric field in the $x-z$ plane, ${\bf E}_{xz}$, is dominated by the Hall electric fields (${\bf E}_{\rm Hall}={\bf J}\times {\bf B}/enc$), which is set up due to the charge separation between the faster-moving electrons and slower ions, with the $z$ component pointing towards the mid-plane and the $x$ component away from the X-line; its effect is to slow down electrons while speeding up ions. It is, therefore, useful to further decompose ${\bf J}\cdot {\bf E}$ by considering the current density of each species. 

\begin{figure}[h]
\centering
\includegraphics[width=10cm]{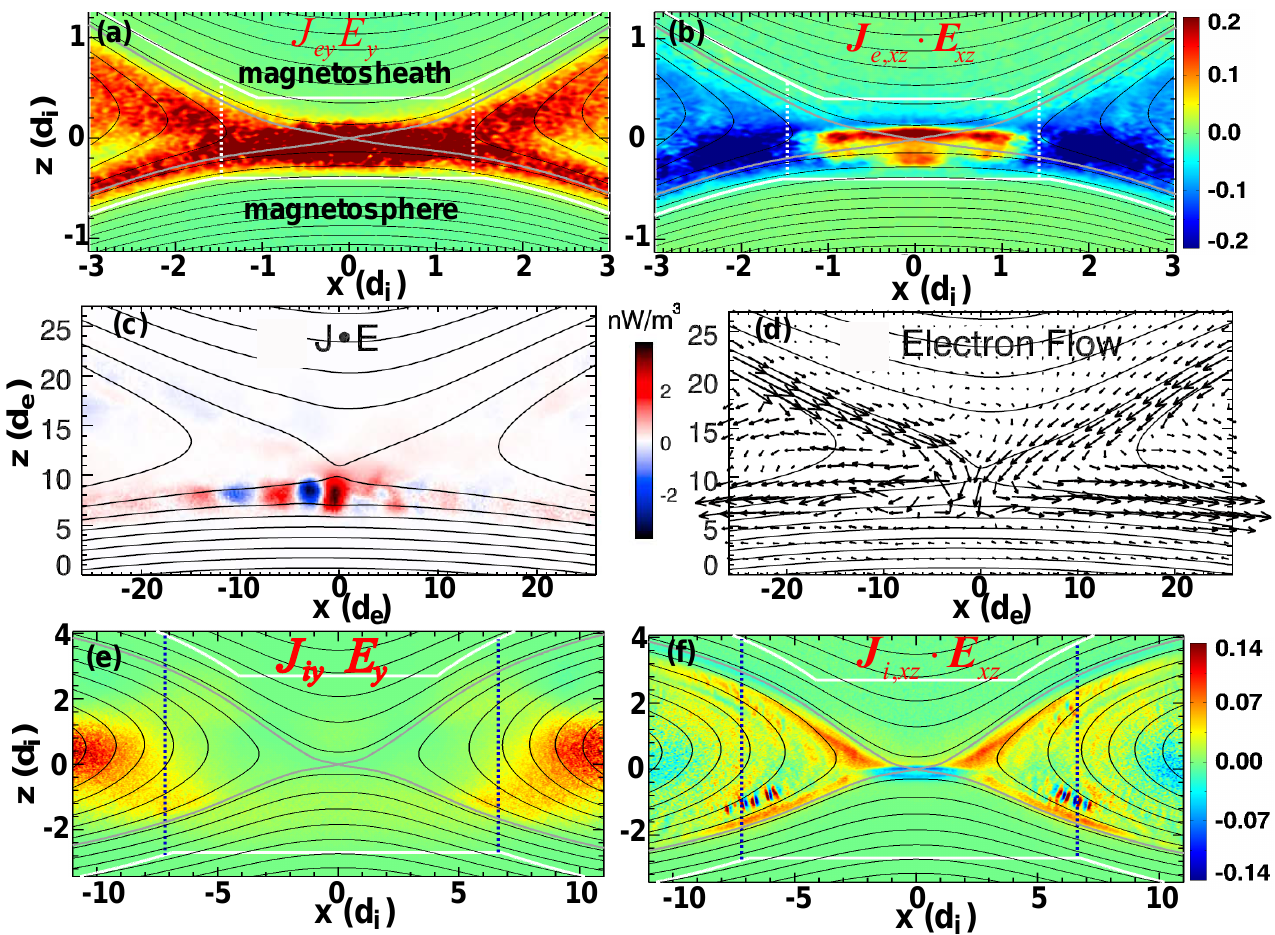} 
\caption {${\bf J}_s\cdot {\bf E}$ {\bf decomposition by the electric field components in PIC simulations} of asymmetric reconnection with zero guide field. Positive $J_{ey}E_y$ dominates the energy conversion to electrons in the EDR (a), and the ${\bf E}_{xz}$ does negative work outside of EDR (b). (c) In certain parameter regimes, ${\bf J}_e\cdot {\bf E}$  can exhibit significant oscillations due to oscillating $V_{ez}$ (d). (e)-(f) energy conversion for ions, where the Hall fields dominate while $E_y$ has a positive contribution in a broad region. Panels (a), (b), (e), (f) are modified from \cite{SWang18a}; (c)-(d) are adapted from \cite{swisdak17a}, with the permission of Wiley.} 
\label{3je}
\end{figure}

Figure~\ref{3je} shows such decomposition for asymmetric reconnection in PIC simulations, and we will discuss the electron energization first, then ion energization in the next paragraph. Around the EDR, $J_{ey} E_y$ is dominantly positive (Fig.~\ref{3je}(a)), such that electrons gain energy from $E_y$ during the meandering motion. ${\bf J}_{e,xz}\cdot {\bf E}_{xz}$ is mainly negative, especially further than $\sim 1 d_i$ downstream of the X-line (Fig.~\ref{3je}(b)). Such features also exist for symmetric reconnection, and ${\bf J}_{e,xz}\cdot {\bf E}_{xz}$ within the EDR has a much smaller amplitude than $J_{ey}E_y$, as, e.g., shown in \cite{payne21a}.  Their study further shows that for a well-developed reconnection layer, a region may develop around the end of the EDR with negative $J_{ey} E_y$ (not shown here) and positive $J_{ex} E_x$, as the electron flow turns from the $y$ to the $x$ direction and the electrons become re-magnetized. For asymmetric reconnection, because the stagnation point is on the magnetospheric side of the X-line (seen from the electron flow lines in Fig.~\ref{3je}(d)) and the magnetosheath-pointing $E_z$ extends to the magnetosheath side of the X-line (e.g., \cite{Shay16,LJChen16a}), a region with positive ${\bf J}_{e,xz}\cdot {\bf E}_{xz}$ exists near the X-line that contributes additional electron energy gain (Fig.~\ref{3je}(b)). The ${\bf J}_{e,xz}\cdot {\bf E}_{xz}$ profile exhibits fluctuations, and in certain parameter regimes, the fluctuations can be more significant and dominate the total ${\bf J}\cdot {\bf E}$ profile (Fig. 5(c), \cite{swisdak17a}). As electrons bounce within the current sheet, they gain a velocity along $x$ by turning around $B_z$ and $B_y$, so most electrons cannot bounce many times at the same $x$ location to maintain similar densities for populations at positive and negative $v_z$, leading to non-zero and fluctuating bulk $V_{ez}$ (Fig.~\ref{3je}(d)) and hence oscillating ${\bf J}\cdot {\bf E}$. Adding a guide field, the amplitude of ${\bf J}_{e,xz}\cdot {\bf E}_{xz}$ in the central EDR becomes smaller compared to $J_{ey} E_y$ \citep{Cassak17,SWang18a}, as electrons have less freedom to bounce across the current sheet.

Within the IDR (but outside the EDR), the electric field ${\bf E}=-{\bf V}_e\times {\bf B}/c$, thus the rate of the electron energy gain ${\bf E}\cdot{\bf J}_e$, vanishes. The rate of the ion energy gain is ${\bf E}\cdot {\bf J}_i =(-{\bf V}_i\times{\bf B}/c+{\bf J}\times{\bf B}/nec)\cdot {\bf J}_i$$={\bf E}_{\rm Hall}\cdot {\bf J}_i$. Thus, the Hall electric field dominates the energization of ions overall (Fig.~\ref{3je}(e)-(f)). Since the Hall field is set up due to the ion-electron decoupling, it plays opposite roles in the energization of two species.  We may understand the Hall field as a pathway to transfer energies between the two species without energy exchange between fields and particles, as quantified by ${\bf E}_{\rm Hall}\cdot {\bf J}=0$. The Hall electromagnetic fields lead to the diverging Poynting flux streamline patterns around the x-line, which is critical in facilitating fast reconnection (Sec.~\ref{localization}, \cite{yhliu22a}). For asymmetric reconnection, ${\bf J}_{i,xz}\cdot {\bf E}_{xz}$  is negative in a localized region near the X-line (Fig.~\ref{3je}(f)), and coincides with the positive ${\bf J}_{e,xz}\cdot {\bf E}_{xz}$ in the similar region (Fig.~\ref{3je}(b)). $J_{iy}E_y$ has a smaller net contribution than ${\bf J}_{i,xz}\cdot {\bf E}_{xz}$ when integrating over the entire diffusion region \citep{SWang18a}. However, $J_{iy}E_y$ dominates close to the X-line and has positive values in a broad region over $z$ due to $J_{iy}$ from the finite Larmor radius effect of meandering ions near the boundary of the ion current layer (Fig.~\ref{3je}(f)). 

 Turning to observations more specifically, \cite{genestreti18c} demonstrated that ${\bf J}_e\cdot {\bf E}\sim J_{ey} E_y$ in a symmetric reconnection EDR, while ${\bf J}_e\cdot {\bf E}$ at dayside asymmetric reconnection exhibits significant fluctuations that may be associated with fluctuating upstream conditions \citep{genestreti22a} beyond the scope of the simulation discussions here. \cite{genestreti18c} and \cite{payne20a} also used MMS to further evaluate the balance between $\partial u_{EM}/\partial t$ and $-{\bf J}\cdot {\bf E}- \nabla\cdot {\bf S}$ in Poynting's theorem for magnetopause and magnetotail EDRs, respectively. The time-derivative term $\partial u_{EM}/\partial t$ in the X-line frame was calculated based on $du_{EM}/dt=\partial u_{EM}/\partial t+{\bf V}_{\rm X}\cdot \nabla u_{EM}$, where $du_{EM}/dt$ is the temporal evolution in the spacecraft frame, and ${\bf V}_{\rm X}$ is the X-line velocity. The results indicate that $\partial u_{EM}/\partial t$ is close to zero near the X-line, but it has more variations away from the X-line. The 2D PIC simulation exhibits an overall consistent pattern (e.g., \cite{payne20a}), while detailed comparisons suggest that events observed by MMS may be at a locally more unsteady state than what is seen in 2D simulations \citep{genestreti18c}. A relevant quantity is ${\bf J}\cdot ({\bf E}+{\bf V}_e\times {\bf B}/c)$, that is the energy conversion rate measured in the local bulk electron frame \citep{zenitani11c}; this useful quantity is often used to identify EDRs.

%\subsection{Further Decomposition and ${\bf V}\cdot (\nabla\cdot {\bf P})$} 
\subsection{Further Decomposition and $({\bf P}\cdot \nabla)\cdot {\bf V}$} 
\label{VGP}
We now discuss the evolution of plasma energy, and we treat this by considering two equations that describe the bulk and thermal forms separately. By dotting the momentum equation with ${\bf V}_s$, one can write the governing equation of bulk flow kinetic energy $u_{{\rm bulk},s}\equiv (1/2)nm_s V_s^2$ in conservative form,
\begin{equation}
\frac{\partial u_{{\rm bulk},s}}{\partial t}+\nabla\cdot {\bf K}_s={\bf J}_s\cdot {\bf E}-{\bf V}_s\cdot (\nabla\cdot {\bf P}_s).
\label{u_bulk}
\end{equation}
Subtracting Eq. (\ref{u_bulk}) from Eq. (\ref{energy}), the equation for the thermal energy $u_{th}\equiv (1/2)\mbox{Tr}({\bf P}_s)$ is obtained,
\begin{equation}
\frac{\partial u_{{\rm th},s}}{\partial t}+\nabla\cdot {\bf H}_s+\nabla\cdot {\bf q}_s={\bf V}_s\cdot (\nabla\cdot {\bf P}_s).
\label{u_th}
\end{equation}
Note that the sum of Eq. (\ref{u_bulk}) and (\ref{u_th}) gives the overall particle energy equation that is the direct counterpart of Poynting's theorem. Interestingly, from the source terms on the right-hand side of these two equations, we can tell that the ${\bf V}_s\cdot (\nabla\cdot {\bf P}_s)$ term re-distributes the energy stored in the bulk and thermal forms. 

\begin{figure}[h]
\centering
\includegraphics[width=8cm]{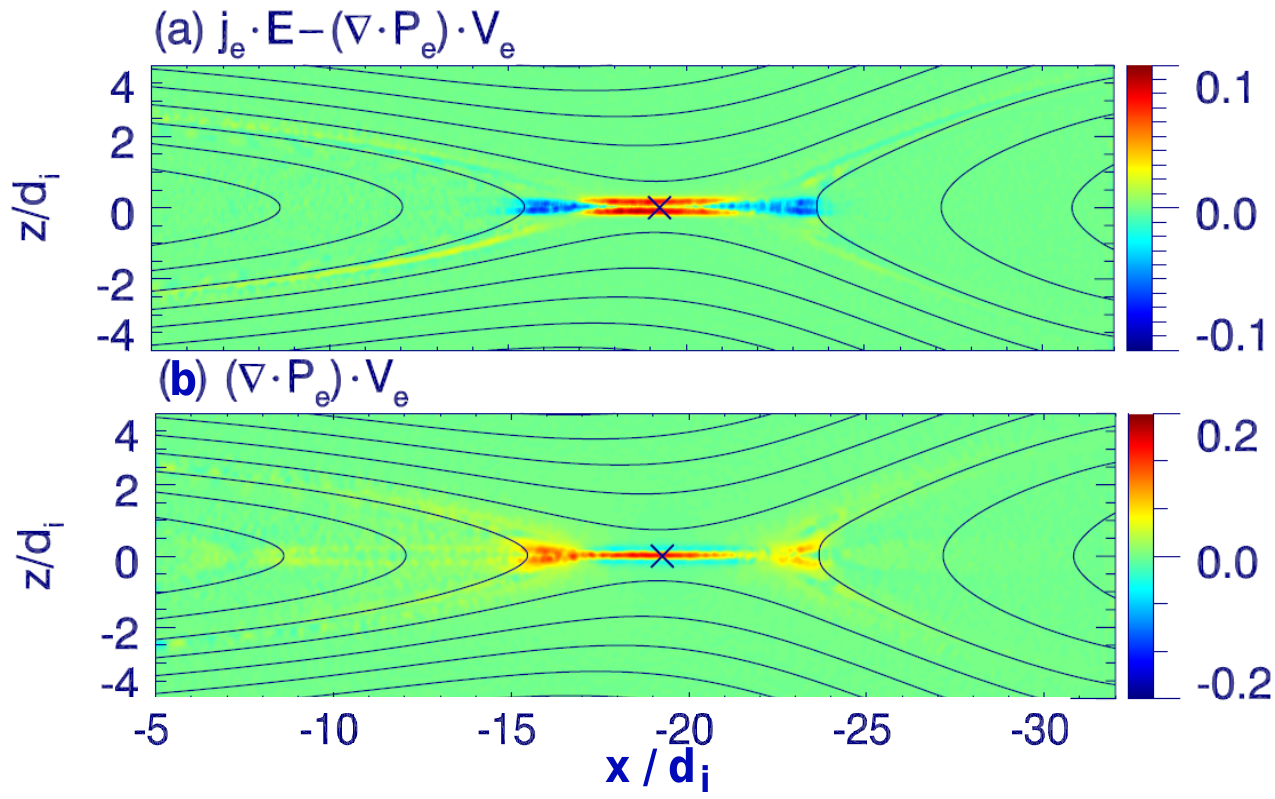} 
\caption {Profiles of the source terms of the electron bulk energy equation ${\bf J}_e\cdot {\bf E}-{\bf V}_e\cdot (\nabla\cdot {\bf P}_e)$  (a) and thermal energy equation ${\bf V}_e\cdot (\nabla\cdot {\bf P}_e)$ in PIC. Reprinted from \cite{SLu18b}, with the permission of AIP Publishing.} 
\label{2JeE_decompose}
\end{figure}

Figure~\ref{2JeE_decompose} shows the source terms on the right-hand side of Eqs. (\ref{u_bulk}) and (\ref{u_th}). Electrons gain both significant bulk (Fig.~\ref{2JeE_decompose}a) and thermal energies (Fig.~\ref{2JeE_decompose}b) within the EDR. Around the end of the EDR, the bulk energy gain is negative, and the thermal energy gain is positive, indicating the conversion from bulk to thermal energies \citep{SLu18b}, which from the kinetic perspective is associated with electron re-magnetization through gyro-turning around the reconnected magnetic field \citep{shuster15a, payne21a}. For ions, comparable bulk and thermal energy gains occur throughout the reconnection region \citep{SLu18b}. The source term for the thermal energy gain can be further decomposed in different forms (e.g., \cite{hesse18a,lapenta20a}), and it has been demonstrated, using both simulations \citep{hesse18a} and MMS observations \citep{holmes21a}, that the ``quasi-viscous'' term associated with the off-diagonal components of the pressure tensor has a dominant contribution in describing electron heating from inflow to outflow regions.

There is another form of the equation that is used to quantify the energy conversion between bulk flow kinetic energy and thermal energy of a species $s$. A brief calculation shows that $\nabla \cdot {\bf H}_s = \nabla \cdot (u_{{\rm th},s} {\bf V}_s) + {\bf V}_s \cdot (\nabla \cdot {\bf P}_s) + ({\bf P}_s \cdot \nabla) \cdot {\bf V}_s$. Substituting this expression into Eq.~(\ref{u_th}) gives
\begin{equation}
\frac{\partial u_{{\rm th},s}}{\partial t}+\nabla \cdot \left(u_{{\rm th},s}{\bf V}_s +{\bf q}_s \right)= - ({\bf P}_s \cdot \nabla) \cdot {\bf V}_s.
\label{u_th2}
\end{equation}
Similarly, rearranging Eq.~(\ref{u_bulk}) gives
\begin{equation}
\frac{\partial u_{{\rm bulk},s}}{\partial t}+\nabla\cdot ({\bf K}_s+ {\bf V}_s\cdot {\bf P}_s)={\bf J}_s\cdot {\bf E}+ ({\bf P}_s \cdot \nabla) \cdot {\bf V}_s.
\label{u_bulk2}
\end{equation}
In this form, it is readily apparent that $-({\bf P}_s \cdot \nabla) \cdot {\bf V}_s$ is a source of internal energy. Since the same term appears in the kinetic energy equation with the opposite sign, this term also describes the conversion between bulk kinetic energy and internal energy. This term, with the minus sign, is called ``the pressure-strain interaction.''
A relevant discussion of this term to reconnection electric field can be found in Sec.~\ref{nature_of_reconnection_electric_field}, where this term is explicitly related to the non-gyrotropic plasma pressure within the EDR \citep{hesse18a}

The pressure-strain interaction has undergone significant study in the MMS era because MMS is uniquely capable of making reliable {\it in situ} measurements of it.  One special property of the pressure-strain interaction is that if one has a closed (infinite or isolated) system, the volume integral over the whole domain ${\rm V}$ of Eq.~(\ref{u_th2}) reveals \citep{Yang2017a,yang_pressurestrain_2022_ApJ},
\begin{equation}
\frac{dU_{{\rm th},s}}{dt} = \int_{\rm V} d^3r [- ({\bf P}_s \cdot \nabla) \cdot {\bf V}_s],
\label{u_th2volume}
\end{equation}
where $U_{{\rm th},s} = \int_{\rm V}u_{{\rm th},s} d^3r$ is the total thermal energy in the system.  Thus, in a collisionless closed system, the volume-integrated pressure-strain interaction is the {\it only} source of thermal energy. It is important to emphasize, however, that it is not the only source of thermal energy locally at any given position \citep{song_forcebalance_2020, Du_energy_2020_PRE,barbhuiya24a}; Eq.~(\ref{u_th2}) shows that other terms (the thermal energy flux and the heat flux) can also change the local internal energy. Also, for systems that are not closed (such as any system in space or astrophysical settings), the other fluxes can lead to a non-zero source or sink for internal energy.
%{\color{red} [making it clearer why not using Eq.~(\ref{u_th}) instead?]}

The pressure-strain interaction has been further decomposed to isolate the key physics causing the change in internal energy.  One decomposition is to write \citep{Yang2017a,Yang2017b}
\begin{equation}
    -({\bf P}_s \cdot \nabla) \cdot {\bf V}_s = -P_s (\nabla \cdot {\bf V}_s) - \boldsymbol{\Pi}_{s} : {\bf D}_s, \label{pressstrain}
\end{equation}
%where ${\cal P}_s = (1/3) {\rm Tr}({\bf P}_s)$ is the effective 
where $P_s \equiv (1/3) {\rm Tr}({\bf P}_s)$ is the effective
(scalar) pressure, $\boldsymbol{\Pi}_s = {\bf P}_s - P_s {\bf I}$ is the deviatoric pressure tensor which describes the departure of the pressure tensor from being isotropic, and $D_{s,jk} = (1/2) (\partial V_{s,j}/\partial r_k + \partial V_{s,k}/\partial r_j) - (1/3) \delta_{jk} (\nabla \cdot {\bf V}_s)$ is the ``traceless strain rate tensor'' which describes the incompressible portion of the flow.  Thus, the first term on the right of Eq.~(\ref{pressstrain}) describes heating or cooling via compression or expansion, and the second term on the right describes incompressible deformation of fluid elements \citep{del_sarto_pressure_2016,Yang2017a,del_sarto_pressure_2018}. The second term can be further decomposed into incompressible deformation due to normal flow and incompressible deformation due to flow shear \citep{Cassak_PiD1_2022}. The latter decomposition can be useful for reconnection studies because it isolates the effect of converging flow and flow shear. The pressure-strain interaction has also been written in magnetic field-aligned coordinates \citep{Cassak_PiD2_2022}, which allows one to determine if the compression, deformation, or shear is parallel or perpendicular to the magnetic field. The pressure-strain interaction and its decompositions have been studied in numerical simulations of magnetic reconnection \citep{Sitnov18,Du18,song_forcebalance_2020,Fadanelli21,Barbhuiya_PiD3_2022} and turbulence \citep{Parashar18,Pezzi19,yang_scale_2019,Hellinger22} and in MMS observations \citep{Chasapis18,Zhong19,Bandyopadhyay20,bandyopadhyay_energy_2021,zhou_measurements_2021,wang21}.  
%See also Chapter~\ref{otherchapter} %for other details.

\subsection{Describing Changes to Internal Moments Beyond Internal Energy} 

Equation~(\ref{energy}) and its subsequent decompositions discussed in Secs.~\ref{Energy_Conservation_and_Energy_Fluxes}-\ref{VGP} follow from the second moment of the Vlasov equation and contain a complete description of the information about energy conversion associated with the number density (the zeroth moment of the distribution function), bulk flow (the first moment), and the thermal energy (the trace of the second moments). However, the distribution function has an infinite number of moments, and the evolution of the other moments is not described by Eq.~(\ref{energy}). For systems close to local thermodynamic equilibrium (LTE), {\it i.e.,} the distribution function is close to being Maxwellian, the other moments are small and their evolution is typically ignored. Any systems of interest for space and astrophysical environments, however, are far from LTE because they are weakly collisional or essentially collisionless. For such systems, it has been unclear how to quantify changes to the higher-order internal moments beyond density, bulk flow, and temperature. The wealth of particle distribution data from MMS, in particular, is now bringing these questions to the fore.

Recently, an approach to quantify changes associated with higher-order internal moments was suggested \citep{Cassak_FirstLaw_2023,barbhuiya24a}.  The key quantity is the so-called relative entropy density $s_{s,{\rm rel}}$, given by
\begin{equation}
    s_{s,{\rm rel}} = -k_B \int f_s \ln \left( \frac{f_s}{f_{sM}} \right) d^3v_s,
\end{equation}
where the integral is over all of the velocity space. Here, $f_{sM}$ is the ``Maxwellianized'' distribution associated with the distribution function $f_s$, given by a Maxwellian distribution with the number density $n_s$, the bulk flow ${\bf V}_s$ and temperature $T_s = (1/3) {\rm Tr}({\bf P}_s)/n_s k_B$ \citep{Grad65}. This quantity is a measure of how non-Maxwellian a distribution function is, with $s_{s,{\rm rel}} = 0$ if $f_s$ is a Maxwellian distribution and it being negative-definite if $f_s$ is anything non-Maxwellian.  

Because $s_{s,{\rm rel}}$ is a measure of how non-Maxwellian a distribution is, its time derivative describes how rapidly the shape of the distribution is changing to become more or less Maxwellian \citep{Cassak_FirstLaw_2023}.  In particular, if $(d/dt) (s_{s,{\rm rel}}/n_s) > 0$, then $f_s$ is becoming more Maxwellian in the comoving (Lagrangian) reference frame, while $(d/dt) (s_{s,{\rm rel}}/n_s) < 0$ implies $f_s$ is becoming less Maxwellian. Dividing by $n_s$ to give the relative entropy per particle is done to not include compression, which is described in the energy equation. It was argued \citep{Cassak_FirstLaw_2023} that scaling $(d/dt) (s_{s,{\rm rel}}/n_s)$ by the temperature gives an effective energy per particle associated with changes to any (and all) of the higher order moments, called the change of relative energy per particle $d{\cal E}_{s,{\rm rel}}$ and given by
\begin{equation}
    \frac{d{\cal E}_{s,{\rm rel}}}{dt} = T_s \frac{d(s_{s,{\rm rel}}/n_s)}{dt}.
\end{equation}
It is important to note that ${\cal E}_{s,{\rm rel}}$ is not a form of energy and, therefore, does not appear in the second moment of the Vlasov equation (Eq.~(\ref{energy})), but it does have the same dimensions and therefore is a quantitative measure of the changes to the higher order internal moments of the distribution that can be directly compared to the standard forms of energy.

Understanding the interplay of changes of all of the higher-order internal moments and the lower-order moments is in its infancy. In a single simulation of reconnection using a particle-in-cell code with 25,600 particles per grid cell, it was shown \citep{Cassak_FirstLaw_2023} that the relative energy change can locally be important or even dominate the changes of internal. How relative energy and entropy depend on ambient plasma parameters and the time evolution of reconnection remains unknown.  Entropy-related quantities have been measured with MMS \citep{Argall22}, but relative entropy has yet to be measured with MMS.

\subsection{Energy Partition between Ions and Electrons} 
Understanding the energy partition between species is also desirable, particularly for understanding reconnection in settings where data may be incomplete (for example, in remote observations or planetary missions where the experimental payload is not optimized for plasma physics). Most incoming electromagnetic energy is eventually converted to the enthalpy flux at locations away from the X-line, as discussed in Sec.~\ref{Energy_Conservation_and_Energy_Fluxes}. In the more general asymmetric reconnection case, the thermal energy gain of each species was modeled as \citep{SWang18a, shay14a} 
\begin{equation}
\frac{\Delta U_{{\rm th},s}}{U_{\rm in}} =\frac{\gamma}{\gamma-1}\frac{T_{{\rm out},s}-T_{{\rm in},s}}{m_i V_{A,{\rm asym}}^2} 
\label{energy_partition}
\end{equation}
where $\Delta U_{{\rm th},s}\equiv\int {\bf J}_s\cdot {\bf E} d^3r$ and $U_{\rm in}$ is the input field energy available for conversion. $T_{{\rm in},s}=(n_1 T_{1,s} B_2+n_2 T_{2,s} B_1)/(n_1 B_2+n_2 B_1)$ represents the inflow temperature, $V_{A,{\rm asym}}=(B_1 B_2/(4\pi m_i )(B_1+B_2)/(n_1 B_2+n_2 B_1 ))^{1/2}$ is the hybrid Alfv\'en speed for asymmetric reconnection (Eq.~(\ref{eq:asymoutflow})), and $\gamma=5/3$ is the ratio of specific heats. $T_{{\rm out},s}\equiv\left<nV_{x,s} T_s\right>/\left<nV_{x,s}\right>$ can be regarded as the outflow temperature averaged over the outflow exhaust with a weighting factor of $nV_{x,s}$. 
The outflow temperature $T_{{\rm out},s}$ was further approximated to be the temperature averaged using $n$ as the weighting factor. 
The observations suggest that the heating rate of $(T_{{\rm out},s}-T_{{\rm in},s})/(m_i V_{A,{\rm asym}}^2 )$ is $1.7\%$ for electrons \citep{phan2013electron} and $13\%$ for ions \citep{phan2014ion}, evaluated using the $n$-weighted $T_{{\rm out},s}$ across exhausts at the far downstream region. PIC simulations show similar results \citep{shay14a}. A test using PIC indicates that the heating rate based on the $n$-weighted $T_{\rm out}$ is nearly constant at varying distances from the X-line \citep{SWang18a}. A caveat is that while $T_{\rm out}$ is insensitive to the approximated forms at distances well away from the EDR, the original $nV_x$-weighted $T_{\rm out}$ should be used around the EDR when using Eq. (\ref{energy_partition}). Close to the EDR, the heating rate is only a few percent while the electron enthalpy flux gain is tens of percent of the incoming Poynting flux as the dominant form of energy conversion. The application of Eq. (\ref{energy_partition}) to a magnetopause reconnection event observed by MMS in between the EDR and IDR boundaries suggests comparable energy partitions between ions and electrons, consistent with the trend predicted by PIC \citep{SWang18a}. We note that the calculation of $nV_{x,s}$-weighted $T_{{\rm out},s}$ has significant uncertainties, so quantitative values need to be treated with caution.

Particle energization mechanisms provide insight into understanding the scaling laws of heating. For magnetized ions or electrons within outflow exhausts, the particles can be roughly described as moving along field lines at $\sim V_A$ in the Alfv\'enic outflow frame. Thus, the superposition of particles from two inflow regions leads to counter-streaming beams in the distribution, so that the effective temperature scales with $V_A^2$ (e.g., \cite{yhliu11a,shay14a}). A parallel potential exists in the exhaust, which modulates the beam speeds, and hence modifies the temperature profile and affects the overall $\Delta T_i/\Delta T_e$ \citep{haggerty2015competition}. Such modulations of the ion beam speeds have been observed by MMS \citep{SWang19a}. 

\begin{figure}[h]
\centering
\includegraphics[width=12cm]{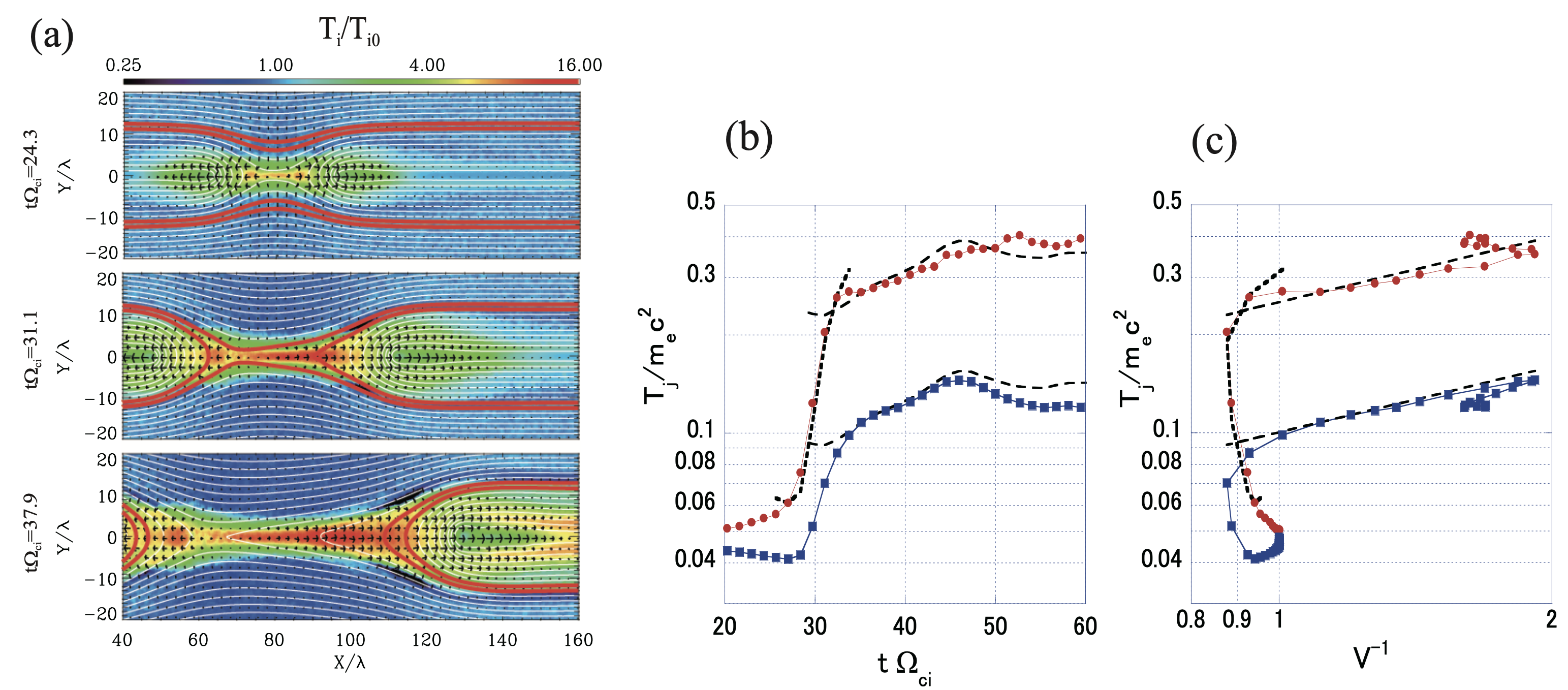} 
\caption {(a) Time evolution of ion temperature in two-dimensional PIC simulation. (b) Time history of ion (red) and electron (blue) temperatures confined in the magnetic flux tube. The dashed lines are the adiabatic relation with $T_s {\rm V}^{2/3} = const.$, and the dotted line is obtained using the effective Ohmic heating model of $T_i/T_e = (m_i/m_e)^{1/4}$ (Eq.~\ref{T_ratio}). (c) Relationship between the reciprocal of the flux tube volume (${\rm V}^{-1}$) and the temperatures ($T_s$) for ions (red) and electrons (blue). The dashed and dotted lines are the same as those in panel (b)). Adapted from \cite{hoshino18a}, reproduced by permission of the AAS.} 
\label{hoshino18}
\end{figure}

Inside the diffusion region, the acceleration by $E_z$ (reconnection electric field in Fig.~\ref{hoshino18}) during the meandering motion was considered to be the primary energization mechanism. \cite{hoshino18a} estimated the ratio of the ion-to-electron temperature enhancement $\Delta T_i/\Delta T_e$ using the effective Ohmic heating rates $E_z J_{zs}{\rm V}_s$ of the two species, where $J_{zs}$ and ${\rm V}_s$ are the ion/electron electric current density and the volume of the diffusion region (e.g., \cite{Coppi66,Coroniti85}), respectively. This leads to  
\begin{equation}
\frac{\Delta \left<T_i\right>_{\rm flux}}{\Delta \left<T_e\right>_{\rm flux}}\simeq \left(\frac{m_i}{m_e}\right)^{1/4}\left( \frac{T_{i0}}{T_{e0}}\right)^{1/4},
\label{T_ratio}
\end{equation}
where $ \left<T_s\right>_{\rm flux}$ is the species temperature averaged over the flux tubes and $T_{s0}$ is the far upstream temperature. This scaling is supported by PIC simulations, as shown in Fig.~\ref{hoshino18}. It is also interesting to note that the averaged temperature follows the adiabatic heating law (i.e., $P/n^{5/3}=\rm const$) during the contraction of reconnected flux tubes (Fig.~\ref{hoshino18}(c)). 

\section{Concluding Remarks and Future Prospects}\
 
In this tutorial review article, we have presented the basics of collisionless magnetic reconnection and highlighted some recent progress in understanding the generalized Ohm's law, the reconnection rate, and the energy conversion around the diffusion region. We also showed supporting evidence from local kinetic simulations and in-situ spacecraft observations, particularly from NASA's ongoing Magnetospheric Multiscale (MMS) mission, which is capable of performing multi-point measurements on electron-kinetic scale physics within Earth's magnetosphere. 

The discussion of theories in this article focuses mostly on 2D models, originating from the classical Sweet-Parker \citep{parker57a,sweet58a} and Petschek \citep{petschek64a} solutions, where the spatial variation scale along the reconnection X-line is assumed to be much longer than the in-plane spatial scale (i.e., near translational invariance along the out-of-plane direction). The difference from these classical models is that we treat the collisionless limit as it is relevant to most applications of reconnection in space plasmas. 
%While shock physics is inherently a one-dimensional (1D) problem, reconnection is inherently a two-dimensional (2D) problem; 
It is interesting to note that an initially three-dimensional (3D), short reconnection X-line within a uniform current sheet is inclined to spread linearly out of the reconnection plane, making the local geometry two-dimensional \citep{huba02b,shay03a,karimabadi04b,lapenta06a,TKMNakamura12a,shepherd12a,TCLi20a}.
One should always be aware that Nature works in three-dimensional spatial space, often accompanied by a high degree of complexity, even if insights from reduced dimensions enable one to extract essential physics.
%; such a simplification is the tactic enjoyed by physicists.
Understanding these 2D limits remains indispensable when seeking to single out the inherently 3D effects that only exist in a 3D system. 
%One needs to carefully discern the ``cause'' from the ``consequences'' of a given question.

In the following, we discuss potential topics critical to the further understanding of magnetic reconnection. From the {\bf local perspective}, a complete theory of the rate of collisionless reconnection, similar to that discussed for standard symmetric reconnection in Sec. \ref{localization}, is still missing for most regimes discussed in Sec.~\ref{reconnection_rate_models} and deserves further development. The descriptions proposed thus far are primarily based on the moments of Vlasov equations, only minimally considering kinetic effects. Kinetic features not included here could play important roles and are highlighted in {\color{blue}Norgren et al. (2024, this issue)}. 
%Theories in this review only look for a steady-state solution, but time-dependent dynamics can also be important. For instance, the plasmoid instability/secondary tearing instability is shown to regulate the diffusion region length, providing the ``localization mechanism'' in the uniform-resistivity MHD model \citep{Huang10, Bhattacharjee09,loureiro07a,Shibata01}. 

The study of the three-dimensional nature of reconnection X-lines is also important. For instance, how does the reconnection X-line orient itself within an asymmetric current sheet \citep{sonnerup74a,swisdak07a,hesse13a,aunai16a, yhliu13a,yhliu18b}? How does an X-line spread \citep{huba03b,shay03a,lapenta06a,TKMNakamura12a,shepherd12a,jain17a,yhliu19a,TCLi20a,Arencibia22}, and what is its minimal length \citep{shay03a,yhliu19a,KHuang20a,pyakurel21a}? Over a larger spatial scale, the implication of these ``local'' 3D X-line properties to global solutions \citep{trattner07a} and 3D MHD reconnection theories \citep{priest03a,pontin22a,TLi21a}, such as fan-spine reconnection, remains unclear. In terms of future observations, the ESA's SMILE mission \citep{raab16a}, NASA's LEXI telescope \citep{walsh20a}, and TRACERS \citep{kletzing19a} will provide the intriguing possibility of imaging magnetopause dynamics for the first time, providing truly novel experimental data for addressing these questions.
 
A full 3D system also introduces additional players that are suppressed in two dimensions, including instabilities \citep{che11a,daughton11a,che11c,roytershteyn12a,yhliu13a}, waves \citep{khotyaintsev19a,yoo20a,Graham2022Nat,ng23a}, and turbulence \citep{ergun18a,stawarz19a}, either at MHD or kinetic scales. The impact of these fundamental plasma processes on the reconnection rate and particle energization will continue to be an active direction for research.
In particular, while the idea of ``anomalous resistivity and transport'' is appealing to MHD modeling of magnetic reconnection \citep{kulsrud01a,SCLin21a,Jimenez22a}, concrete evidence that links this idea to fast reconnection in collisionless plasmas \citep{davidson75a,yoo20a,Graham2022Nat,yoo24a} remains elusive.  
Another relevant open question is the existence of turbulent reconnection that has a thick diffusion region (TDR) on the MHD scales [as theorized in, e.g.,  \cite{Lazarian99}] with well-defined global inflows and outflows. To address this problem in first-principle simulations, it would be ideal to have open boundaries (for more discussion in \citep{ji22a}), avoiding the exaggerated turbulence levels caused by the recycling of particles and magnetic structures from small periodic boundary conditions \citep{yhliu18b}.
More discussion on waves and turbulence associated with reconnection can be found in {\color{blue}  Stawarz et al. (2024, this issue)} and {\color{blue} Graham et al. (2024, this issue)}.

While we only discussed collisionless reconnection in this review, reconnection also occurs in collisional \citep{daughton09a,stanier19a} and partially ionized plasmas \citep{zweibel89a,zweibel11a,murphy15a,LNi18a,jara_almonte19a,LNi20a}. The study of such reconnection could be important to understand the heating in the lower atmosphere of the Sun [e.g., jetlets in \cite{shibata07a,raouafi23a}] or the production of precipitating energetic electrons in ionospheres [e.g., aurora spirals in \cite{KHuang22a}]. The transition from the collisional to collisionless limits could also be critical in understanding the onset problem of reconnection on the Sun, where the initial current sheet can be collisional and as thick as $\sim 10^6 d_i$. The plasmoid instability \citep{biskamp82a,Shibata01,Bhattacharjee09,loureiro07a,pucci14a,comisso16b} in collisional plasmas may enable a transition into the collisionless regime \citep{Shibata01,daughton09a,YMHuang17a,stanier19a,jara_almonte21a}. In contrast, the plasmas in Earth's magnetotail are nearly collisionless, and the onset study of tail reconnection relevant to substorms is concerned more with the stability of a 2D magnetotail geometry \citep{schindler74a,Lembege82,hesse01a,pritchett05a,sitnov09a,yhliu14b,bessho14b}, where the collisionless tearing instability can be suppressed by the magnetic field normal to the current sheet (because electrons remain magnetized). The question in this context is under what conditions reconnection onset can be triggered in collisionless plasmas, enabling the energy release of geomagnetic substorms. More discussion on the onset problem can be found in the Outlook paper {\color{blue} (R. Nakamura, 2024, this issue)}.

From the {\bf global perspective}, it is critical to integrate our understanding of the local reconnection physics into the macroscale phenomena of a given system.
The multiscale nature of reconnection makes this process interesting but also challenging for both first-principles numerical simulations and analytical theory. 
While the theoretical framework in Sec.~\ref{standard_symmetric_reconnection} had coupled the mesoscale MHD region upstream of the ion diffusion region (IDR) to the electron diffusion region (EDR) in the steady state, 
it is assumed that the flux-breaking mechanism within the EDR can ``passively'' match (presumably by thinning) the reconnection electric field dictated by the outer region. A detailed coupling between the EDR particle kinetics (as discussed in {\color{blue} Norgren et al. (2023, this issue)}) and the IDR solution has not yet been established.
On the other hand, it remains unclear how one can couple this locally steady-state solution to the global macroscale in general settings, not to mention the difficulty in modeling the full macro-micro coupling in a time-dependent, dynamical system. Important progress may be made through existing state-of-art simulations [e.g., embedded PIC simulations \citep{daldorff14a,toth16a}] and the development of other novel numerical techniques ({\color{blue} Shay et al., 2023, this issue}). Different macro-micro couplings are summarized in reconnection phase diagrams~\citep{ji11,ji22a} based on the previously mentioned plasmoid instability of long current sheets. During macro-micro coupling, key questions to ask are where and how a current sheet forms in a given global context, when reconnection can be triggered, and how efficiently it works. Such macro-micro coupling, for instance, includes reconnection within Kelvin-Helmholtz vortices [e.g., \cite{TKMNakamura22a, blasl23a}], other MHD-scale instabilities [e.g., \cite{kliem06a,zuccarello14a}], solar wind-magnetosphere coupling [e.g., \cite{dorelli19a}], solar flares [e.g., \cite{wyper17a,dahlin22a}], etc.
 
The growing effort in space exploration [e.g., BepiColombo \citep{heyner21a} at Mercury, Juno \citep{bolton17a} at Jupiter...etc] provides exciting opportunities to perform comparative studies of planetary magnetospheric reconnection; more discussion can be found in \cite{gershman24a} and \cite{fuselier24a}. Both ground and spaceborne remote sensing/imaginary will further enable our understanding of solar flares, the coronal heating problem, and solar wind drivers; more discussion can be found in {\color{blue} Drake et al. (2024, this issue)}. Meanwhile, terrestrial laboratory experiments [e.g., MRX \citep{yamada97a}, TS-3/4 \citep{ono93a}, TREX \citep{olson16a}, PHASMA \citep{PShi22a}, FLARE \citep{ji18a}, etc.] provide invaluable studies performed in a controlled, repeatable manner; more discussion can be found in \cite{ji23a}.  Our hope is that what we have learned from magnetic reconnection within our solar system can also be used to understand other astrophysical objects in the Universe, such as the magnetospheres of stars, exoplanets, and the extreme plasmas near compact objects, including black holes and neutron stars; more discussion can be found in {\color{blue} Guo et al., (2024, this issue)}. Going forward, continuous communication across disciplines will be the key to making breakthroughs in understanding this fundamental, universal plasma process.  

%It is interesting to note that an initially three-dimensional, short reconnection X-line is inclined to spread into a straight line within a uniform current sheet, making the local geometry two-dimensional \citep{yhliu18b,cassak,TKMNakamura}. The three-dimensional effect starts to suppress reconnection if the length of the X-line is shorter than $\mathcal{O}(10 d_i)$.
%While this minimum dimensionality of 2D is required to describe the reconnection process, 3D physics could also play an important role \citep{daughton11a,yhliu13a}, and this is an ongoing active research topic. 
%Having said that, a complete rate theory like that discussed in the regular symmetric reconnection (Sec. \ref{standard_symmetric_reconnection}) is still missing for most regimes discussed here. Understanding these 2D limits is a critical step to contrast potentially important 3D effects in a more general three-dimensional system.

%\paragraph{What we did not discuss--}
%Collisional and partially ionized plasmas. Turbulence Reconnection. Relativistic Reconnection. The macro-micro coupling \citep{hesse20a}... The 3D nature of reconnection X-line (e.g., instabilities, spreading, orientation, short X-line)
%After all, a complete rate theory like that discussed in the regular symmetric reconnection (Sec. \ref{standard_symmetric_reconnection}) is still missing for most regimes.

\section{Acknowledgements}

 YL thanks Shan Wang for the useful discussion on the section of energy conversion within the diffusion region.  YL is supported by NASA MMS Theory and Modeling grant 80NSSC21K2048 and by NASA grant 80NSSC20K1316 and NSF grant 214230. PAC acknowledges support from NASA Grants 80NSSC24K0172, 80NSSC23K0409, 80NSSC22K0323, and 80NSSC19M0146, NSF Grant PHY-2308669, and DoE Grant DE-SC0020294. JPE acknowledges UKRI/STFC grant ST/W001071/1. STR acknowledges support of MCIN/AEI/PRTR 10.13039/501100011033 (Grant PID2020-471 112805GA-I00) and Seneca Agency from Region of Murcia (Grant 21910/PI/22). HJ acknowledges support by NASA under Grants No. NNH15AB29I and 80HQTR21T0105. CN acknowledges support from the Research Council of Norway under Contract No. 300865, and the Swedish National Space Board under Grant 2022-00121.

\bibliographystyle{sn-basic} 
%\bibliography{paper_MMS5}

\end{document}